%% file: stress-diffusion.tex
\documentclass[reqno, a4paper]{amsart}
\usepackage{amsmath}
\usepackage{amssymb}
\usepackage{amsthm}

\usepackage[scale=0.9]{geometry}

\usepackage{natbib}
\usepackage{bibentry} 

\usepackage[english]{babel}
\usepackage[utf8]{inputenc}

\usepackage{subfig}
\usepackage{graphicx}

\usepackage[unicode, breaklinks]{hyperref}
\usepackage[active]{srcltx} 

\usepackage{mathbbol}
\usepackage{bm} 
\usepackage{stmaryrd}
\usepackage{MnSymbol} 

\usepackage{units}
\usepackage{tensor}
\usepackage{accents}

\usepackage{enumitem}

\usepackage[mathlines]{lineno}

\usepackage{xcolor}
\usepackage{mdframed}
\usepackage{newfloat}

\usepackage{booktabs}

\input{vit-prusa-macros-experimental}

\input{stress-diffusion-macros}

\input{stability-macros}

\input{derivation-mechanics-of-fluids-macros}
\input{summary-style}

\numberwithin{equation}{section}

\title{Thermodynamics of viscoelastic rate-type fluids with stress diffusion}

\author{Josef M\'alek}
\date{\today}
\address{
Faculty of Mathematics and Physics\\
Charles University in Prague\\
Sokolovsk\'a 83\\
Praha 8 -- Karl\'{\i}n\\
CZ 186\;75\\
Czech Republic
}
\email{malek@karlin.mff.cuni.cz}

\author{V\'{\i}t Pr\r{u}\v{s}a}
\date{\today}
\address{
Faculty of Mathematics and Physics\\
Charles University in Prague\\
Sokolovsk\'a 83\\
Praha 8 -- Karl\'{\i}n\\
CZ 186\;75\\
Czech Republic
}
\email{prusv@karlin.mff.cuni.cz}

\author{Tom\'a\v{s} Sk\v{r}ivan}
\address{
Faculty of Mathematics and Physics\\
Charles University in Prague\\
Sokolovsk\'a 83\\
Praha 8 -- Karl\'{\i}n\\
CZ 186\;75\\
Czech Republic
}
\email{skrivantomas@seznam.cz}

\author{Endre S\"uli}
\address{
Mathematical Institute\\
University of Oxford\\
Radcliffe Observatory Quarter\\
Woodstock Road\\
Oxford OX2 6GG\\
United Kingdom
}
\email{suli@maths.ox.ac.uk}

\thanks{Josef M\'alek, V\'{\i}t Pr\r{u}\v{s}a and Endre S\"uli acknowledge the support of the Project LL1202 in the programme ERC-CZ funded by the Ministry of Education, Youth and Sports of the Czech Republic.}
\keywords{stress diffusion, viscoelasticity, nonlocal entropy production, nonlocal energy storage, heat equation}
\subjclass[2000]{76A05, 
76A10, 
74A15
}

\begin{document}

\begin{abstract}
We propose thermodynamically consistent models for viscoelastic fluids with a stress diffusion term. In particular, we derive variants of compressible/incompressible Maxwell/Oldroyd-B models with a stress diffusion term in the evolution equation for the extra stress tensor. It is shown that the stress diffusion term can be interpreted either as a consequence of a nonlocal energy storage mechanism or as a consequence of a nonlocal entropy production mechanism, while different interpretations of the stress diffusion mechanism lead to different evolution equations for the temperature. The benefits of the knowledge of the thermodynamical background of the derived models are documented in the study of nonlinear stability of equilibrium rest states. The derived models open up the possibility to study fully coupled thermomechanical problems involving viscoelastic rate-type fluids with stress diffusion.
\end{abstract}

\maketitle

\tableofcontents


\input{text/stress-diffusion-paper-body}

\bibliographystyle{chicago}
\bibliography{vit-prusa}


\end{document}

%% file: vit-prusa-macros-experimental.tex
\DeclareMathOperator{\divergence}{div}

\DeclareMathOperator{\Tr}{Tr}

\newcommand{\reference}{\mathrm{ref}}

\newcommand{\bydefinition}{\mathrm{def}}
\newcommand{\traceless}[1]{{#1}_{\delta}}

\newcommand{\diff}{\mathrm{d}}

\renewcommand{\vec}[1]{\ensuremath{\mathbf{#1}}}

\makeatletter
\@ifpackageloaded{bm}%
{\renewcommand{\vec}[1]{\ensuremath{\bm{#1}}}%
}{%
\relax
}
\makeatother
\newcommand{\tensorq}[1]{\ensuremath{\mathbb{#1}}}      
\newcommand{\tensorc}[1]{\ensuremath{\mathrm{#1}}}      

\newcommand{\transpose}[1]{#1^\top}
\newcommand{\transposei}[1]{#1^{-\top}}
\newcommand{\inverse}[1]{#1^{-1}}

\newcommand{\identity}{\ensuremath{\tensorq{I}}}

\newcommand{\cstress}{\tensorq{T}}

\newcommand{\ecstress}{\tensorq{S}}

\newcommand{\fgrad}{\tensorq{F}}

\newcommand{\rcg}{\tensorq{C}}

\newcommand{\lcg}{\tensorq{B}}

\makeatletter
\@ifpackageloaded{bm}%
{%
}{%

}

\@ifpackageloaded{bm}%
{%
 
}{%

}

\@ifpackageloaded{bm}%
{%
}{%

}
\makeatother

\newcommand{\generictensor}{{\tensorq{A}}}

\newcommand{\gradsym}{\ensuremath{\tensorq{D}}}

\newcommand{\gradvl}{\ensuremath{\tensorq{L}}}

\newcommand{\kdelta}[1]{\tensor{\delta}{#1}}


\newcommand{\ienergy}{\ensuremath{e}} 
\newcommand{\fenergy}{\ensuremath{\psi}} 
\newcommand{\entropy}{\ensuremath{\eta}} 

\newcommand{\temp}{\ensuremath{\theta}} 
\newcommand{\thpressure}{\ensuremath{p_{\mathrm{th}}}} 
\newcommand{\mns}{\ensuremath{m}} 

\newcommand{\nettenergy}{\ensuremath{E}_{\mathrm{tot}}} 
\newcommand{\netentropy}{\ensuremath{S}} 

\newcommand{\cheatvol}{\ensuremath{c_{\mathrm{V}}}}

\newcommand{\efluxc}{\vec{j}_{e}} 

\newcommand{\entfluxc}{\vec{j}_{\entropy}} 

\newcommand{\entprodc}{\xi} 

\newcommand{\pd}[2]{\ensuremath{\frac{\partial {#1}}{\partial {#2}}}}
\newcommand{\ppd}[2]{\ensuremath{\frac{\partial^2 {#1}}{\partial {#2^2}}}}
\newcommand{\dd}[2]{\ensuremath{\frac{\diff {#1}}{\diff {#2}}}}

\newcommand{\ddd}[2]{\ensuremath{\frac{\diff^2 {#1}}{\diff {#2}^2}}}

\newcommand{\fid}[1]{\ensuremath{\accentset{\triangledown}{#1}}}

\newcommand{\lfid}[1]{\ensuremath{\accentset{\meddiamond}{#1}}}

\makeatletter
\@ifpackageloaded{tensor}
{

}{%

}
\makeatother

\makeatletter
\@ifpackageloaded{tensor}
{

}{%

}
\makeatother

\newcommand{\absnorm}[1]{\ensuremath{\left|#1\right|}}

\makeatletter
\@ifundefined{volume}{%
}%
{%
}
\makeatother

\newcommand{\cvolumee}{\diff \mathrm{v}}

\newcommand{\csurfacees}{\diff \mathrm{s}}

\newcommand{\tensortensor}[2]{\ensuremath{#1 \otimes #2}}
\makeatletter
\@ifpackageloaded{MnSymbol} 
{
\newcommand{\tensordot}[2]{\ensuremath{#1 \vdotdot #2}} 
}{%
\newcommand{\tensordot}[2]{\ensuremath{#1 : #2}} 
}
\makeatother
\newcommand{\vectordot}[2]{\ensuremath{#1 \bullet #2}}



%% file: stress-diffusion-macros.tex
\makeatletter
\@ifpackageloaded{MnSymbol} 
{
\newcommand{\tensorddot}[2]{\ensuremath{#1 \vdots #2}} 
}{%
\newcommand{\tensorddot}[2]{\ensuremath{#1 \vdots #2}} 
}
\makeatother

\newcommand{\lcgncc}{\ensuremath{{\tensorc{B}_{\nplacer}}}}

\newcommand{\thpressureNSE}{\ensuremath{p^{\mathrm{NSE}}_{\mathrm{th}}}}
\newcommand{\thpressuredM}{\ensuremath{p^{\mathrm{dM}}_{\mathrm{th}}}}

\newcommand{\thpressuredMTr}{\ensuremath{p^{\mathrm{dM}, \mathrm{A}}_{\mathrm{th}}}}

\newcommand{\thpressuredMdiff}{\ensuremath{p^{\mathrm{dM}, \mathrm{B}}_{\mathrm{th}}}}

\newcommand{\cheatvolNSE}{\ensuremath{\cheatvol^{\mathrm{NSE}}}}

%% file: stability-macros.tex
\newcommand{\tempeq}{\temp_{\mathrm{eq}}}

\newcommand{\tempref}{\ensuremath{\temp_{\reference}}}



%% file: derivation-mechanics-of-fluids-macros.tex
\newcommand{\nplacer}{\kappa_{\mathnormal{p}(\mathnormal{t})}}  

\newcommand{\fgradrng}{\ensuremath{\tensorq{G}}} 

\newcommand{\gradvlrn}{\ensuremath{\gradvl_{\nplacer}}} 
\newcommand{\gradsymrn}{\ensuremath{\gradsym_{\nplacer}}} 

\newcommand{\lcgnc}{\ensuremath{\lcg_{\nplacer}}} 
\newcommand{\rcgnc}{\ensuremath{\rcg_{\nplacer}}} 

\newcommand{\fgradnc}{\ensuremath{\fgrad_{\nplacer}}} 
\newcommand{\fgradrc}{\ensuremath{\fgrad}} 

%% file: summary-style.tex
\DeclareFloatingEnvironment[fileext=sum,placement={!ht},name=Summary]{summaryflt}

\newenvironment{summary}[1][]
    {
    \begin{summaryflt}[tb]
        \begin{summarycont}[#1] 
    }
    {
        \end{summarycont}
        \end{summaryflt}
    }

\mdfdefinestyle{summarystyle}{%
linecolor=black,linewidth=3pt,%
frametitlerule=true,%
frametitlebackgroundcolor=gray!20,
innertopmargin=\topskip,
}

\mdtheorem[style=summarystyle]{summarycont}{Summary}

\newmdenv[style=scholionstyle]{scholion}
\mdfdefinestyle{scholionstyle}{
  backgroundcolor=gray!10,
  roundcorner=10pt
}

%% file: text/stress-diffusion-paper-body.tex
\section{Introduction}
\label{sec:introduction-1}
Standard models for viscoelastic fluids such as the Oldroyd-B model, see~\cite{oldroyd.jg:on}, the Johnson--Segalman model, see~\cite{johnson.mw.segalman.d:model}, or the Giesekus model, see~\cite{giesekus.h:simple}, are frequently modified by the addition of the so-called \emph{stress diffusion term}. Such a term usually takes the form $\Delta \ecstress$, where $\Delta$ denotes the Laplace operator and $\ecstress$ denotes the extra stress tensor, and the term is added to the evolution equation for the extra stress tensor. For example, the governing equations for a viscoelastic fluid described by the incompressible Oldroyd-B model with a stress diffusion term read
\begin{subequations}
  \label{eq:71}
  \begin{align}
    \label{eq:76}
    \divergence \vec{v} &= 0, \\
    \label{eq:72}
    \rho \dd{\vec{v}}{t} &= \divergence \cstress, \\
    \label{eq:73}
    \frac{\nu_1}{\mu} \fid{\ecstress} + \ecstress &= 2 \nu_1 \gradsym + \varepsilon \Delta \ecstress, \\
    \label{eq:75}
    \cstress &= - p \identity + 2 \nu \gradsym + \ecstress,
  \end{align}
\end{subequations}
where $\fid{\ecstress}$ denotes the upper convected time derivative, $\vec{v}$, $\gradsym$, $\rho$, $\cstress$ and $p$ denote the velocity field, the symmetric part of the velocity gradient, the density, the Cauchy stress tensor and the pressure respectively, and $\nu$, $\nu_1$, $\mu$ and $\varepsilon$ are positive material constants. 
  
The presence of the stress diffusion term $\varepsilon \Delta \ecstress$ in the governing equations is important for various reasons. First, it improves, to a certain extent, qualitative mathematical properties of the governing equations, see for example~\cite{el-kareh.aw.leal.lg:existence}, \cite{barrett.jw.boyaval.s:existence} or \cite{chupin.l.martin.s:stationary}. Second, the presence of the diffusive term has a significant impact on the dynamical behaviour predicted by the given system of governing equations. This is, for example, exploited in the modelling of the shear banding phenomenon, see the reviews by~\cite{cates.me.fielding.sm:rheology}, \cite{fielding.sm:complex}, \cite{dhont.jg.briels.w:gradient}, \cite{olmsted.pd:perspectives}, \cite{subbotin.av.malkin.ay.ea:self-organization}, \cite{fardin.ma.ober.tj.ea:potential}, \cite{fardin.m.radulescu.o.ea:stress} and~\cite{divoux.t.fardin.ma.ea:shear}, to name a few.

The presence of the diffusive term can be justified by appealing to a kinetic-theory-based approach to the rheology of dilute polymer solutions. If the classical kinetic-theory-based approach is employed, and if the inhomogeneities of the velocity and stress fields are carefully taken into account, then the diffusive term naturally appears in the evolution equation for the extra stress tensor $\ecstress$, see for example~\cite{el-kareh.aw.leal.lg:existence} and~\cite{bhave.av.armstrong.rc.ea:kinetic}. The weakness of the existing kinetic-theory-based approaches is that they do not provide a full set of mutually consistent governing equations for the fluid of interest. Indeed, the focus is solely on the governing equations for the \emph{mechanical quantities}, while the evolution equation for the \emph{temperature} is not formulated, or even thought of.

The first drawback of the focus on mechanical aspects is that the \emph{thermodynamical consistency} of the models is not justified. In particular, the consistency of the model with the second law of thermodynamics remains questionable. Clearly, the consistency with the second law of thermodynamics can not be analysed without the complete characterisation of the energy transfers in the fluid. Since the energy of a viscoelastic fluid can take the form of the kinetic energy, the thermal energy and the energy accumulated in the ``elastic'' part of the fluid, the appropriate description of the energy transfer mechanisms is conceptually a difficult task. In the case of viscoelastic rate-type fluids \emph{with stress diffusion}, the energy transfer mechanisms are anticipated to be even more complex due to the presence of the stress diffusion term. However, the impact of the stress diffusion term on the energy transfer mechanisms has not yet been analysed.

\begin{figure}[h]
  \centering
  \includegraphics[width=0.4\textwidth]{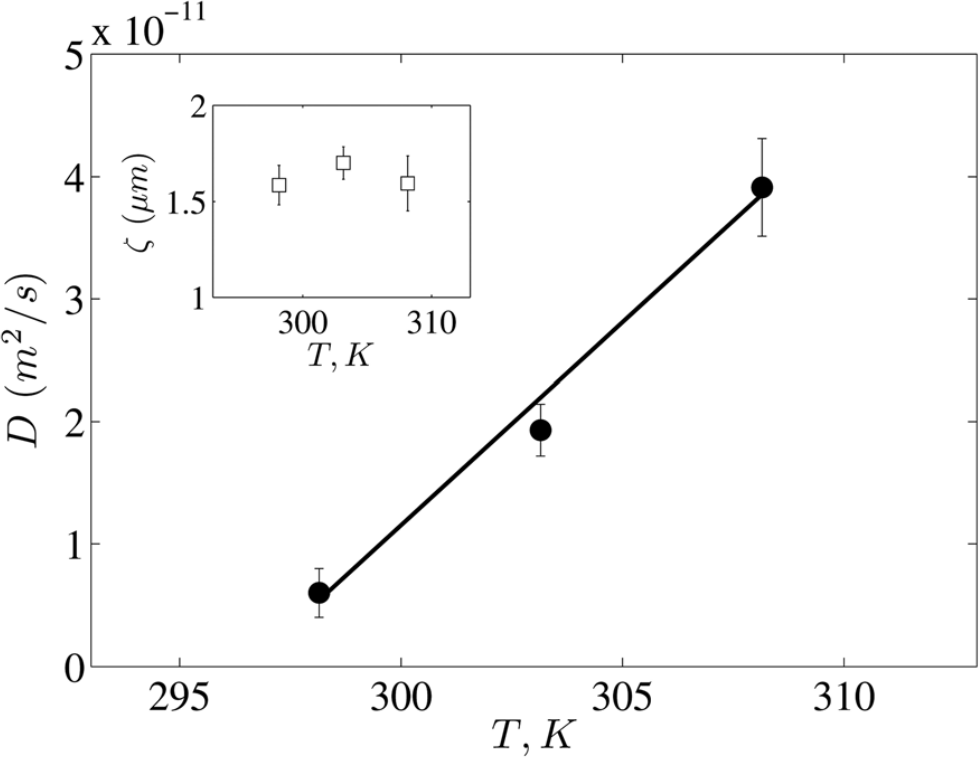}
  \caption{Experimental data for stress diffusion coefficient $D$ versus temperature for the $\mathrm{CTAB}-\mathrm{NaNO}_3$ system, Figure 7 in~\cite{mohammadigoushki.h.muller.sj:flow}. (The authors have considered the diffusive Johnson--Segalman model. The Cauchy stress tensor is given by the formula $\cstress = - p \identity + 2 \eta \gradsym + \tensorq{\Sigma}$, while the extra stress tensor $\tensorq{\Sigma}$ is governed by the equation $ \protect\lfid{\tensorq{\Sigma}} = \frac{2 \mu}{\tau_R} \gradsym - \frac{1}{\tau_R}\tensorq{\Sigma} + D \Delta \tensorq{\Sigma}$. The symbol $\protect\lfid{\tensorq{\Sigma}}$ denotes the Gordon--Schowalter convected derivative, $\eta$, $\mu$ and $\tau_R$ are constant material parameters.) Reproduced from \cite{mohammadigoushki.h.muller.sj:flow} with permission from The Royal Society of Chemistry.}
  \label{fig:temeprature-stress-diffusion}
\end{figure}

The second drawback of the prevailing focus on mechanical aspects is the inability to deal with viscoelastic rate-type fluids with \emph{temperature-dependent material coefficients}. This is a serious drawback, since the response of most viscoelastic rate-type fluids is strongly temperature-dependent. In particular, the stress diffusion coefficient can depend on the temperature, see for example Figure~\ref{fig:temeprature-stress-diffusion}, that reproduces experimental data obtained by~\cite{mohammadigoushki.h.muller.sj:flow}. 
If the material coefficients in the mechanical part of the system depend on the temperature, then the correct prediction of the values of the mechanical quantities requires one to formulate an evolution equation for the temperature. Since the interplay between the thermal energy and other forms of the energy in a viscoelastic fluid with a stress diffusion mechanism can be rather complex, the temperature evolution equation is expected to be markedly different from the standard heat equation used in the case of a compressible/incompressible Navier--Stokes--Fourier fluid. However, the correct temperature evolution equation has not yet been formulated for viscoelastic rate-type fluids \emph{with stress diffusion}.

In what follows we propose a phenomenological \emph{thermodynamical framework for Maxwell/Oldroyd-B type viscoelastic models with a stress diffusion term in the evolution equation for the extra stress tensor}. Both compressible and incompressible variants of the models are considered, and the stress diffusion coefficient is considered to be a temperature-dependent quantity. Using this thermodynamical framework, we derive a complete set of governing equations in the full thermomechanical setting. In particular, we formulate the corresponding temperature evolution equation, and we show that it is compatible with the constitutive relations for the mechanical quantities. Further the whole system of governing equations is shown to be compatible with the second law of thermodynamics.

The stress diffusion term is interpreted in two ways: either as a consequence of a \emph{nonlocal energy storage mechanism}, or as a consequence of a \emph{nonlocal entropy production mechanism}. These different interpretations of the stress diffusion mechanism lead to different evolution equations for the temperature. The derived models open up the possibility to study fully coupled thermomechanical problems involving viscoelastic rate-type fluids with stress diffusion terms.

\section{Phenomenological approach to rate-type viscoelastic models}
\label{sec:phen-appr-visc}
Phenomenological non-equilibrium thermodynamics offers plenty of approaches for the derivation of thermodynamically consistent rate-type viscoelastic models, see for example~\cite{marrucci.g:free}, \cite{leonov.ai:nonequilibrium}, \cite{grmela.m.carreau.pj:conformation}, \cite{mattos.hsc:thermodynamically}, \cite{wapperom.p.hulsen.ma:thermodynamics} or~\cite{dressler.m.edwards.bj.ea:macroscopic}, to name a few. (See especially~\cite{dressler.m.edwards.bj.ea:macroscopic} for a thorough discussion and a rich bibliography on the subject matter.) The approach that is used below conceptually follows the approach that was introduced by~\cite{rajagopal.kr.srinivasa.ar:thermodynamic} and~\cite{rajagopal.kr.srinivasa.ar:on*7}, and was later fruitfully followed in other works, see for example~\cite{rao.ij.rajagopal.kr:thermodynamic}, \cite{kannan.k.rao.ij.ea:thermomechanical} and~\cite{malek.j.rajagopal.kr.ea:on}. The advantages of this approach are that it transparently handles the incompressibility constraint and that it works exclusively on the phenomenological level.

The approach is based on the idea that a material is fully characterised by the way it stores the energy and produces the entropy, which is an idea that was, in a similar form\footnote{See~\cite{rajagopal.kr.srinivasa.ar:on*7} and~\cite{rajagopal.kr.srinivasa.ar:on*8} for comments on the relation of the procedure considered herein to that suggested by~\cite{ziegler.h.wehrli.c:derivation}.}, articulated even earlier, see~\cite{ziegler.h.wehrli.c:derivation}. The first advantage of such an approach is that the energy storage and production mechanisms are specified in terms of two \emph{scalar} quantities, the specific Helmholtz free energy $\fenergy$ and the entropy production $\entprodc$, say. The complex relations between the \emph{tensorial} quantities such as the Cauchy stress tensor $\cstress$ and the symmetric part of the velocity gradient $\gradsym$ then follow from the choice of the formula for the energy and entropy production.

The other ingredient of the approach by~\cite{rajagopal.kr.srinivasa.ar:thermodynamic} is the concept of evolving natural configuration. In the case of viscoelastic fluids this concept in fact reflects the interpretation of the viscoelastic response as a composition of a viscous (dissipative) and an elastic (nondissipative) response. In a sense, this concept can be seen as an extensive generalisation of the  well known one-dimensional spring-dashpot analogues, see~for example~\cite{wineman.as.rajagopal.kr:mechanical}, to a fully nonlinear three-dimensional setting.

The approach by~\cite{rajagopal.kr.srinivasa.ar:thermodynamic} has been successfully used in the derivation of the classical Maxwell and Oldroyd-B viscoelastic models, as well as in the derivation of advanced rate-type viscoelastic models for complex substances such as asphalt, see for example~\cite{krishnan.jm.rajagopal.kr:thermodynamic} or~\cite{malek.j.rajagopal.kr.ea:thermodynamically}. In what follows we deviate from the approach by~\cite{rajagopal.kr.srinivasa.ar:thermodynamic} in two ways.

First, the procedure used by~\cite{rajagopal.kr.srinivasa.ar:thermodynamic} in the isothermal setting is extended to the non-isothermal setting. In particular, the evolution equation for the entropy
\begin{equation}
  \label{eq:1}
  \rho \dd{\entropy}{t} + \divergence \entfluxc = \entprodc,
\end{equation}
is exploited not only in the derivation of the thermodynamically compatible constitutive relations, but also in the formulation of the evolution equation for the temperature. (Here $\entropy$ denotes the specific entropy, $\entfluxc$ stands for the entropy flux, and $\entprodc$ denotes the entropy production, see Section~\ref{sec:evol-equat-entr-1} for details.) Indeed, once the entropy production $\entprodc$ is known, the evolution equation for the temperature is easy to obtain. It suffices to realise that the entropy can be obtained by the differentiation of the free energy with respect to the temperature, and use the explicit formula for the free energy. The application of the chain rule then in fact converts~\eqref{eq:1} into an evolution equation for the temperature. This modification of the original procedure by~\cite{rajagopal.kr.srinivasa.ar:thermodynamic} basically follows the subsequent works by~\cite{rao.ij.rajagopal.kr:thermodynamic}, \cite{kannan.k.rao.ij.ea:thermomechanical}, \cite{kannan.k.rajagopal.kr:thermomechanical}, \cite{prusa.rajagopal.kr:on*1} and especially~\cite{hron.j.milos.v.ea:on}.

Second, the Helmholtz free energy $\fenergy$ that characterises the energy storage mechanisms may in our case include a \emph{higher order gradient} of the tensor $\lcgnc$. (The tensor $\lcgnc$ is the left Cauchy--Green tensor $\lcgnc$ associated with the elastic part of the total mechanical response of the fluid. See Section~\ref{sec:kinematics} for the definition of $\lcgnc$.) The inclusion of a higher gradient into the Helmholtz free energy is a common practice in the theory of elasticity, see for example~\cite{eringen.ac:nonlocal} and the newer contributions by~\cite{fried.e.gurtin.me:tractions}, \cite{polizzotto.c:gradient}, \cite{javili.a..f.ea:geometrically}, \cite{borino.g.polizzotto.c:method} and \cite{silhavy.m:higher}, and references therein. Note, however, that the approach reported below \emph{avoids} the usage of additional concepts such as hyperstress, which is an important notion in higher-order gradient theories of elasticity.

\section{Kinematics of the evolving natural configuration}
\label{sec:kinematics}

Let us apply the proposed phenomenological approach in the case of Maxwell/Oldroyd-B type models. In order to do so, we need to investigate the underlying kinematics\footnote{Most of the calculations and algebraic manipulations used in this section are the same as in~\cite{malek.j.rajagopal.kr.ea:on} and~\cite{hron.j.milos.v.ea:on} where the isothermal and non-isothermal Maxwell/Oldroyd-B models were analysed, respectively. Consequently, we comment in depth only on the results that are specific to the case of models with stress diffusion. The reader interested in the results for standard Maxwell/Oldroyd-B models is referred to~\cite{malek.j.rajagopal.kr.ea:on} and~\cite{hron.j.milos.v.ea:on}. In what follows the symbol $\dd{}{t}$ denotes the material time derivative, $\dd{}{t}=_{\bydefinition} \pd{}{t} + \vectordot{\vec{v}}{\nabla}$, where $\vec{v}$ is the Eulerian velocity field.}
that is motivated by a one-dimensional spring-dashpot model for the behaviour of a Maxwell type viscoelastic fluid, see for example~\cite{wineman.as.rajagopal.kr:mechanical}. The deformation from the initial configuration to the current configuration is \emph{virtually} split into the deformation of the natural configuration and the instantaneous elastic deformation from the natural configuration to the current configuration, see Figure~\ref{fig:viscoelastic-kinematics}. The evolution of the natural configuration is understood as an entropy producing process, while the energy storage ability is attributed to the elastic deformation from the natural configuration to the current configuration, see \cite{rajagopal.kr.srinivasa.ar:thermodynamic} and also~\cite{prusa.rajagopal.kr:on*1}, \cite{malek.j.rajagopal.kr.ea:on} and \cite{malek.j.prusa.v:derivation} for details.

\begin{figure}[h]
  \centering
  \includegraphics[width=0.4\textwidth]{figures-viscoelastic/configurations}
  \caption{Viscoelastic fluid -- kinematics.}
  \label{fig:viscoelastic-kinematics}
\end{figure}

If the total deformation is seen as a composition of the two deformations, then the total deformation gradient $\fgrad$ can be written~as
\begin{equation}
  \label{eq:2}
  \fgradrc = \fgradnc \fgradrng,
\end{equation}
where $\fgradnc$ and $\fgradrng$ are the deformation gradients of the deformation from the reference to the natural configuration and the deformation from the natural configuration to the current configuration. The standard relation $\dd{\fgrad}{t} = \gradvl \fgrad$ between the spatial velocity gradient $\gradvl=_{\bydefinition} \nabla \vec{v}$ and the deformation gradient $\fgrad$, then motivates the introduction of new tensorial quantities~$\gradvlrn$ and $\gradsymrn$ defined as
\begin{equation}
  \label{eq:3}
  \gradvlrn =_{\bydefinition} \dd{\fgradrng}{t} \inverse{\fgradrng}
  ,
  \qquad
  \gradsymrn = _{\bydefinition} \frac{1}{2} \left(\gradvlrn + \transpose{\gradvlrn}\right).
\end{equation}
Using~\eqref{eq:2} and the definition of $\gradvlrn$, the material time derivative of $\fgradnc$ can be expressed as
\begin{equation}
  \label{eq:4}
  \dd{\fgradnc}{t}
  =
  \gradvl \fgradnc - \fgradnc \gradvlrn
  .
\end{equation}
Further, the material time derivative of the left Cauchy--Green tensor $\lcgnc =_{\bydefinition} \fgradnc \transpose{\fgradnc}$ associated with the instantaneous elastic (non-dissipative) response then reads
\begin{equation}
  \label{eq:5}
  \dd{\lcgnc}{t}
  =
  \gradvl \lcgnc
  +
  \lcgnc \transpose{\gradvl}
  -
  2 \fgradnc \gradsymrn \transpose{\fgradnc}.
\end{equation}
Note that the last formula reduces, using the definition of the upper convected derivative,
$
   \fid{\generictensor}=_{\bydefinition} \dd{\generictensor}{t} - \gradvl \generictensor - \generictensor \transpose{\gradvl}
$%
, to the formula
\begin{equation}
  \label{eq:6}
  \fid{\overline{\lcgnc}}
  =
  -
  2 \fgradnc \gradsymrn \transpose{\fgradnc}.
\end{equation}
Further, using the formula for the time derivative of $\lcgnc$, one can show that
\begin{subequations}
  \label{eq:7}
  \begin{align}
    \label{eq:8}
    \dd{}{t} \Tr \lcgnc
    &=
    2 \tensordot{\lcgnc}{\gradsym}
    -
    2 \tensordot{\rcgnc}{\gradsymrn}
    ,
    \\
    \label{eq:9}
    \dd{}{t}
    \left(
      \ln  \left[ \det \lcgnc \right]
    \right)
    &=
    2 \tensordot{\identity}{\gradsym}
    -
    2 \tensordot{\identity}{\gradsymrn}
    ,
  \end{align}
  where $\tensordot{\generictensor}{\tensorq{B}} =_{\bydefinition} \Tr \left(\generictensor \transpose{\tensorq{B}}\right)$ denotes the standard scalar product on the space of matrices. The symbol $\rcgnc =_{\bydefinition} \transpose{\fgradnc} \fgradnc$ denotes the right Cauchy--Green   tensor associated with the non-dissipative response. Further, we see that the material time derivative of the term
  $
  \absnorm{
    \nabla \Tr \lcgnc
  }^2
  $
  reads
  \begin{multline}
  \label{eq:10}
    \dd{}{t}
    \absnorm{
      \nabla \Tr \lcgnc
    }^2
    =
    2
    \divergence
    \left[
      \left[
        \nabla \left( \Tr \lcgnc \right)
      \right]
      \dd{}{t}
      \left(
        \Tr \lcgnc
      \right)
    \right]
    -
    2
    \left[
      \Delta \left( \Tr \lcgnc \right)
    \right]
    \dd{}{t}
    \left(
      \Tr \lcgnc
    \right)
    \\
    -
    2
    \tensordot
    {
      \left[
        \tensortensor{\nabla \left( \Tr \lcgnc \right)}{\nabla \left( \Tr \lcgnc \right)}
      \right]
    }
    {\gradsym}
    .
  \end{multline}
\end{subequations}
The reason for writing the time derivative in the form shown in~\eqref{eq:10} will become clear later on, see Section~\ref{sec:evol-equat-entr-1}.

In principle, the aim is to use the identity $\vectordot{\nabla \phi}{\nabla \psi} = \divergence\left[ \left(\nabla \phi\right) \psi\right] - \left(\Delta \phi\right) \psi$ and move all the gradients to one of the quantities~$\psi$ and~$\phi$ at the expense of adding a flux term. The manipulation is loosely motivated by an analogous manipulation used by~\cite{heida.m.malek.j:on} in the thermodynamics-based derivation of the constitutive relations for compressible Korteweg type fluids.

\section{Helmholtz free energy}
\label{sec:helmh-free-energy}
Let us now make the first step in the thermodynamical procedure by specifying the Helmholtz free energy of the material of interest.
Naturally, the non-dissipative (elastic) part of the response should somehow enter into the formula for the Helmholtz free energy. The quantity that characterises the non-dissipative response is $\lcgnc$. The reason is that the energy storage ability is in finite elasticity theory described in terms of the left Cauchy--Green tensor $\lcg =_{\bydefinition} \fgrad \transpose{\fgrad}$. In our case, however, only a part of the total deformation gradient $\fgrad$ is attributed to a non-dissipative/elastic response. Consequently, only $\lcgnc =_{\bydefinition} \fgradnc \transpose{\fgradnc}$, rather than $\lcg$, plays the role of an additional variable in the formula for the Helmholtz free energy.

If one deals with a compressible/incompressible viscoelastic fluid, then the following \emph{ansatz} for the Helmholtz free energy~$\fenergy$,
\begin{equation}
  \label{eq:11}
  \fenergy
    =_{\bydefinition}
    \widetilde{\fenergy} \left(\temp, \rho\right)
    +
    \frac{\mu}{2\rho}
    \left(
      \Tr \lcgnc
      -
      3
      -
      \ln \det \lcgnc
    \right)
    ,
\end{equation}
where $\mu$ is a constant and $\temp$ denotes the temperature, is known to generate variants of the compressible/incompressible\footnote{In the incompressible case the density $\rho$ in~\eqref{eq:11} is a constant.} Maxwell/Oldroyd-B type models, see for example~\cite{malek.j.rajagopal.kr.ea:on}. (The early investigations of the Helmholtz free energy for standard viscoelastic rate type fluids date back to~\cite{marrucci.g:free}, see also~\cite{dressler.m.edwards.bj.ea:macroscopic} for an extensive bibliography.) The deviation from the standard thermodynamical procedure that leads to the standard Maxwell/Oldroyd-B viscoelastic model is the possibility of the presence of the \emph{gradient} of $\lcgnc$ in the \emph{ansatz} for the Helmholtz free energy. As we have already noted, the inclusion of the higher deformation gradient is motivated by the same idea as in the theory of finite elasticity.

Consequently, we consider two variants of the \emph{ansatz} for the Helmholtz free energy, namely
\begin{subequations}
  \label{eq:helmholtz-free-energy-ansatz-stress-diffusion}
  \begin{align}
    \tag{A}
    \label{eq:12}
    \fenergy
    &=_{\bydefinition}
    \widetilde{\fenergy} \left(\temp, \rho\right)
    +
    \frac{\mu}{2\rho}
    \left(
      \Tr \lcgnc
      -
      3
      -
      \ln \det \lcgnc
    \right)
    +
    \frac{\tilde{\mu}(\temp)}{2\rho}
    \absnorm{ \nabla \Tr \lcgnc}^2,
    \\
    \tag{B}
    \label{eq:13}
    \fenergy
    &=_{\bydefinition}
    \widetilde{\fenergy} \left(\temp, \rho\right)
    +
    \frac{\mu}{2\rho}
    \left(
      \Tr \lcgnc
      -
      3
      -
      \ln \det \lcgnc
    \right)
    ,
  \end{align}
\end{subequations}
where $\mu$ is a constant and $\tilde{\mu}$ is a function of the temperature $\temp$. The material parameter $\tilde{\mu}$ is referred to as the stress diffusion coefficient. Note that, in both cases, the \emph{ansatz} for the free energy has the form
\begin{equation}
  \label{eq:14}
  \fenergy = \widetilde{\fenergy} \left(\temp, \rho\right) + \frac{1}{\rho} \widetilde{\widetilde{\fenergy}} \left(\temp, \lcgnc, \nabla \lcgnc\right).
\end{equation}

As we shall see later, variant~\eqref{eq:12} indeed leads to a viscoelastic rate-type model with a \emph{stress diffusion term}, see Section~\ref{sec:expl-deriv-cons}. In this case the stress diffusion term in the evolution equation for the extra stress tensor is a consequence of the presence of the additional term $\absnorm{ \nabla \Tr \lcgnc}^2$ in the Helmholtz free energy \emph{ansatz}. In this sense, the stress diffusion term is interpreted as a consequence of a non-standard \emph{energy storage mechanism} in the fluid of interest.

On the other hand variant~\eqref{eq:13} is identical to the Helmholtz free energy~\emph{ansatz} for the standard Maxwell/Oldroyd-B fluid. In this case the stress diffusion term in the evolution equation for the extra stress tensor is a consequence of the presence of an additional term in the \emph{entropy production}, see Section~\ref{sec:deriv-cons-relat}. In this sense, the stress diffusion term is not interpreted as a consequence of a non-standard \emph{energy storage mechanism} but rather as a consequence of a non-standard \emph{entropy production mechanism}.

Naturally, the coefficient $\mu$ can also be taken to be temperature-dependent, but we will, for the sake of simplicity of the presentation, consider it to be a constant. The reader interested in the impact of a temperature-dependent coefficient $\mu$ on the dynamics of viscoelastic fluids \emph{without} stress diffusion is referred to~\cite{hron.j.milos.v.ea:on}. The methods presented in~\cite{hron.j.milos.v.ea:on} can be, if necessary, applied even in the case of fluids with stress diffusion.

\section{General evolution equation for the entropy}
\label{sec:evol-equat-entr-1}
The general evolution equation for the specific internal energy $\ienergy$ in a single continuous medium reads
\begin{equation}
  \label{eq:15}
  \rho \dd{\ienergy}{t} = \tensordot{\cstress}{\gradvl} - \divergence \efluxc,
\end{equation}
where $\ienergy$ denotes the specific internal energy, $\cstress$ is the Cauchy stress tensor\footnote{We assume that the Cauchy stress tensor is a symmetric tensor.}, and $\efluxc$ denotes the energy flux. This equation can be exploited in the derivation of the evolution equation for the entropy. Indeed, if the energetic equation of state is given in the form
\begin{equation}
  \label{eq:70}
  \ienergy = \ienergy(\entropy, y_1, \dots, y_n),
\end{equation}
where $\entropy$ denotes the entropy and the symbols $\left\{ y_i \right\}_{i=1}^n$ denote other variables such as the density, then the definition of the temperature $\temp =_{\bydefinition} \pd{\ienergy}{\entropy} (\entropy, y_1, \dots, y_n)$, \eqref{eq:15}, and the chain rule immediately lead to the evolution equation for the entropy. However, we do not follow this path. The energetic equation of state~\eqref{eq:70} is inconvenient from the practical point of view, since it includes the \emph{entropy} as a variable.

From the practical point of view one prefers to work with the specific Helmholtz free energy $\fenergy$ instead of the specific internal energy $\ienergy$. The reason is that the specific Helmholtz free energy $\fenergy$ is a function of the \emph{temperature} $\temp$ and other variables\footnote{The notion of temperature can be subtle in the context of systems out of equilibrium, and in the context of complex fluids, see for example~\cite{grmela.m:letter} and comments therein. We follow a pragmatic approach that is based on the idea that the standard formulae are valid even in the presence of additional field variables, that is we assume that we can still define temperature as $\temp = \pd{\ienergy}{\entropy}(\entropy, y_1, \dots, y_n)$ and so forth. See also comments in Section~\ref{sec:container-flows}.}. The Helmholtz free energy is defined, see~\cite{callen.hb:thermodynamics}, as the Legendre transform of internal energy with respect to the entropy
\begin{equation}
  \label{eq:16}
  \fenergy(\temp, y_1, \dots, y_n)
  =
  \ienergy(\entropy(\temp, y_1, \dots, y_n), y_1, \dots, y_n)
  -
  \temp \entropy(\temp, y_1, \dots, y_n),
\end{equation}
where $\entropy(\temp, y_1, \dots, y_n)$ is a function obtained by solving the equation
\begin{equation}
  \label{eq:123}
  \temp = \pd{\ienergy}{\entropy}(\entropy, y_1, \dots, y_n)
\end{equation}
for the entropy $\entropy$. Note also that~\eqref{eq:16} and~\eqref{eq:123} imply the standard thermodynamical identity
\begin{equation}
  \label{eq:124}
  \pd{\fenergy}{\temp}(\temp, y_1, \dots, y_n) = - \entropy(\temp, y_1, \dots, y_n). 
\end{equation}
Taking the time derivative of~\eqref{eq:16} and using the chain rule then yields a formula for the time derivative of the internal energy $\ienergy$ in terms of the partial derivatives of the Helmholtz free energy $\fenergy$ and the time derivative of the entropy $\entropy$ and other variables
\begin{equation}
  \label{eq:17}
  \dd{\ienergy}{t}
  =
  \temp
  \dd{\entropy}{t}
  +
  \pd{\fenergy}{y_1}
  \dd{y_1}{t}
  +
  \dots
  +
  \pd{\fenergy}{y_n}
  \dd{y_n}{t}
  .
\end{equation}
Using this formula on the left-hand side of~\eqref{eq:15} leads to the evolution equation for the entropy $\entropy$. In particular, if the Helmholtz free energy is given as $\fenergy = \fenergy \left(\temp, \rho, \Tr \lcgnc, \ln \det \lcgnc, \absnorm{\nabla \Tr \lcgnc}^2 \right)$, then one gets
\begin{multline}
  \label{eq:18}
  \rho \temp \dd{\entropy}{t}
  =
  -
  \rho
  \pd{\fenergy}{\rho} \dd{\rho}{t}
  -
  \rho
  \pd{\fenergy}{\Tr \lcgnc} \dd{\Tr \lcgnc}{t}
  -
  \rho
  \pd{\fenergy}{\ln \left[\det \lcgnc\right]} \dd{}{t} \left(\ln \left[ \det \lcgnc \right] \right)
  -
  \rho
  \pd{\fenergy}{\absnorm{\nabla \Tr \lcgnc}^2} \dd{}{t} \absnorm{\nabla \Tr \lcgnc}^2
  \\
  +
  \tensordot{\cstress}{\gradvl} - \divergence \efluxc,
\end{multline}

Now we are in a position to exploit the evolution equations for $\rho$ and $\lcgnc$. The balance of mass equation,
\begin{equation}
  \label{eq:19}
  \dd{\rho}{t} + \rho \divergence \vec{v} = 0,
\end{equation}
allows us to substitute for $\dd{\rho}{t}$. Concerning the time derivatives of the left Cauchy--Green tensor $\lcgnc$ the kinematical considerations imply fomulae~\eqref{eq:7}, which allow us to substitute for the remaining time derivatives on the right-hand side of~\eqref{eq:18}. The first term on the right-hand side of~\eqref{eq:18} can be, thanks to~\eqref{eq:19}, rewritten as
\begin{equation}
  \label{eq:20}
  -
  \rho
  \pd{\fenergy}{\rho} \dd{\rho}{t}
  =
  \rho^2
  \pd{\fenergy}{\rho} \divergence \vec{v}
  ,
\end{equation}
which upon denoting
\begin{subequations}
  \label{eq:thermodynamic-pressure-definition}
  \begin{align}
    \label{eq:pressure-NSE}
    \thpressureNSE &=_{\bydefinition} \rho^2 \pd{\widetilde{\fenergy}}{\rho}, \\
    \label{eq:pressure-dM}
    \thpressuredM &=_{\bydefinition} \rho^2 \pd{\fenergy}{\rho},
  \end{align}
\end{subequations}
yields $\thpressuredM = \thpressureNSE - \widetilde{\widetilde{\fenergy}}$ and
\begin{equation}
  \label{eq:21}
  -
  \rho
  \pd{\fenergy}{\rho} \dd{\rho}{t}
  =
  \thpressuredM \divergence{\vec{v}}.
\end{equation}
(See \eqref{eq:14} for the notation $\widetilde{\widetilde{\fenergy}}$ and $\widetilde{\fenergy}$.) Apparently, the definition \eqref{eq:thermodynamic-pressure-definition} mimics/generalises the classical formula for the relation between the free energy $\fenergy$ and the thermodynamic pressure~$\thpressure$, $\thpressure = \rho^2 \pd{\fenergy}{\rho}$, see for example~\cite{callen.hb:thermodynamics}.

The thermodynamical pressure $\thpressuredM$ has two contributions. The first contribution is $\thpressureNSE$, and it comes from the part of the free energy that is independent of $\lcgnc$. This contribution is tantamount to the classical contribution known from a compressible Navier--Stokes fluid. The second contribution to the thermodynamical pressure $\thpressuredM$ is the term $-\widetilde{\widetilde{\fenergy}}$, which is a contribution due to the ``elastic'' part of the fluid.

Substituting for all time derivatives into~\eqref{eq:18} yields, after some manipulation, the explicit evolution equation for the entropy. This equation can be used in the derivation of the constitutive relations for $\cstress$ and $\efluxc$ from the knowledge of the entropy production $\entprodc$.

\section{Entropy production}
\label{sec:entropy-production}

Following~\cite{rajagopal.kr.srinivasa.ar:thermodynamic} and~\cite{rajagopal.kr.srinivasa.ar:on*7} we are now in a position to specify how the material produces the entropy. In our case, we use for compressible fluids the entropy production $\entprodc$ \emph{ansatz} in the form
\begin{subequations}
  \label{eq:entropy-production-ansatz-stress-diffusion}
  \begin{align}
    \tag{C}
    \label{eq:22}
    \widetilde{\entprodc}
    &=
    \frac{1}{\temp}
    \left(
      \frac{2 \nu + 3 \lambda}{3} \left(\divergence \vec{v}\right)^2
      +
      2 \nu \tensordot{\traceless{\gradsym}}{\traceless{\gradsym}}
      +
      \frac{\nu_1}{2}
      \Tr
      \left(
        \fid{\overline{\lcgnc}}\inverse{\lcgnc}\fid{\overline{\lcgnc}}
      \right)
      +
      \frac
      {
        \kappa \absnorm{\nabla \temp}^2
      }
      {
        \temp
      }
    \right)
    ,
    \\
    \tag{D}
    \label{eq:23}
    \widetilde{\entprodc}
    &=
    \frac{1}{\temp}
    \left(
      \frac{2 \nu + 3 \lambda}{3} \left(\divergence \vec{v}\right)^2
      +
      2 \nu \tensordot{\traceless{\gradsym}}{\traceless{\gradsym}}
      +
      \frac{\mu^2}{2\nu_1}
      \left(
        \Tr \lcgnc
        +
        \Tr \inverse{\lcgnc}
        -
        6
      \right)
      +
      \frac{\mu \tilde{\mu}(\temp)}{2\nu_1}
      \tensorddot{
        \nabla \lcgnc
      }
      {
        \nabla \lcgnc
      }
      +
      \frac
      {
        \kappa \absnorm{\nabla \temp}^2
      }
      {
        \temp
      }
    \right)
    .
  \end{align}
\end{subequations}
The entropy production \emph{ansatz}~\eqref{eq:22} is used for a fluid specified by the Helmholtz free energy \emph{ansatz}~\eqref{eq:12}, while the entropy production \emph{ansatz}~\eqref{eq:23} in used for a fluid specified by the Helmholtz free energy \emph{ansatz}~\eqref{eq:13}. See Section~\ref{sec:expl-deriv-cons} and Section~\ref{sec:deriv-cons-relat} for the rationale behind these formulae.

Having the formulae for the entropy production~$\entprodc$, the derived entropy evolution equation~\eqref{eq:18} can be compared with the general evolution equation for the entropy that takes the form
\begin{equation}
  \label{eq:24}
  \rho \dd{\entropy}{t} + \divergence \entfluxc = \entprodc,
\end{equation}
where $\entfluxc$ denotes the entropy flux. The ``comparison'' of the two equations in principle allows one to identify the sought constitutive relations for $\cstress$ and $\efluxc$. Details are given for each case separately; see Section~\ref{sec:expl-deriv-cons} for the Helmholtz free energy \emph{ansatz}~\eqref{eq:12} and Section~\ref{sec:deriv-cons-relat} for the Helmholtz free energy \emph{ansatz}~\eqref{eq:13}.

\section{Derivation of constitutive relations -- stress diffusion as a consequence of a nonstandard energy storage mechanism}
\label{sec:expl-deriv-cons}
In this section we derive a model for viscoelastic fluids in which the stress diffusion term is attributed to a \emph{nonstandard energy storage mechanism}. The nonstandard energy storage mechanism is characterised by the presence of a gradient (nonlocal) term in the \emph{ansatz} for Helmholtz free energy~\eqref{eq:12}.

\subsection{Evolution equation for the entropy}
\label{sec:evol-equat-entr-3}
If the Helmholtz free energy \emph{ansatz} takes the specific form~\eqref{eq:12}, then the evolution equation for the entropy~\eqref{eq:18} reads
\begin{multline}
  \label{eq:evolution-equation-entropy-var-a}
  \rho \temp \dd{\entropy}{t}
  =
  \bigg\{
  \mns
  +
  \thpressuredMTr
  -
  \frac{\mu}{3}
  \Tr \lcgnc
  +
  \mu
  +
  \frac{\tilde{\mu}}{3}
  \Tr
  \left[
    \tensortensor{\left( \nabla  \Tr \lcgnc \right)}{\left( \nabla  \Tr \lcgnc \right)}
  \right]
  +
  \frac{2}{3}\tilde{\mu}
  \Tr \lcgnc
  \left(
    \Delta
    \Tr \lcgnc
  \right)
  \bigg\}
  \divergence{\vec{v}}
  \\
  +
  \tensordot{
    \bigg\{
    \traceless{\cstress}
    -
    \mu \traceless{\left( \lcgnc \right)}
    +
    \tilde{\mu}
    \traceless{
      \left[
        \tensortensor{\left( \nabla  \Tr \lcgnc \right)}{\left( \nabla  \Tr \lcgnc \right)}
      \right]
    }
    +
    2
    \tilde{\mu}
    \left( \Delta \Tr \lcgnc \right)
    \traceless{
      \left[
        \lcgnc
      \right]
    }
    \bigg\}
  }
  {
    \traceless{\gradsym}
  }
  \\
  +
  \tensordot{
    \bigg\{
    \mu
    \left(\rcgnc - \identity\right)
    -
    2
    \tilde{\mu}
    \left( \Delta \Tr \lcgnc \right)
    \rcgnc
    \bigg\}
  }
  {
    \gradsymrn
  }
  -
  \divergence
  \efluxc
  -
  \tilde{\mu}
  \divergence
  \left[
    \left( \nabla \Tr \lcgnc \right)  \dd{}{t} \left( \Tr \lcgnc \right)
  \right]
  ,
\end{multline}
where the time derivatives in~\eqref{eq:18} have been evaluated using the explicit formulae~\eqref{eq:7}. The thermodynamic pressure defined via~\eqref{eq:pressure-dM} reads
\begin{equation}
  \label{eq:25}
  \thpressuredMTr
  =
  \rho^2
  \pd{\widetilde{\fenergy}}{\rho}
  -
  \frac{\mu}{2}
  \left(
    \Tr \lcgnc
    -
    3
    -
    \ln \det \lcgnc
  \right)
  -
  \frac{\tilde{\mu}}{2}
  \absnorm{\nabla \Tr \lcgnc}^2.
\end{equation}
The superscript $\mathrm{A}$ in $\thpressuredMTr$ indicates that the thermodynamic pressure depends on the choice of the Helmholtz free energy, and that we work with the Helmholtz free energy in the form~\eqref{eq:12}. Further, in~\eqref{eq:evolution-equation-entropy-var-a} we have split the corresponding tensor fields into their spherical and traceless parts,
\begin{equation}
  \label{eq:26}
  \traceless{\cstress} =_{\bydefinition} \cstress - \frac{1}{3} \left( \Tr \cstress \right) \identity, \qquad
  \traceless{\left[ \lcgnc \right]} =_{\bydefinition} \lcgnc - \frac{1}{3} \left( \Tr \lcgnc \right) \identity, \qquad
  \traceless{\gradsym} =_{\bydefinition} \gradsym - \frac{1}{3} \left(\divergence \vec{v}\right) \identity,
\end{equation}
and we have introduced the notation
\begin{equation}
  \label{eq:27}
  \mns =_{\bydefinition} \frac{1}{3} \Tr \cstress
\end{equation}
for the mean normal stress. The splitting allows one to identify the entropy production mechanisms that are associated with different stimuli. The entropy production due to volume changes is captured by the first term on the right-hand side of~\eqref{eq:evolution-equation-entropy-var-a}. The entropy production due to volume preserving mechanisms, such as shearing, is captured by the second term on the right-hand side of~\eqref{eq:evolution-equation-entropy-var-a}.

The first three terms on the right-hand side of~\eqref{eq:evolution-equation-entropy-var-a} have the desired form of the product of thermodynamic affinities and fluxes. It remains to manipulate the last two terms in~\eqref{eq:evolution-equation-entropy-var-a}. We see that
\begin{multline}
  \label{eq:28}
  \frac{1}{\temp}
  \left\{
    \divergence
    \efluxc
    +
    \tilde{\mu}
    \divergence
    \left[
      \left( \nabla \Tr \lcgnc \right)  \dd{}{t} \left( \Tr \lcgnc \right)
    \right]
  \right\}
  =
  \divergence
  \left\{
    \frac{\efluxc + \tilde{\mu} \tensordot{\left( \nabla \Tr \lcgnc \right)}{ \dd{}{t} \left( \Tr \lcgnc \right)}}{\temp}
  \right\}
  \\
  +
  \frac{1}{\temp^2}
  \vectordot{
    \left\{
      \efluxc
      +
      \left[
        \tilde{\mu}
        -
        \temp
        \dd{\tilde{\mu}}{\temp}
      \right]
      \left( \nabla \Tr \lcgnc \right)  \dd{}{t} \left( \Tr \lcgnc \right)
    \right\}
  }
  {
    \nabla \temp
  }
  ,
\end{multline}
which allows us to rewrite the entropy evolution equation~\eqref{eq:evolution-equation-entropy-var-a} in the form
\begin{multline}
  \label{eq:29}
  \rho \dd{\entropy}{t}
  +
  \divergence
  \left\{
    \frac{\efluxc + \tilde{\mu} \tensordot{\left( \nabla \Tr \lcgnc \right)}{ \dd{}{t} \left( \Tr \lcgnc \right)}}{\temp}
  \right\}
  =
  \\
  \frac{1}{\temp}
  \underbrace{
    \bigg\{
    \thpressuredMTr
    -
    \frac{\mu}{3}
    \Tr \lcgnc
    +
    \mu
    +
    \mns
    +
    \frac{\tilde{\mu}}{3}
    \Tr
    \left[
      \tensortensor{\left( \nabla  \Tr \lcgnc \right)}{\left( \nabla  \Tr \lcgnc \right)}
    \right]
    +
    \frac{2}{3}\tilde{\mu}
    \Tr \lcgnc
    \left(
      \Delta
      \Tr \lcgnc
    \right)
    \bigg\}
    \divergence{\vec{v}}
  }_A
  \\
  +
  \frac{1}{\temp}
  \underbrace{
    \tensordot{
      \bigg\{
      \traceless{\cstress}
      -
      \mu \traceless{\left( \lcgnc \right)}
      +
      \tilde{\mu}
      \traceless{
        \left[
          \tensortensor{\left( \nabla  \Tr \lcgnc \right)}{\left( \nabla  \Tr \lcgnc \right)}
        \right]
      }
      +
      2
      \tilde{\mu}
      \left( \Delta \Tr \lcgnc \right)
      \traceless{
        \left[
          \lcgnc
        \right]
      }
      \bigg\}
    }
    {
      \traceless{\gradsym}
    }
  }_B
  \\
  +
  \frac{1}{\temp}
  \underbrace{
    \tensordot{
      \bigg\{
      \mu
      \left(\rcgnc - \identity\right)
      -
      2
      \tilde{\mu}
      \left( \Delta \Tr \lcgnc \right)
      \rcgnc
      \bigg\}
    }
    {
      \gradsymrn
    }
  }_C
  \\
  -
  \frac{1}{\temp^2}
  \underbrace{
    \vectordot{
      \left\{
        \efluxc
        +
        \left[
          \tilde{\mu}
          -
          \temp
          \dd{\tilde{\mu}}{\temp}
        \right]
        \left( \nabla \Tr \lcgnc \right)  \dd{}{t} \left( \Tr \lcgnc \right)
      \right\}
    }
    {
      \nabla \temp
    }
  }_D
  .
\end{multline}
Note that once we have evaluated the time derivative $\dd{}{t} \Tr \lcgnc$, then the last term $D$ can be interpreted in several ways. First, we recall the formula for the time derivative of $\Tr \lcgnc$,
\begin{equation}
  \label{eq:30}
  \dd{}{t} \Tr \lcgnc
  =
  2 \tensordot{\lcgnc}{\gradsym}
  -
  2 \tensordot{\rcgnc}{\gradsymrn}
  =
  2
  \tensordot{
    \traceless{\left[ \lcgnc \right]}
  }
  {
    \traceless{\gradsym}
  }
  +
  \frac{2}{3}
  \left(
    \Tr \lcgnc
  \right)
  \divergence \vec{v}
  -
  2
  \tensordot{\rcgnc}{\gradsymrn},
\end{equation}
see~\eqref{eq:8} and~\eqref{eq:26}. Substituting~\eqref{eq:30} into
$
\vectordot{
  \left\{
    \efluxc
    +
    \left[
      \tilde{\mu}
      -
      \temp
      \dd{\tilde{\mu}}{\temp}
    \right]
    \left( \nabla \Tr \lcgnc \right)  \dd{}{t} \left( \Tr \lcgnc \right)
  \right\}
}
{
  \nabla \temp
}$
yields
\begin{multline}
  \label{eq:31}
  \vectordot{
    \left\{
      \efluxc
      +
      \left[
        \tilde{\mu}
        -
        \temp
        \dd{\tilde{\mu}}{\temp}
      \right]
      \left( \nabla \Tr \lcgnc \right)  \dd{}{t} \left( \Tr \lcgnc \right)
    \right\}
  }
  {
    \nabla \temp
  }
  \\
  =
  \vectordot{\efluxc}{\nabla \temp}
  +
  2 \tilde{\mu}
  \left(
    \tensordot{
      \traceless{\left[ \lcgnc \right]}
    }
    {
      \traceless{\gradsym}
    }
    +
    \frac{1}{3}
    \left(
      \Tr \lcgnc
    \right)
    \divergence \vec{v}
    -
    \tensordot{\rcgnc}{\gradsymrn}
  \right)
  \vectordot{\left(\nabla \Tr \lcgnc\right)}{\nabla \temp}
  \\
  -
  2
  \temp
  \dd{\tilde{\mu}}{\temp}
  \left(
    \tensordot{
      \traceless{\left[ \lcgnc \right]}
    }
    {
      \traceless{\gradsym}
    }
    +
    \frac{1}{3}
    \left(
      \Tr \lcgnc
    \right)
    \divergence \vec{v}
    -
    \tensordot{\rcgnc}{\gradsymrn}
  \right)
  \vectordot{\left(\nabla \Tr \lcgnc\right)}{\nabla \temp}
  .
\end{multline}

The penultimate term in~\eqref{eq:31} can be read as
\begin{multline}
  \label{eq:32}
  2 \tilde{\mu}
  \left(
    \tensordot{
      \traceless{\left[ \lcgnc \right]}
    }
    {
      \traceless{\gradsym}
    }
    +
    \frac{1}{3}
    \left(
      \Tr \lcgnc
    \right)
    \divergence \vec{v}
    -
    \tensordot{\rcgnc}{\gradsymrn}
  \right)
  \vectordot{\left(\nabla \Tr \lcgnc\right)}{\nabla \temp}
  \\
  =
  \begin{cases}
    \vectordot{
      \bigg[
        2 \tilde{\mu}
        \left(
          \tensordot{
            \traceless{\left[ \lcgnc \right]}
          }
          {
            \traceless{\gradsym}
          }
          +
          \frac{1}{3}
          \left(
            \Tr \lcgnc
          \right)
          \divergence \vec{v}
          -
          \tensordot{\rcgnc}{\gradsymrn}
        \right)
        \left(\nabla \Tr \lcgnc\right)
      \bigg]
    }
    {
      \nabla \temp
    }
    ,
    \\
    \\
    2 \tilde{\mu}
    \bigg[
    \vectordot{
      \left(\nabla \Tr \lcgnc\right)
    }
    {
      \nabla \temp
    }
    \bigg]
    \bigg[
      \tensordot{
        \traceless{\left[ \lcgnc \right]}
      }
      {
        \traceless{\gradsym}
      }
      +
      \frac{1}{3}
      \Tr \lcgnc
      \divergence \vec{v}
      -
      \tensordot{
        \rcgnc
      }
      {
        \gradsymrn
      }
    \bigg]
    .
  \end{cases}
\end{multline}
The first option in~\eqref{eq:32} suggests that the term
\begin{equation}
  \label{eq:33}
  2 \tilde{\mu}
  \left(
    \tensordot{
      \traceless{\left[ \lcgnc \right]}
    }
    {
      \traceless{\gradsym}
    }
    +
    \frac{1}{3}
    \left(
      \Tr \lcgnc
    \right)
    \divergence \vec{v}
    -
    \tensordot{\rcgnc}{\gradsymrn}
  \right)
  \vectordot{\left(\nabla \Tr \lcgnc\right)}{\nabla \temp}
\end{equation}
should be interpreted as a flux associated with the affinity $\nabla \temp$, and consequently it should stay as a factor in the term~$D$. On the other hand, the second option in~\eqref{eq:32} suggests that the term should be split as
\begin{equation}
  \label{eq:34}
  2 \tilde{\mu}
  \bigg[
  \vectordot{
    \left(\nabla \Tr \lcgnc\right)
  }
  {
    \nabla \temp
  }
  \bigg]
  \bigg[
  \tensordot{
    \traceless{\left[ \lcgnc \right]}
  }
  {
    \traceless{\gradsym}
  }
  +
  \frac{1}{3}
  \Tr \lcgnc
  \divergence \vec{v}
  -
  \tensordot{
    \rcgnc
  }
  {
    \gradsymrn
  }
  \bigg]
  ,
\end{equation}
and interpreted as a sum of fluxes associated with the affinities $\traceless{\gradsym}$, $\divergence \vec{v}$ and $\gradsymrn$, and hence grouped with the terms $A$, $B$ and $C$ in~\eqref{eq:29}. Similarly, the last term in~\eqref{eq:31} can be read as
\begin{multline}
  \label{eq:35}
  2
  \temp
  \dd{\tilde{\mu}}{\temp}
  \left(
    \tensordot{
      \traceless{\left[ \lcgnc \right]}
    }
    {
      \traceless{\gradsym}
    }
    +
    \frac{1}{3}
    \left(
      \Tr \lcgnc
    \right)
    \divergence \vec{v}
    -
    \tensordot{\rcgnc}{\gradsymrn}
  \right)
  \vectordot{\left(\nabla \Tr \lcgnc\right)}{\nabla \temp}
  \\
  =
  \begin{cases}
    \vectordot{
      \bigg[
        2 \temp \dd{\tilde{\mu}}{\temp}
        \left(
          \tensordot{
            \traceless{\left[ \lcgnc \right]}
          }
          {
            \traceless{\gradsym}
          }
          +
          \frac{1}{3}
          \left(
            \Tr \lcgnc
          \right)
          \divergence \vec{v}
          -
          \tensordot{\rcgnc}{\gradsymrn}
        \right)
        \left(\nabla \Tr \lcgnc\right)
      \bigg]
    }
    {
      \nabla \temp
    }
    ,
    \\
    \\
    2 \temp \dd{\tilde{\mu}}{\temp}
    \bigg[
    \vectordot{
      \left(\nabla \Tr \lcgnc\right)
    }
    {
      \nabla \temp
    }
    \bigg]
    \bigg[
      \tensordot{
        \traceless{\left[ \lcgnc \right]}
      }
      {
        \traceless{\gradsym}
      }
      +
      \frac{1}{3}
      \Tr \lcgnc
      \divergence \vec{v}
      -
      \tensordot{
        \rcgnc
      }
      {
        \gradsymrn
      }
    \bigg]
    .
  \end{cases}
\end{multline}
The first option in~\eqref{eq:35} again suggests that the corresponding term should be interpreted as a flux associated with the affinity~$\nabla \temp$, and consequently it should stay as a factor in the term~$D$. On the other hand, the second option in~\eqref{eq:35} suggests that the corresponding term should be interpreted as a sum of fluxes associated with the affinities $\traceless{\gradsym}$, $\divergence \vec{v}$ and~$\gradsymrn$, and hence grouped with the terms $A$, $B$ and $C$ in~\eqref{eq:29}.

Let us now leave the question of suitable splitting open, and let us formally split both terms using weights $\alpha, \beta \in [0,1]$, that is
\begin{multline}
  \label{eq:36}
  \vectordot{
    \left\{
      \efluxc
      +
      \left[
        \tilde{\mu}
        -
        \temp
        \dd{\tilde{\mu}}{\temp}
      \right]
      \left( \nabla \Tr \lcgnc \right)  \dd{}{t} \left( \Tr \lcgnc \right)
    \right\}
  }
  {
    \nabla \temp
  }
  \\
  =
  \vectordot{
    \left\{
      \efluxc
      +
      \left[
        \alpha
        \tilde{\mu}
        -
        \beta
        \temp
        \dd{\tilde{\mu}}{\temp}
      \right]
      \left( \nabla \Tr \lcgnc \right)  \dd{}{t} \left( \Tr \lcgnc \right)
    \right\}
  }
  {
    \nabla \temp
  }
  \\
  +
  2 \left[(1-\alpha) \tilde{\mu} - (1 - \beta)\temp \dd{\tilde{\mu}}{\temp} \right]
  \bigg[
  \vectordot{
    \left(\nabla \Tr \lcgnc\right)
  }
  {
    \nabla \temp
  }
  \bigg]
  \bigg[
  \tensordot{
    \traceless{\left[ \lcgnc \right]}
  }
  {
    \traceless{\gradsym}
  }
  +
  \frac{1}{3}
  \Tr \lcgnc
  \divergence \vec{v}
  -
  \tensordot{
    \rcgnc
  }
  {
    \gradsymrn
  }
  \bigg]
  .
\end{multline}
Note that the splitting has no influence on the right-hand side of the evolution equation for the entropy~\eqref{eq:29}: the entropy production remains the same regardless of the value of the splitting parameters $\alpha$ and $\beta$.

We conclude that the evolution equation for the entropy reads
\begin{subequations}
  \label{eq:evolution-equation-entorpy-helmholtz-A-split}
  \begin{equation}
    \label{eq:37}
    \rho \dd{\entropy}{t}
    +
    \divergence
    \entfluxc
    =
    \frac{1}{\temp}
    \left\{
      \left( J_{\divergence \vec{v}} \right)
      \divergence{\vec{v}}
      +
      \tensordot{
        \tensorq{J}_{\traceless{\gradsym}}
      }
      {
        \traceless{\gradsym}
      }
      +
      \tensordot{
        \tensorq{J}_{\gradsymrn}
      }
      {
        \gradsymrn
      }
      -
      \vectordot{
        \vec{J}_{\nabla \temp}
      }
      {
        \frac{\nabla \temp}{\temp}
      }
    \right\}
    ,
  \end{equation}
  where the entropy flux has been identified as
  \begin{equation}
    \label{eq:38}
    \entfluxc =_{\bydefinition}
    \frac{\efluxc + \tilde{\mu} \left( \nabla \Tr \lcgnc \right) \dd{}{t} \left( \Tr \lcgnc \right)}{\temp}
    ,
  \end{equation}
  which by virtue of the explicit formula for~$\dd{}{t} \Tr \lcgnc$, see~\eqref{eq:8}, means that the explicit formula for the entropy flux~$\entfluxc$ reads
  \begin{equation}
    \label{eq:39}
    \entfluxc
    =
    \frac{
      \efluxc
      +
      2
      \tilde{\mu}
      \left[
        \left(
          \tensordot{\lcgnc}{\gradsym}
        \right)
        -
        \left(
          \tensordot{\rcgnc}{\gradsymrn}
        \right)
      \right]
      \nabla \Tr \lcgnc
    }
    {
      \temp
    }
    .
  \end{equation}
  The flux terms $J_{\divergence \vec{v}}$, $\tensorq{J}_{\traceless{\gradsym}}$, $\tensorq{J}_{\gradsymrn}$ and $\vec{J}_{\nabla \temp}$ are given by the formulae
  \begin{multline}
    \label{eq:40}
    J_{\divergence \vec{v}}
    =_{\bydefinition}
    \mns
    +
    \thpressuredMTr
    -
    \frac{\mu}{3}
    \Tr \lcgnc
    +
    \mu
    +
    \frac{\tilde{\mu}}{3}
    \Tr
    \left[
      \tensortensor{\left( \nabla  \Tr \lcgnc \right)}{\left( \nabla  \Tr \lcgnc \right)}
    \right]
    +
    \frac{2}{3}\tilde{\mu}
    \Tr \lcgnc
    \left(
      \Delta
      \Tr \lcgnc
    \right)
    \\
    -
    \frac{2}{3}
    \left[(1-\alpha) \tilde{\mu} - (1 - \beta)\temp \dd{\tilde{\mu}}{\temp} \right]
    \bigg[
    \vectordot{
      \left(\nabla \Tr \lcgnc\right)
    }
    {
      \frac{\nabla \temp}{\temp}
    }
    \bigg]
    \Tr \lcgnc
    ,
  \end{multline}
  and
  \begin{multline}
    \label{eq:41}
    \tensorq{J}_{\traceless{\gradsym}}
    =_{\bydefinition}
    \traceless{\cstress}
    -
    \mu \traceless{\left( \lcgnc \right)}
    +
    \tilde{\mu}
    \traceless{
      \left[
        \tensortensor{\left( \nabla  \Tr \lcgnc \right)}{\left( \nabla  \Tr \lcgnc \right)}
      \right]
    }
    +
    2
    \tilde{\mu}
    \left(\Delta \Tr \lcgnc\right)
    \traceless{
      \left(
        \lcgnc
      \right)
    }
    \\
    -
    2
    \left[(1-\alpha) \tilde{\mu} - (1 - \beta)\temp \dd{\tilde{\mu}}{\temp} \right]
    \bigg[
    \vectordot{
      \left(\nabla \Tr \lcgnc\right)
    }
    {
      \frac{\nabla \temp}{\temp}
    }
    \bigg]
    \traceless{
      \left(
        \lcgnc
      \right)
    }
    ,
  \end{multline}
  and
  \begin{align}
    \label{eq:42}
    \tensorq{J}_{\gradsymrn}
    &=_{\bydefinition}
    \mu
    \left(\rcgnc - \identity\right)
    -
    2
    \tilde{\mu}
    \left( \Delta \Tr \lcgnc \right)
    \rcgnc
    +
    2
    \left[(1-\alpha) \tilde{\mu} - (1 - \beta)\temp \dd{\tilde{\mu}}{\temp} \right]
    \bigg[
    \vectordot{
      \left(\nabla \Tr \lcgnc\right)
    }
    {
      \frac{\nabla \temp}{\temp}
    }
    \bigg]
    \rcgnc
    ,
    \\
    \label{eq:43}
    \vec{J}_{\nabla \temp}
    &=_{\bydefinition}
    \efluxc
    +
    \left[
      \alpha
      \tilde{\mu}
      -
      \beta
      \temp
      \dd{\tilde{\mu}}{\temp}
    \right]
    \left( \nabla \Tr \lcgnc \right)  \dd{}{t} \left( \Tr \lcgnc \right)
    ,
  \end{align}
\end{subequations}
where $\alpha$ and $\beta$ are the splitting parameters.

The derived entropy evolution equation~\eqref{eq:37} can now be compared with a generic evolution equation for the entropy that takes the form~\eqref{eq:24}. Following~\cite{rajagopal.kr.srinivasa.ar:thermodynamic} and~\cite{rajagopal.kr.srinivasa.ar:on*7} we are now in a position to specify how the material produces the entropy. We choose a specific $\entprodc$ in~\eqref{eq:24} and compare the desired entropy production $\entprodc$ with the right-hand side of~\eqref{eq:37}, which is the form implied by the choice of the Helmholtz free energy. This leads to the identification of the relations between the flux terms $J_{\divergence \vec{v}}$, $\tensorq{J}_{\traceless{\gradsym}}$, $\tensorq{J}_{\gradsymrn}$ and $\vec{J}_{\nabla \temp}$ and the corresponding kinematical/thermal quantities.

In principle, the ``comparison'' of the desired entropy production $\entprodc$ and the entropy production structure dictated by~\eqref{eq:evolution-equation-entorpy-helmholtz-A-split} can be made more precise by appealing to the maximisation of entropy production procedure,~\cite{rajagopal.kr.srinivasa.ar:on*7}, or to some other thermodynamics-based argument\footnote{The relation of the entropy production maximisation procedure to other procedures used in the development of mathematical models for dissipative processes is discussed in~\cite{jane-cka.a.pavelka.m:gradient}, see also~\cite{grmela.m:externally}.}. In the present case, we limit ourselves to a simple ``comparison'' of the two formulae for the entropy production. This provides us with a simple argument that, in the present case, effectively leads to the same result as more involved thermodynamics-based arguments.

\subsection{Entropy production and constitutive relations}
\label{sec:entr-prod-const}
If the \emph{ansatz} for the Helmholtz free energy is chosen as in~\eqref{eq:12}, then the evolution equation for the entropy is~\eqref{eq:evolution-equation-entorpy-helmholtz-A-split}. On the other hand, the entropy production $\entprodc$ in the material is assumed to take the form
\begin{equation}
  \label{eq:44}
  \widetilde{\entprodc}
  =
  \frac{1}{\temp}
  \left(
    \frac{2 \nu + 3 \lambda}{3} \left(\divergence \vec{v}\right)^2
    +
    2 \nu \tensordot{\traceless{\gradsym}}{\traceless{\gradsym}}
    +
    \nu_1
    \tensordot{
      \left(
        \rcgnc
        \gradsymrn
        +
        \gradsymrn
        \rcgnc
      \right)
    }
    {
        \gradsymrn
    }
    +
    \frac{1}{\temp} \kappa \absnorm{\nabla \temp}^2
  \right)
  ,
\end{equation}
where $\nu$, $\nu_1$, $\lambda$ and $\kappa$ are constants\footnote{See Section~\ref{sec:remarks} for the discussion of physical meanings of the constants.} such that
\begin{equation}
  \label{eq:45}
  \frac{2 \nu + 3 \lambda}{3} \geq 0, \qquad \nu \geq 0, \qquad \nu_1 \geq 0, \qquad \kappa \geq 0.
\end{equation}
The definition of $\rcgnc = \transpose{\fgradnc} \fgradnc$ implies that the entropy production $ \widetilde{\entprodc}$ is non-negative\footnote{Indeed, by virtue of $\rcgnc = \transpose{\fgradnc} \fgradnc$ we see that
$
\tensordot{
  \left(
    \rcgnc
    \gradsymrn
    +
    \gradsymrn
    \rcgnc
  \right)
}
{
  \gradsymrn
}
=
2
\absnorm{\fgradnc \gradsymrn}^2
\geq 0
$.}
.
 Consequently, the second law of thermodynamics is automatically satisfied. Apparently, the \emph{ansatz} for the entropy production is motivated by the knowledge of the entropy production formulae for a compressible Navier--Stokes--Fourier fluid and a Maxwell/Oldroyd-B incompressible viscoelastic fluid, see~\cite{malek.j.rajagopal.kr.ea:on}, \cite{malek.j.prusa.v:derivation} and~\cite{hron.j.milos.v.ea:on}.

If we compare the generic entropy evolution equation~\eqref{eq:24} with desired entropy production $\entprodc =_{\bydefinition} \widetilde{\entprodc}$, that is
\begin{equation}
  \label{eq:46}
  \rho \dd{\entropy}{t} + \divergence \entfluxc
  =
  \frac{1}{\temp}
  \left(
    \frac{2 \nu + 3 \lambda}{3} \left(\divergence \vec{v}\right)^2
    +
    2 \nu \tensordot{\traceless{\gradsym}}{\traceless{\gradsym}}
    +
    \nu_1
    \tensordot{
      \left(
        \rcgnc
        \gradsymrn
        +
        \gradsymrn
        \rcgnc
      \right)
    }
    {
        \gradsymrn
    }
    +
    \frac{1}{\temp} \kappa \absnorm{\nabla \temp}^2
  \right),
\end{equation}
with the entropy evolution equation~\eqref{eq:29} implied by the \emph{ansatz} for the Helmholtz free energy, that is
\begin{equation}
  \label{eq:47}
  \rho \dd{\entropy}{t}
  +
  \divergence
  \entfluxc
  =
  \frac{1}{\temp}
  \left\{
    \left( J_{\divergence \vec{v}} \right)
    \divergence{\vec{v}}
    +
    \tensordot{
      \tensorq{J}_{\traceless{\gradsym}}
    }
    {
      \traceless{\gradsym}
    }
    +
    \tensordot{
      \tensorq{J}_{\gradsymrn}
    }
    {
      \gradsymrn
    }
    -
    \vectordot{
      \vec{J}_{\nabla \temp}
    }
    {
      \frac{\nabla \temp}{\temp}
    }
  \right\}
  ,
\end{equation}
we see that the flux terms $J_{\divergence \vec{v}}$, $\tensorq{J}_{\traceless{\gradsym}}$, $\tensorq{J}_{\gradsymrn}$ and $\vec{J}_{\nabla \temp}$ must satisfy the equalities
\begin{subequations}
  \label{eq:constitutive-relations-ansatz-a-split-a}
  \begin{align}
    \label{eq:48}
    J_{\divergence \vec{v}}
    &=
    \frac{2 \nu + 3 \lambda}{3} \divergence \vec{v}
    ,
    \\
    \label{eq:49}
    \tensorq{J}_{\traceless{\gradsym}}
    &=
    2 \nu \traceless{\gradsym}
    ,
    \\
    \label{eq:50}
    \tensorq{J}_{\gradsymrn}
    &=
    \nu_1
    \left(
      \rcgnc
      \gradsymrn
      +
      \gradsymrn
      \rcgnc
    \right)
    ,
    \\
    \label{eq:51}
    \vec{J}_{\nabla \temp}
    &=
    -\kappa \nabla \temp.
  \end{align}
\end{subequations}

The equations~\eqref{eq:constitutive-relations-ansatz-a-split-a} are in fact the sought constitutive relations for the Cauchy stress tensor $\cstress = \mns \identity + \traceless{\cstress}$ and the energy flux~$\efluxc$; in particular~\eqref{eq:50} is, as we shall show below, a rate-type equation for $\lcgnc$.

Indeed, if we recall the definition of~$J_{\divergence \vec{v}}$, see~\eqref{eq:40}, then equation~\eqref{eq:48} can be solved for the mean normal stress~$\mns$, which yields
\begin{subequations}
  \label{eq:consitutive-relations-ansatz-a-split-a-explicit}
  \begin{multline}
    \label{eq:52}
    \mns
    =
    -
    \thpressuredMTr
    +
    \frac{\mu}{3}
    \Tr \lcgnc
    -
    \mu
    -
    \frac{\tilde{\mu}}{3}
    \Tr
    \left[
      \tensortensor{\left( \nabla  \Tr \lcgnc \right)}{\left( \nabla  \Tr \lcgnc \right)}
    \right]
    -
    \frac{2}{3}\tilde{\mu}
    \Tr \lcgnc
    \left(
      \Delta
      \Tr \lcgnc
    \right)
    \\
    +
    \frac{2}{3}
    \left[(1-\alpha) \tilde{\mu} - (1 - \beta)\temp \dd{\tilde{\mu}}{\temp} \right]
    \bigg[
    \vectordot{
      \left(\nabla \Tr \lcgnc\right)
    }
    {
      \frac{\nabla \temp}{\temp}
    }
    \bigg]
    \Tr \lcgnc
    +
    \frac{2 \nu + 3 \lambda}{3} \divergence \vec{v}
    .
  \end{multline}
  The constitutive relation for the traceless part of the Cauchy stress tensor $\traceless{\cstress}$ can be, by virtue of the definition of the flux term~$\tensorq{J}_{\traceless{\gradsym}}$, see~\eqref{eq:41}, read from~\eqref{eq:49} as
  \begin{multline}
    \label{eq:53}
    \traceless{\cstress}
    =
    2 \nu \traceless{\gradsym}
    +
    \mu \traceless{\left( \lcgnc \right)}
    -
    \tilde{\mu}
    \traceless{
      \left[
        \tensortensor{\left( \nabla  \Tr \lcgnc \right)}{\left( \nabla  \Tr \lcgnc \right)}
      \right]
    }
    -
    2
    \tilde{\mu}
    \left(\Delta \Tr \lcgnc\right)
    \traceless{
      \left(
        \lcgnc
      \right)
    }
    \\
    +
    2
    \left[(1-\alpha) \tilde{\mu} - (1 - \beta)\temp \dd{\tilde{\mu}}{\temp} \right]
    \bigg[
    \vectordot{
      \left(\nabla \Tr \lcgnc\right)
    }
    {
      \frac{\nabla \temp}{\temp}
    }
    \bigg]
    \traceless{
      \left(
        \lcgnc
      \right)
    }
    .
  \end{multline}
  The last equation~\eqref{eq:51} yields, by virtue of the definition~\eqref{eq:43}, the following formula for the energy flux $\efluxc$:
  \begin{equation}
    \label{eq:54}
    \efluxc
    =
    -
    \kappa \nabla \temp
    -
    \left[
      \alpha
      \tilde{\mu}
      -
      \beta
      \temp
      \dd{\tilde{\mu}}{\temp}
    \right]
    \left( \nabla \Tr \lcgnc \right)  \dd{}{t} \left( \Tr \lcgnc \right)
    .
\end{equation}

Finally, equation~\eqref{eq:50} and the definition of the flux term $\tensorq{J}_{\gradsymrn}$, see~\eqref{eq:42}, imply that $\rcgnc$ and $\gradsymrn$ commute. (The proof is the same as in the classical case $\tilde{\mu}=0$, see \cite{rajagopal.kr.srinivasa.ar:thermodynamic}.) Once we know that $\rcgnc$ and~$\gradsymrn$ commute, we can multiply~\eqref{eq:50} by $\transpose{\fgradnc}$ from the right and by $\transposei{\fgradnc}$ from the left, which yields
\begin{multline}
  \label{eq:55}
  \transposei{\fgradnc}
  \left\{
    \mu
    \left(\rcgnc - \identity\right)
    -
    2
    \tilde{\mu}
    \left( \Delta \Tr \lcgnc \right)
    \rcgnc
    +
    2
    \left[(1-\alpha) \tilde{\mu} - (1 - \beta)\temp \dd{\tilde{\mu}}{\temp} \right]
    \bigg[
    \vectordot{
      \left(\nabla \Tr \lcgnc\right)
    }
    {
      \frac{\nabla \temp}{\temp}
    }
    \bigg]
    \rcgnc
  \right\}
  \transpose{\fgradnc}
  \\
  =
  2
  \nu_1
  \fgradnc
  \gradsymrn
  \transpose{\fgradnc}
  .
\end{multline}
Now we recall the definitions $\rcgnc =_{\bydefinition} \transpose{\fgradnc} \fgradnc$ and $\lcgnc =_{\bydefinition} \fgradnc \transpose{\fgradnc}$ and the fact that the right-hand side can be identified with the upper convected derivative of $\lcgnc$, see~\eqref{eq:6}. This yields the following evolution equation for $\lcgnc$:
\begin{equation}
  \label{eq:56}
  \nu_1 \fid{\overline{\lcgnc}}
  +
  \mu \left(\lcgnc - \identity\right)
  =
  2
  \tilde{\mu}
  \left( \Delta \Tr \lcgnc \right)
  \lcgnc
  -
  2
  \left[(1-\alpha) \tilde{\mu} - (1 - \beta)\temp \dd{\tilde{\mu}}{\temp} \right]
  \bigg[
  \vectordot{
    \left(\nabla \Tr \lcgnc\right)
  }
  {
    \frac{\nabla \temp}{\temp}
  }
  \bigg]
  \lcgnc
  .
\end{equation}
\end{subequations}
Once we have this equation, we can determine the evolution of $\lcgnc$, which is the quantity that appears in the formulae for the Cauchy stress tensor $\cstress$ and the energy flux $\efluxc$, see~\eqref{eq:52}--\eqref{eq:54}. This completes the formulation of the constitutive relations for $\cstress$ and $\efluxc$.

\subsection{Entropy production in terms of the primitive variables}
\label{sec:entr-prod-terms}
The entropy production specified in~\eqref{eq:44} contains the quantities~$\rcgnc$ and $\gradsymrn$ that are not convenient since we do not have explicit evolution equations for these variables. However, the term
$
\tensordot{
  \left(
    \rcgnc
    \gradsymrn
    +
    \gradsymrn
    \rcgnc
  \right)
}
{
  \gradsymrn
}
$
can be easily rewritten in terms of $\lcgnc$ and its time derivatives. Indeed, the critical term reads
$
\tensordot{
  \left(
    \rcgnc
    \gradsymrn
    +
    \gradsymrn
    \rcgnc
  \right)
}
{
  \gradsymrn
}
=
\absnorm{\fgradnc \gradsymrn}^2
$
,
which can be converted, by virtue of~\eqref{eq:6}, to
\begin{equation}
  \label{eq:57}
  \tensordot{
    \left(
      \rcgnc
      \gradsymrn
      +
      \gradsymrn
      \rcgnc
    \right)
  }
  {
    \gradsymrn
  }
  =
  \frac{1}{4}
  \Tr
  \left(
    \fid{\overline{\lcgnc}}\inverse{\lcgnc}\fid{\overline{\lcgnc}}
  \right)
  .
\end{equation}
(See also~\cite{hron.j.milos.v.ea:on} for a similar manipulation in the case of classical viscoelastic rate-type models.) Further, one can exploit the evolution equation for $\lcgnc$, see~\eqref{eq:56}, and convert the right-hand side of~\eqref{eq:57} to a form that does not include time derivatives,
\begin{subequations}
  \label{eq:58}
\begin{equation}
  \label{eq:59}
  \tensordot{
    \left(
      \rcgnc
      \gradsymrn
      +
      \gradsymrn
      \rcgnc
    \right)
  }
  {
    \gradsymrn
  }
  =
  \frac{1}{4}
  \Tr
  \left(
    \generictensor
    \inverse{\lcgnc}
    \generictensor
  \right)
  ,
\end{equation}
where
\begin{equation}
  \label{eq:60}
  \generictensor
  =_{\bydefinition}
  \frac{2 \tilde{\mu}}{\nu_1}
  \left(
    \Delta \Tr \lcgnc
  \right)
  \lcgnc
  -
  \frac{2}{\nu_1}
  \left[(1-\alpha) \tilde{\mu} - (1 - \beta)\temp \dd{\tilde{\mu}}{\temp} \right]
  \bigg[
  \vectordot{
    \left(\nabla \Tr \lcgnc\right)
  }
  {
    \frac{\nabla \temp}{\temp}
  }
  \bigg]
  \lcgnc
  -
  \frac{\mu}{\nu_1}
  \left(
    \lcgnc - \identity
  \right)
  .
\end{equation}
\end{subequations}

Using~\eqref{eq:57}, we see that the entropy production~\eqref{eq:44} can be rewritten as
\begin{equation}
  \label{eq:61}
  \widetilde{\entprodc}
  =
  \frac{1}{\temp}
  \left(
    \frac{2 \nu + 3 \lambda}{3} \left(\divergence \vec{v}\right)^2
    +
    2 \nu \tensordot{\traceless{\gradsym}}{\traceless{\gradsym}}
    +
    \frac{\nu_1}{2}
    \Tr
    \left(
      \fid{\overline{\lcgnc}}\inverse{\lcgnc}\fid{\overline{\lcgnc}}
    \right)
    +
    \frac
    {
      \kappa \absnorm{\nabla \temp}^2
    }
    {
      \temp
    }
  \right)
  ,
\end{equation}
hence the entropy production contains the same quantities as the constitutive relations for the Cauchy stress tensor, the energy flux and the evolution equation for $\lcgnc$. If necessary, identity~\eqref{eq:59} can be used as well, which would yield yet another reformulation of the entropy production. In particular, if $\tilde{\mu}=0$, one would get
\begin{equation}
  \label{eq:62}
  \frac{\nu_1}{2}
  \Tr
  \left(
    \fid{\overline{\lcgnc}}\inverse{\lcgnc}\fid{\overline{\lcgnc}}
  \right)
  =
  \frac{\mu^2}{2 \nu_1}
  \left(
    \Tr \lcgnc
    +
    \Tr \inverse{\lcgnc}
    -
    6
  \right)
  .
\end{equation}

\subsection{Evolution equation for the temperature}
\label{sec:evol-equat-temp}
Having expressed the entropy production in terms of the primitive variables $\vec{v}$, $\temp$, $\rho$ and $\lcgnc$, we are ready to formulate the evolution equation for the temperature. The temperature evolution equation follows from the entropy evolution equation. The entropy $\entropy$ is given as the derivative of the Helmholtz free energy~$\fenergy$ with respect to the temperature,
\begin{equation}
  \label{eq:63}
  \entropy = - \pd{\fenergy}{\temp}(\temp, \rho, \lcgnc, \nabla \Tr \lcgnc).
\end{equation}
Using the decomposition~\eqref{eq:14} and the chain rule, we see that the Helmoltz free energy \emph{ansatz}~\eqref{eq:12} leads to
\begin{equation}
  \label{eq:64}
  \dd{\entropy}{t}
  =
  \dd{}{t}
  \left(
    -
    \pd{
      \widetilde{\fenergy}
    }{\temp}
    -
    \frac{1}{\rho} \pd{\widetilde{\widetilde{\fenergy}}}{\temp}
  \right)
  =
  \left(
    -
    \ppd{
      \widetilde{\fenergy}
    }{\temp}
    -
    \frac{1}{\rho}
    \ppd{
      \widetilde{\widetilde{\fenergy}}
    }{\temp}
  \right)
  \dd{\temp}{t}
  +
  \pd{}{\temp}
  \left(
    -
    \pd{
      \widetilde{\fenergy}
    }{\rho}
    +
    \frac{\widetilde{\widetilde{\fenergy}}}{\rho^2}
  \right)
  \dd{\rho}{t}
  -
  \frac{1}{\rho}
  \pd{^2\widetilde{\widetilde{\fenergy}}}{\temp \partial \absnorm{\nabla \Tr \lcgnc}^2}
  \dd{}{t} \absnorm{\nabla \Tr \lcgnc}^2
  .
\end{equation}
Introducing the notation
\begin{equation}
  \label{eq:65}
  \cheatvolNSE
  =_{\bydefinition}
  -
  \temp
  \ppd{
    \widetilde{\fenergy}
  }{\temp}
  ,
\end{equation}
and using the definition of the thermodynamic pressure $\thpressuredMTr$, see~\eqref{eq:pressure-dM}, we can rewrite~\eqref{eq:64} as
\begin{equation}
  \label{eq:66}
  \dd{\entropy}{t}
  =
  \frac{1}{\temp}
  \left(
    \cheatvolNSE
    -
    \frac{\temp}{2 \rho}
    \ddd{\tilde{\mu}}{\temp}
    \absnorm{\nabla \Tr \lcgnc}^2
  \right)
  \dd{\temp}{t}
  -
  \frac{1}{\rho^2}
  \pd{\thpressuredMTr}{\temp}
  \dd{\rho}{t}
  -
  \frac{1}{2\rho}
  \dd{\tilde{\mu}}{\temp}
  \dd{}{t} \absnorm{\nabla \Tr \lcgnc}^2.
\end{equation}
This provides us with a relation between the time derivative of the entropy and the time derivative of the temperature. Clearly, the notation $\cheatvolNSE$ is motivated by the classical formula for the specific heat at constant volume. Concerning the time derivatives $\dd{\rho}{t}$ and $\dd{}{t} \absnorm{\nabla \Tr \lcgnc}^2$ on the right-hand side of~\eqref{eq:66}, we can exploit the balance of mass~\eqref{eq:19} and the kinematic identity~\eqref{eq:10}, which yield
\begin{multline}
  \label{eq:67}
  \rho \dd{\entropy}{t}
  =
  \frac{1}{\temp}
  \left\{
    \left(
      \rho
      \cheatvolNSE
      -
      \frac{\temp}{2}
      \ddd{\tilde{\mu}}{\temp}
      \absnorm{\nabla \Tr \lcgnc}^2
    \right)
    \dd{\temp}{t}
    +
    \temp
    \pd{\thpressuredMTr}{\temp}
    \divergence \vec{v}
  \right\}
  \\
  +
  \dd{\tilde{\mu}}{\temp}
  \left\{
    \left[
      \Delta \left( \Tr \lcgnc \right)
    \right]
    \dd{}{t}
    \left(
      \Tr \lcgnc
    \right)
    +
    \tensordot{\gradvl}
    {
      \left[
        \tensortensor{\nabla \left( \Tr \lcgnc \right)}{\nabla \left( \Tr \lcgnc \right)}
      \right]
    }
  \right\}
  \\
  -
  \dd{\tilde{\mu}}{\temp}
  \divergence
  \left(
    \left(
      \nabla \Tr \lcgnc
    \right)
    \dd{}{t}
    \left(
      \Tr \lcgnc
    \right)
  \right)
  .
\end{multline}

Now we substitute the explicit formula~\eqref{eq:67} for the time derivative of the entropy into the entropy evolution equation
\begin{equation}
  \label{eq:68}
  \rho \dd{\entropy}{t} + \divergence \entfluxc = \widetilde{\entprodc}.
\end{equation}
By virtue of the entropy production \emph{ansatz}~\eqref{eq:44}, see also~\eqref{eq:61}, the formula for the entropy flux~\eqref{eq:38}, and the kinematic identity \eqref{eq:30}, we can rewrite~\eqref{eq:68} as
\begin{multline}
  \label{eq:69}
  \left(
    \rho
    \cheatvolNSE
    -
    \frac{\temp}{2}
    \ddd{\tilde{\mu}}{\temp}
    \absnorm{\nabla \Tr \lcgnc}^2
  \right)
  \dd{\temp}{t}
  \\
  +
  \temp
  \pd{}{\temp}
  \left\{
    \thpressuredMTr
    +
    \frac{\tilde{\mu}}{3}
    \Tr
    \left[
      \tensortensor{\left( \nabla  \Tr \lcgnc \right)}{\left( \nabla  \Tr \lcgnc \right)}
    \right]
    +
    \frac{2}{3}\tilde{\mu}
    \Tr \lcgnc
    \left(
      \Delta
      \Tr \lcgnc
    \right)
  \right\}
  \divergence{\vec{v}}
  \\
  +
  \tensordot{
    \left\{
      \temp
      \pd{}{\temp}
      \left[
        \tilde{\mu}
        \traceless{\left[ \tensortensor{\left( \nabla  \Tr \lcgnc \right)}{\left( \nabla  \Tr \lcgnc \right)} \right]}
        +
        2
        \tilde{\mu}
        \left(\Delta \Tr \lcgnc\right)
        \traceless{
          \left(
            \lcgnc
          \right)
        }
      \right]
    \right\}
  }
  {
    \traceless{\gradsym}
  }
  \\
  +
    \tensordot{
    \left\{
      \temp
      \pd{}{\temp}
      \left[
        2
        \tilde{\mu}
        \left( \Delta \Tr \lcgnc \right)
        \rcgnc
      \right]
    \right\}
  }
  {
    \gradsymrn
  }
  \\
  -
  \temp
  \dd{\tilde{\mu}}{\temp}
  \divergence
  \left(
    \left(
      \nabla \Tr \lcgnc
    \right)
    \dd{}{t}
    \left(
      \Tr \lcgnc
    \right)
  \right)
  +
  \temp
  \divergence
  \left(
    \frac{\efluxc + \tilde{\mu} \left( \nabla \Tr \lcgnc \right) \dd{}{t} \left( \Tr \lcgnc \right)}{\temp}
  \right)
  \\
  =
  \frac{2 \nu + 3 \lambda}{3} \left(\divergence \vec{v}\right)^2
  +
  2 \nu \tensordot{\traceless{\gradsym}}{\traceless{\gradsym}}
  +
  \frac{\nu_1}{2}
  \Tr
  \left(
    \fid{\overline{\lcgnc}}\inverse{\lcgnc}\fid{\overline{\lcgnc}}
  \right)
  +
  \frac
  {
    \kappa \absnorm{\nabla \temp}^2
  }
  {
    \temp
  }
  .
\end{multline}
Note that the term $\tensordot{\rcgnc}{\gradsymrn}$ can be expressed in terms of~$\lcgnc$; it suffices to take the trace of~\eqref{eq:50} and use the definition of $\tensorq{J}_{\gradsymrn}$. Further, the energy flux is given by~\eqref{eq:54}. Consequently~\eqref{eq:69} is the sought evolution equation for the temperature in terms of the primitive variables $\temp$, $\rho$, $\vec{v}$ and $\lcgnc$.

\subsection{Full system of governing equations for primitive variables -- compressible fluid}
\label{sec:full-syst-govern-1}
If we fix the splitting parameters $\alpha=1$ and $\beta=0$, then the energy flux $\efluxc$ is given as
\begin{equation}
  \label{eq:105}
  \efluxc
  =
  -
  \kappa \nabla \temp
  -
  \tilde{\mu}
  \left( \nabla \Tr \lcgnc \right)  \dd{}{t} \left( \Tr \lcgnc \right),
\end{equation}
see~\eqref{eq:54}, while the formula for the entropy flux $\entfluxc$ reads
\begin{equation}
  \label{eq:106}
  \entfluxc
  =
  -
  \frac{
    \kappa \nabla \temp
  }
  {
    \temp
  }
  ,
\end{equation}
see~\eqref{eq:38}. The temperature evolution equation~\eqref{eq:69} simplifies to
\begin{multline}
  \label{eq:107}
  \left(
    \rho
    \cheatvolNSE
    -
    \frac{\temp}{2}
    \ddd{\tilde{\mu}}{\temp}
    \absnorm{\nabla \Tr \lcgnc}^2
  \right)
  \dd{\temp}{t}
  +
  \temp
  \pd{}{\temp}
  \left\{
    \thpressuredMTr
    +
    \frac{\tilde{\mu}}{3}
    \Tr
    \left[
      \tensortensor{\left( \nabla  \Tr \lcgnc \right)}{\left( \nabla  \Tr \lcgnc \right)}
    \right]
    +
    \frac{2}{3}\tilde{\mu}
    \Tr \lcgnc
    \left(
      \Delta
      \Tr \lcgnc
    \right)
  \right\}
  \divergence{\vec{v}}
  \\
  +
  \tensordot{
    \left\{
      \temp
      \pd{}{\temp}
      \left[
        \tilde{\mu}
        \traceless{\left[ \tensortensor{\left( \nabla  \Tr \lcgnc \right)}{\left( \nabla  \Tr \lcgnc \right)} \right]}
        +
        2
        \tilde{\mu}
        \left(\Delta \Tr \lcgnc\right)
        \traceless{
          \left(
            \lcgnc
          \right)
        }
      \right]
    \right\}
  }
  {
    \traceless{\gradsym}
  }
  +
    \tensordot{
    \left\{
      \temp
      \pd{}{\temp}
      \left[
        2
        \tilde{\mu}
        \left( \Delta \Tr \lcgnc \right)
        \rcgnc
      \right]
    \right\}
  }
  {
    \gradsymrn
  }
  \\
  =
  \divergence\left(\kappa \nabla \temp \right)
  +
  \frac{2 \nu + 3 \lambda}{3} \left(\divergence \vec{v}\right)^2
  +
  2 \nu \tensordot{\traceless{\gradsym}}{\traceless{\gradsym}}
  +
  \frac{\nu_1}{2}
  \Tr
  \left(
    \fid{\overline{\lcgnc}}\inverse{\lcgnc}\fid{\overline{\lcgnc}}
  \right)
  \\
  +
  \temp
  \dd{\tilde{\mu}}{\temp}
  \divergence
  \left[
    \left(\nabla \Tr \lcgnc \right)
    \dd{}{t}
    \Tr \lcgnc
  \right]
  .
\end{multline}

This equation can be further transformed into a more convenient form. First, we return to the constitutive relations for the Cauchy stress tensor $\cstress = \mns \identity + \traceless{\cstress}$, where the mean normal stress $\mns$ and the traceless part are given by~\eqref{eq:52} and \eqref{eq:53}. The Cauchy stress tensor can be decomposed as
\begin{equation}
  \label{eq:108}
  \cstress
  =
  -
  p_{\mathrm{eq}}
  \identity
  -
  \tensorq{P}_{\mathrm{eq}}
  +
  \cstress_{\mathrm{vis}}
  -
  2
  \dd{\tilde{\mu}}{\temp}
  \left[
    \vectordot{\left(\nabla \Tr \lcgnc\right)}{\nabla \temp}
  \right]
  \lcgnc
  ,
\end{equation}
where
\begin{subequations}
  \label{eq:109}
  \begin{align}
    \label{eq:110}
    \tensorq{P}_{\mathrm{eq}}
    &=_{\bydefinition}
    -
    \mu \traceless{\left( \lcgnc \right)}
    +
    \tilde{\mu}
    \traceless{
      \left[
        \tensortensor{\left( \nabla  \Tr \lcgnc \right)}{\left( \nabla  \Tr \lcgnc \right)}
      \right]
    }
    +
    2
    \tilde{\mu}
    \left(\Delta \Tr \lcgnc\right)
    \traceless{
      \left(
        \lcgnc
      \right)
    }
    ,
    \\
    \label{eq:111}
    p_{\mathrm{eq}}
    &=_{\bydefinition}
    \thpressuredMTr
    -
    \frac{\mu}{3}
    \Tr \lcgnc
    +
    \mu
    +
    \frac{\tilde{\mu}}{3}
    \Tr
    \left[
      \tensortensor{\left( \nabla  \Tr \lcgnc \right)}{\left( \nabla  \Tr \lcgnc \right)}
    \right]
    +
    \frac{2}{3}\tilde{\mu}
    \Tr \lcgnc
    \left(
      \Delta
      \Tr \lcgnc
    \right)
    ,
    \\
    \label{eq:112}
    \cstress_{\mathrm{vis}}
    &=_{\bydefinition}
    \lambda \left( \divergence \vec{v} \right)
    \identity
    +
    2
    \nu
    \gradsym
    ,
  \end{align}
\end{subequations}
and using this notation, we can rewrite~\eqref{eq:107} as\footnote{Recall that $\mu$ is assumed to be a constant.}
\begin{multline}
  \label{eq:113}
  \left(
    \rho
    \cheatvolNSE
    -
    \frac{\temp}{2}
    \ddd{\tilde{\mu}}{\temp}
    \absnorm{\nabla \Tr \lcgnc}^2
  \right)
  \dd{\temp}{t}
  =
  \tensordot{\cstress_{\mathrm{vis}}}{\gradsym}
  -
  \temp
  \pd{p_{\mathrm{eq}}}{\temp} \divergence \vec{v}
  +
  \divergence\left(\kappa \nabla \temp \right)
  -
  \temp
  \tensordot{
    \pd{
      \tensorq{P}_{\mathrm{eq}}
    }
    {
      \temp
    }
  }
  {
    \traceless{\gradsym}
  }
  +
  \frac{\nu_1}{2}
  \Tr
  \left(
    \fid{\overline{\lcgnc}}\inverse{\lcgnc}\fid{\overline{\lcgnc}}
  \right)
  \\
  -
  2
  \temp
  \dd{\tilde{\mu}}{\temp}
  \left( \Delta \Tr \lcgnc \right)
  \left(
    \tensordot{
      \rcgnc
    }
    {
      \gradsymrn
    }
  \right)
  +
  \temp
  \dd{\tilde{\mu}}{\temp}
  \divergence
  \left[
    \left(\nabla \Tr \lcgnc \right)
    \dd{}{t}
    \Tr \lcgnc
  \right]
  ,
\end{multline}
which is the form that resembles the standard formula
\begin{equation}
  \label{eq:114}
  \rho
  \cheatvolNSE
  \dd{\temp}{t}
  =
  \tensordot{\cstress_{\mathrm{vis}}}{\gradsym}
  -
  \temp
  \pd{p_{\mathrm{eq}}}{\temp} \divergence \vec{v}
  +
  \divergence\left(\kappa \nabla \temp \right)
\end{equation}
for a classical compressible Navier--Stokes--Fourier fluid, see for example~\cite{gurtin.me.fried.e.ea:mechanics}. (Naturally, the formula for the equilibrium pressure $p_{\mathrm{eq}}$  is different in the case of a Navier--Stokes--Fourier fluid. However, if $\mu=0$, $\nu_1=0$ and $\tilde{\mu}=0$, then~\eqref{eq:114} coincides with~\eqref{eq:113}, including the definition of the equilibrium pressure $p_{\mathrm{eq}}$.)  In this sense we have obtained a proper generalisation of the standard temperature evolution equation in the case of a compressible viscoelastic rate-type fluid with stress diffusion.

Note that the time derivative $\dd{}{t} \Tr \lcgnc$ in~\eqref{eq:113} can be explicitly expressed in terms of the primitive variables. It suffices to take the trace of the evolution equation for $\lcgnc$, see~\eqref{eq:56}, which yields
\begin{equation}
  \label{eq:115}
  \dd{}{t} \Tr \lcgnc
  =
  2 \tensordot{\lcgnc}{\gradsym}
  -
  \frac{\mu}{\nu_1}
  \left(
    \Tr \lcgnc - 3
  \right)
  +
  \frac{2 \tilde{\mu}}{\nu_1}
  \left(
    \Delta \Tr \lcgnc
  \right)
  \Tr \lcgnc
  +
  \frac{2}{\nu_1}
  \dd{\tilde{\mu}}{\temp}
  \left[
    \vectordot{\left(\nabla \Tr \lcgnc\right)}{\nabla \temp}
  \right]
  \Tr \lcgnc
  .
\end{equation}
Further, the product $\tensordot{\rcgnc}{\gradsymrn}$ in~\eqref{eq:113} can be also explicitly expressed in terms of the primitive variables. It suffices to take the trace of~\eqref{eq:50} and use the definition of $\tensorq{J}_{\gradsymrn}$, see~\eqref{eq:42}, which yields
\begin{equation}
  \label{eq:116}
  \tensordot{\rcgnc}{\gradsymrn}
  =
  \frac{\mu}{2\nu_1}
  \left(
    \Tr \lcgnc - 3
  \right)
  -
  \frac{\tilde{\mu}}{\nu_1}
  \left(
    \Delta \Tr \lcgnc
  \right)
  \Tr \lcgnc
  -
  \frac{1}{\nu_1}
  \dd{\tilde{\mu}}{\temp}
  \left[
    \vectordot{\left(\nabla \Tr \lcgnc\right)}{\nabla \temp}
  \right]
  \Tr \lcgnc
  .
\end{equation}

Finally, the governing equations for the primitive mechanical variables $\rho$, $\vec{v}$ and $\lcgnc$ are
\begin{subequations}
  \label{eq:117}
  \begin{align}
    \label{eq:118}
    \dd{\rho}{t} + \rho \divergence \vec{v} &= 0, \\
    \label{eq:119}
    \rho \dd{\vec{v}}{t} &= \divergence \cstress + \rho \vec{b}, \\
    \label{eq:120}
    \nu_1 \fid{\overline{\lcgnc}}
    +
    \mu \left(\lcgnc - \identity\right)
    &=
    2
    \tilde{\mu}
    \left( \Delta \Tr \lcgnc \right)
    \lcgnc
    +
    2
    \dd{\tilde{\mu}}{\temp}
    \left[
      \vectordot{
        \left(\nabla \Tr \lcgnc\right)
      }
      {
        \nabla \temp
      }
    \right]
    \lcgnc
    ,
  \end{align}
\end{subequations}
where~\eqref{eq:120} follows from~\eqref{eq:56}. The final full system of governing equations is shown in Summary~\ref{summary:compressible-Tr-alt}.

\input{text/summary-a}

\subsection{Incompressible fluid}
\label{sec:incompressible-fluid-1}

The derivation outlined above can be also used in the case of incompressible fluids. In such a case the procedure is very close to that used by~\cite{malek.j.rajagopal.kr.ea:on} and~\cite{hron.j.milos.v.ea:on} with appropriate changes reflecting the presence of the gradient term in the Helmholtz free energy \emph{ansatz}, see above.

The counterpart of \emph{ansatz}~\eqref{eq:12} in the incompressible case is
\begin{subequations}
  \label{eq:141}
  \begin{equation}
    \label{eq:142}
    \fenergy
    =_{\bydefinition}
    \widetilde{\fenergy} \left(\temp\right)
    +
    \frac{\mu}{2\rho}
    \left(
      \Tr \lcgnc
      -
      3
      -
      \ln \det \lcgnc
    \right)
    +
    \frac{\tilde{\mu}(\temp)}{2\rho}
    \absnorm{ \nabla \Tr \lcgnc}^2,
  \end{equation}
where $\rho$ is a constant. The entropy evolution equation remains almost the same as in~\eqref{eq:47}, except the term
$
\left( J_{\divergence \vec{v}} \right)
\divergence{\vec{v}}
$
that vanishes by virtue of the incompressibility of the fluid. The counterpart of the entropy production \emph{ansatz}~\eqref{eq:44} is in the incompressible case
\begin{equation}
  \label{eq:143}
  \widetilde{\entprodc}
  =_{\bydefinition}
  \frac{1}{\temp}
  \left(
    2 \nu \tensordot{\traceless{\gradsym}}{\traceless{\gradsym}}
    +
    \nu_1
    \tensordot{
      \left(
        \rcgnc
        \gradsymrn
        +
        \gradsymrn
        \rcgnc
      \right)
    }
    {
        \gradsymrn
    }
    +
    \frac{1}{\temp} \kappa \absnorm{\nabla \temp}^2
  \right),
\end{equation}
\end{subequations}
where one can note that $\traceless{\gradsym} = \gradsym$.

\subsection{Full system of governing equations for primitive variables -- incompressible fluid}
\label{sec:full-syst-govern-2}

The Helmholtz free energy \emph{ansatz}~\eqref{eq:142} and entropy production \emph{ansatz}~\eqref{eq:143} then lead to the following governing equations for~$\vec{v}$, $\mns$ and $\lcgnc$:
\begin{subequations}
  \label{eq:168}
  \begin{align}
    \label{eq:169}
    \divergence \vec{v} &= 0, \\
    \label{eq:170}
    \rho \dd{\vec{v}}{t} &= \divergence \cstress + \rho \vec{b}, \\
    \label{eq:171}
    \nu_1 \fid{\overline{\lcgnc}}
    +
    \mu \left(\lcgnc - \identity\right)
    &=
    2
    \tilde{\mu}
    \left( \Delta \Tr \lcgnc \right)
    \lcgnc
    +
    2
    \dd{\tilde{\mu}}{\temp}
    \left[
      \vectordot{
        \left(\nabla \Tr \lcgnc\right)
      }
      {
        \nabla \temp
      }
    \right]
    \lcgnc
    ,
  \end{align}
  and the temperature $\temp$:
  \begin{multline}
    \label{eq:172}
    \left(
      \rho
      \cheatvolNSE
      -
      \frac{\temp}{2}
      \ddd{\tilde{\mu}}{\temp}
      \absnorm{\nabla \Tr \lcgnc}^2
    \right)
    \dd{\temp}{t}
    =
    \tensordot{\tensorq{S}}{\gradsym}
    +
    \divergence\left(\kappa \nabla \temp \right)
    -
    \temp
    \tensordot{
      \pd{
        \tensorq{P}
      }
      {
        \temp
      }
    }
    {
      \gradsym
    }
    +
    \frac{\nu_1}{2}
    \Tr
    \left(
      \fid{\overline{\lcgnc}}\inverse{\lcgnc}\fid{\overline{\lcgnc}}
    \right)
    \\
    -
    2
    \temp
    \dd{\tilde{\mu}}{\temp}
    \left( \Delta \Tr \lcgnc \right)
    \left(
      \tensordot{
        \rcgnc
      }
      {
        \gradsymrn
      }
    \right)
    +
    \temp
    \dd{\tilde{\mu}}{\temp}
    \divergence
    \left[
      \left(\nabla \Tr \lcgnc \right)
      \dd{}{t}
      \Tr \lcgnc
    \right]
    ,
  \end{multline}
  where the constitutive relation for the Cauchy stress tensor reads
  \begin{align}
    \label{eq:173}
    \cstress &= \mns \identity + \tensorq{S}
    -
    2
    \dd{\tilde{\mu}}{\temp}
    \left[
      \vectordot{\left(\nabla \Tr \lcgnc\right)}{\nabla \temp}
    \right]
    \lcgnc
    ,
    \\
    \label{eq:174}
    \tensorq{S}
    &=
    2 \nu \traceless{\gradsym}
    -
    \tensorq{P}
    ,
    \\
    \label{eq:175}
    \tensorq{P}
    &=
    -
    \mu \traceless{\left( \lcgnc \right)}
    +
    \tilde{\mu}
    \traceless{
      \left[
        \tensortensor{\left( \nabla  \Tr \lcgnc \right)}{\left( \nabla  \Tr \lcgnc \right)}
      \right]
    }
    +
    2
    \tilde{\mu}
    \left(\Delta \Tr \lcgnc\right)
    \traceless{
      \left(
        \lcgnc
      \right)
    }
    .
  \end{align}
\end{subequations}

  The definition of the specific heat at constant volume $\cheatvolNSE$ remains the same as in the compressible case, see~\eqref{eq:65}. Further, the formulae for the time derivative $\dd{}{t} \Tr \lcgnc$ and the product $\tensordot{\rcgnc}{\gradsymrn}$, that appear in~the temperature evolution equation~\eqref{eq:172}, are also the same as in the compressible case, see~\eqref{eq:115} and~\eqref{eq:116}. Finally, the formula for the energy flux $\efluxc$ is also the same as in the compressible case, see~\eqref{eq:105}. The final full system of governing equations is shown in Summary~\ref{summary:incompressible-Tr-alt}.

Note that the mean normal stress $\mns$ or the ``pressure'' is in the case of an incompressible fluid a \emph{primitive quantity that must be solved for}. It is not, as in the case of compressible fluid, a known function of the other quantities.

\input{text/summary-b}

\subsection{Remarks}
\label{sec:remarks}
Let us now focus on the model introduced in Section~\ref{sec:full-syst-govern-2} (incompressible fluid, Maxwell/Oldroyd-B, stress diffusion is a consequence of a non-standard energy storage mechanism), and let us consider a \emph{temperature-independent stress diffusion coefficient} $\tilde{\mu}$. A few remarks are in order.

First, the ``nonlocal'' stress diffusion term
$
2
\tilde{\mu}
\left( \Delta \Tr \lcgnc \right)
\lcgnc
$
in~\eqref{eq:171} does not include the gradient of the full tensor~$\lcgnc$, but only the gradient of its trace. However, if one deals with a Johnson--Segalman type model with a stress diffusion term, it is customary to work with a stress diffusion term in the form $\Delta \lcgnc$. (The Laplace operator acts on the full extra stress tensor, not only on its trace.) The question is whether the current type of nonlocal term would also provide a selection criterion for the stress values in viscoelastic models which possess a non-monotonic flow curve, see~\cite{olmsted.pd.radulescu.o.ea:johnson-segalman} and \cite{lu.cyd.olmsted.pd.ea:effects}. We leave this question open, since we are, in any case, dealing with a Maxwell/Oldroyd-B model only.

Second, let us consider $\lcgnc$ in the form $\lcgnc = \identity + \tensorq{b}$, and let us assume that $\tensorq{b}$ is a small quantity. (This corresponds to a close-to-the-equilibrium setting, see Section~\ref{sec:init-value-probl}.) If we decide to use the system~\eqref{eq:168} and keep only the linear terms in $\tensorq{b}$, we see that the right-hand side of~\eqref{eq:171} can be approximated as
\begin{equation}
  \label{eq:193}
  2
  \tilde{\mu}
  \left( \Delta \Tr \lcgnc \right)
  \lcgnc
  \approx
  2
  \tilde{\mu}
  \left( \Delta \Tr \tensorq{b} \right)
  \identity
  .
\end{equation}
Further, the Cauchy stress tensor, see~\eqref{eq:173}, can be approximated as\footnote{Recall that we consider temperature independent stress diffusion coefficient $\tilde{\mu}$.}
\begin{multline}
  \label{eq:194}
  \cstress
  =
  \mns \identity + \tensorq{S}
  -
  2
  \dd{\tilde{\mu}}{\temp}
  \left[
    \vectordot{\left(\nabla \Tr \lcgnc\right)}{\nabla \temp}
  \right]
  \lcgnc
  \\
  =
  \mns \identity
  +
  2 \nu \traceless{\gradsym}
  +
  \mu \traceless{\left( \lcgnc \right)}
  -
  \tilde{\mu}
  \traceless{
    \left[
      \tensortensor{\left( \nabla  \Tr \lcgnc \right)}{\left( \nabla  \Tr \lcgnc \right)}
    \right]
  }
  -
  2
  \tilde{\mu}
  \left(\Delta \Tr \lcgnc\right)
  \traceless{
    \left(
      \lcgnc
    \right)
  }
  \\
  \approx
  \left[
    \mns
    -
    2
    \tilde{\mu}
    \left(\Delta \Tr \tensorq{b}\right)
  \right]\identity
  +
  2 \nu \traceless{\gradsym}
  +
  \mu \traceless{\left( \tensorq{b} \right)}
  =
  \tilde{\mns}\identity
  +
  2 \nu \traceless{\gradsym}
  +
  \mu \traceless{\left( \tensorq{b} \right)}
  ,
\end{multline}
where $\tilde{\mns}$ denotes the modified mean normal stress. 

Consequently, the \emph{first order contribution} of the nonlocal term $\frac{\tilde{\mu}}{2} \absnorm{ \nabla \Tr \lcgnc}^2$ in the free energy \emph{ansatz}~\eqref{eq:12} is from the perspective of the governing equations limited to the presence of the nonlocal term $\Delta \Tr \tensorq{b}$ in the evolution equation~\eqref{eq:171}. The Korteweg type terms
$
\traceless{
  \left[
    \tensortensor{\left( \nabla  \Tr \lcgnc \right)}{\left( \nabla  \Tr \lcgnc \right)}
  \right]
}
$
and so forth in the Cauchy stress tensor $\cstress$, see~\eqref{eq:173}, are of second-order.

Third, the material parameters/functions $\rho$, $\mu$, $\lambda$, $\nu$, $\nu_1$ and $\kappa$ are the standard material parameters/functions that are routinely measured/dealt with for standard viscoelastic fluids \emph{without} stress diffusion, see for example~\cite{leonov.ai.prokunin.an:nonlinear} or~\cite{phan-thien.n:understanding}. The parameter $\kappa$ is referred to as the thermal conductivity, $[\kappa] = \unitfrac{W}{m \cdot K}$, the parameter $\mu$ is referred to as the elastic modulus, $[\mu] = \unit{Pa}$, the combination $\frac{\nu_1}{\mu}$, $\left[ \frac{\nu_1}{\mu} \right] = s$, stands for the relaxation time, and $\nu$ stands for the ``solvent'' shear viscosity, $[\nu] = \unit{Pa \cdot s}$. The symbol $\rho$ denotes the density, which is either a constant (incompressible fluids) or a material function determined by the equation of state (compressible fluids), and, finally, the combination $\frac{2 \nu + 3 \lambda}{2}$ denotes the bulk viscosity, $\left[ \frac{2 \nu + 3 \lambda}{2} \right] = \unit{Pa \cdot s}$  (compressible fluids only).

The only additional parameter in the models is the stress diffusion coefficient $\tilde{\mu}$, $[\tilde{\mu}] = \unit{N}$, which is also a measurable quantity, see~\cite{fardin.m.radulescu.o.ea:stress} and~\cite{cheng.p.burroughs.mc.ea:distinguishing}.  Moreover, the functional dependence of $\tilde{\mu}$ also seems to be experimentally proven, see Figure~\ref{fig:temeprature-stress-diffusion}, where we report experimental data by~\cite{mohammadigoushki.h.muller.sj:flow}, who have discussed a diffusive Johnson--Segalman model. The stress diffusion coefficient $D$ as defined in \cite{mohammadigoushki.h.muller.sj:flow}, see also Figure~\ref{fig:temeprature-stress-diffusion}, corresponds in our notation to the combination $\frac{\tilde{\mu}}{\nu_1}$, $\left[ \frac{\tilde{\mu}}{\nu_1} \right] = \unitfrac{m^2}{s}$.

Finally, we note that the formula for the energy flux $\efluxc$ allows one to identify the boundary conditions that lead to a mechanically and thermally isolated system. Since the evolution equation for total energy reads
\begin{equation}
  \label{eq:195}
  \rho
  \dd{}{t}
  \left(
    \ienergy
    +
    \frac{1}{2}
    \absnorm{\vec{v}}^2
  \right)
  =
  \divergence (\cstress \vec{v})
  -
  \divergence \efluxc
  ,
\end{equation}
we see that the net total energy
$
\int_{\Omega}
\rho
\left(
  \ienergy
  +
  \frac{1}{2}
  \absnorm{\vec{v}}^2
\right)
\,
\cvolumee
$ in the domain $\Omega$ is conserved provided that
$
  \int_{\partial \Omega} \vectordot{\left( \cstress \vec{v} - \efluxc \right)}{\vec{n}} \, \csurfacees = 0
$. This is guaranteed, for example, if the velocity vanishes on the boundary $\partial \Omega$ of domain $\Omega$:
\begin{equation}
  \label{eq:196}
  \left. \vec{v} \right|_{\partial \Omega} = \vec{0},
\end{equation}
and if the energy flux $\efluxc$ vanishes on the boundary:
\begin{equation}
  \label{eq:197}
   \left. \vectordot{\efluxc}{\vec{n}} \right|_{\partial \Omega} = 0.
\end{equation}

If we consider the model derived in Section~\ref{sec:full-syst-govern-1} (compressible fluid) and Section~\ref{sec:full-syst-govern-2} (incompressible fluid), then the energy flux is in both cases given by the formula~\eqref{eq:105}, that is,
\begin{equation}
  \label{eq:198}
  \efluxc
  =
  -
  \kappa \nabla \temp
  -
  \tilde{\mu}
  \left( \nabla \Tr \lcgnc \right)  \dd{}{t} \left( \Tr \lcgnc \right)
  ,
\end{equation}
and~\eqref{eq:197} is fulfilled if one fixes
\begin{subequations}
  \label{eq:199}
  \begin{align}
    \label{eq:200}
    \left. \vectordot{\nabla \temp}{\vec{n}} \right|_{\partial \Omega} &= 0, \\
    \label{eq:283}
    \left. \vectordot{\nabla \Tr \lcgnc}{\vec{n}} \right|_{\partial \Omega} &= 0.
  \end{align}
\end{subequations}
This provides an interpretation of the commonly used boundary condition that is necessary if the nonlocal term $\Delta \Tr \lcgnc$ appears in the governing equations. The natural zero Neumann boundary condition for $\nabla \Tr \lcgnc$ means that the system is closed with respect to the energy flux generated by the stress diffusion.

Finally, we see that the choice of the splitting parameters $\alpha=1$ and $\beta=0$ that was adopted in Section~\ref{sec:full-syst-govern-1} (compressible fluid) and Section~\ref{sec:full-syst-govern-2} (incompressible fluid) leads to the entropy flux in the form
\begin{equation}
  \label{eq:284}
  \entfluxc = -\frac{\kappa \nabla \temp}{\temp}.
\end{equation}
This means that the entropy flux takes in this case the standard form, which is in fact the motivation for the corresponding choice of the splitting parameters.

The reader interested in mathematical properties of a simplified isothermal model of the type~\eqref{eq:168} is referred to~\cite{bulvcek.m.malek.j.ea:pde-analysis}. Note that the manipulations used in the analysis of the mathematical properties of the simplified model are motivated by the thermodynamical underpinnings of the model.

\section{Derivation of constitutive relations -- stress diffusion as a consequence of a nonstandard entropy production mechanism}
\label{sec:deriv-cons-relat}
In this section we derive a model for viscoelastic fluids in which the stress diffusion term is attributed to a \emph{nonstandard entropy production mechanism}. The nonstandard entropy production mechanism is characterised by a gradient (nonlocal) term in the \emph{ansatz} for entropy production, while the \emph{ansatz} for the Helmholtz free energy remains the same as in the classical Maxwell/Oldroyd-B model.

\subsection{Evolution equation for the entropy}
\label{sec:evol-equat-entr} In this case the \emph{ansatz} for the Helmholtz free energy is~\eqref{eq:13}, that is,
\begin{equation}
  \label{eq:201}
  \fenergy
  =_{\bydefinition}
  \widetilde{\fenergy} \left(\temp, \rho\right)
  +
  \frac{\mu}{2\rho}
  \left(
    \Tr \lcgnc
    -
    3
    -
    \ln \det \lcgnc
  \right)
  .
\end{equation}
This is the standard \emph{ansatz} that leads, if the \emph{entropy production} is chosen appropriately, to the Maxwell/Oldroyd-B model, see for example~\cite{malek.j.rajagopal.kr.ea:on}. In this sense, the fluid stores the energy in the same manner as a Maxwell/Oldroyd-B type fluid. Following the same steps as before, we use~\eqref{eq:201} and derive the evolution equation for the entropy, which reads
\begin{subequations}
  \label{eq:202}
  \begin{equation}
    \label{eq:203}
    \rho \dd{\entropy}{t}
    +
    \divergence
    \left(
      \frac{\efluxc}{\temp}
    \right)
    =
    \frac{1}{\temp}
    \left[
      \bigg\{
      \mns
      +
      \thpressuredMdiff
      -
      \frac{\mu}{3}
      \Tr \lcgnc
      +
      \mu
      \bigg\}
      \divergence{\vec{v}}
      +
      \tensordot{
        \bigg\{
        \traceless{\cstress}
        -
        \mu \traceless{\left( \lcgnc \right)}
        \bigg\}
      }
      {
        \traceless{\gradsym}
      }
      +
      \tensordot{
        \bigg\{
        \mu
        \left(\rcgnc - \identity\right)
        \bigg\}
      }
      {
        \gradsymrn
      }
    \right]
    -
    \frac{
      \vectordot{
        \efluxc
      }
      {
        \nabla \temp
      }
    }
    {
      \temp^2
    }
    .
  \end{equation}
  The thermodynamic pressure $\thpressuredMdiff$ is again defined in terms of the Helmholtz free energy~\eqref{eq:201} as
  \begin{equation}
    \label{eq:204}
    \thpressuredMdiff
    =_{\bydefinition}
    \thpressureNSE - \widetilde{\widetilde{\fenergy}},
  \end{equation}
  where $\thpressureNSE$ is defined as in~\eqref{eq:pressure-NSE} and $\widetilde{\widetilde{\fenergy}}$ is the elastic contribution to the Helmholtz free energy, see~\eqref{eq:14}.
\end{subequations}
(Equation \eqref{eq:203} in fact equation~\eqref{eq:29} with $\tilde{\mu}=0$.) The equation has the same form as~\eqref{eq:37}, that is
\begin{subequations}
  \label{eq:205}
  \begin{equation}
    \label{eq:206}
    \rho \dd{\entropy}{t}
    +
    \divergence
    \left(
      \frac{\efluxc}{\temp}
    \right)
    =
    \frac{1}{\temp}
    \left\{
      \left( J_{\divergence \vec{v}} \right)
      \divergence{\vec{v}}
      +
      \tensordot{
        \tensorq{J}_{\traceless{\gradsym}}
      }
      {
        \traceless{\gradsym}
      }
      +
      \tensordot{
        \tensorq{J}_{\gradsymrn}
      }
      {
        \gradsymrn
      }
      -
      \vectordot{
        \vec{J}_{\nabla \temp}
      }
      {
        \frac{\nabla \temp}{\temp}
      }
    \right\}
    ,
  \end{equation}
  where the flux terms $J_{\divergence \vec{v}}$, $\tensorq{J}_{\traceless{\gradsym}}$, $\tensorq{J}_{\gradsymrn}$ and $\vec{J}_{\nabla \temp}$ are given by the formulae
  \begin{align}
    \label{eq:207}
    J_{\divergence \vec{v}}
    &=_{\bydefinition}
    \mns
    +
    \thpressuredMdiff
    -
    \frac{\mu}{3}
    \Tr \lcgnc
    +
    \mu,
    \\
    \label{eq:208}
    \tensorq{J}_{\traceless{\gradsym}}
    &=_{\bydefinition}
    \traceless{\cstress}
    -
    \mu \traceless{\left( \lcgnc \right)},
    \\
    \label{eq:209}
    \tensorq{J}_{\gradsymrn}
    &=_{\bydefinition}
    \mu
    \left(\rcgnc - \identity\right)
    ,
    \\
    \label{eq:210}
    \vec{J}_{\nabla \temp}
    &=_{\bydefinition}
    \efluxc
    .
  \end{align}
\end{subequations}
The task is to exploit~\eqref{eq:205} in the identification of the \emph{constitutive relations for the Cauchy stress tensor} $\cstress = \mns \identity + \traceless{\cstress}$ and the \emph{energy flux} $\efluxc$. In order to do so we rewrite~\eqref{eq:205} in a more convenient form; specifically, we reformulate the product
$
\tensordot{
  \tensorq{J}_{\gradsymrn}
}
{
  \gradsymrn
}
$
in~\eqref{eq:206} as
\begin{equation}
  \label{eq:211}
  \tensordot{
    \tensorq{J}_{\gradsymrn}
  }
  {
    \gradsymrn
  }
  =
  \mu
  \tensordot{
    \left(\rcgnc - \identity\right)
  }
  {
    \gradsymrn
  }
  =
  -
  \frac{\mu}{2}
  \Tr
  \left[
    \fid{\overline{\lcgnc}}
    \left(
      \identity
      -
      \inverse{\lcgnc}
    \right)
  \right]
  ,
\end{equation}
which is a consequence of the identity~\eqref{eq:6} and the definitions $\rcgnc =_{\bydefinition} \transpose{\fgradnc} \fgradnc$ and $\lcgnc =_{\bydefinition} \fgradnc \transpose{\fgradnc}$. This manipulation yields~\eqref{eq:206} in the form
\begin{equation}
  \label{eq:212}
  \rho \dd{\entropy}{t}
  +
  \divergence
  \left(
    \frac{\efluxc}{\temp}
  \right)
  =
  \frac{1}{\temp}
  \bigg\{
    \left[
      \mns
      +
      \thpressuredMdiff
      -
      \frac{\mu}{3}
      \Tr \lcgnc
      +
      \mu
    \right]
    \divergence{\vec{v}}
    +
    \tensordot{
      \left[
        \traceless{\cstress}
        -
        \mu \traceless{\left( \lcgnc \right)}
      \right]
    }
    {
      \traceless{\gradsym}
    }
    -
    \frac{\mu}{2\nu_1}
    \Tr
    \left[
      \nu_1
      \fid{\overline{\lcgnc}}
      \left(
        \identity
        -
        \inverse{\lcgnc}
      \right)
    \right]
    -
    \frac{
      \vectordot{
        \efluxc
      }
      {
        \nabla \temp
      }
    }
    {\temp}
  \bigg\}
  .
\end{equation}

\subsection{Entropy production and constitutive relations}
\label{sec:entr-prod-const-2}
Now we are in a position to specify how the fluid produces the entropy. In other words we need to fix a formula for the entropy production $\widetilde{\entprodc}$. This is the point where we deviate from the approach that would lead to the standard Maxwell/Oldroyd-B fluid flow model: we shall use a different entropy production~$\widetilde{\entprodc}$ than the one that is known to lead to a standard Maxwell/Oldroyd-B fluid. We fix
\begin{equation}
  \label{eq:213}
  \widetilde{\entprodc}
  =
  \frac{1}{\temp}
  \left\{
    \frac{2 \nu + 3 \lambda}{3} \left(\divergence \vec{v}\right)^2
    +
    2 \nu \tensordot{\traceless{\gradsym}}{\traceless{\gradsym}}
    +
    \frac{\mu^2}{2\nu_1}
    \left(
      \Tr \lcgnc
      +
      \Tr \inverse{\lcgnc}
      -
      6
    \right)
    +
    \frac{\mu \tilde{\mu}(\temp)}{2\nu_1}
    \tensorddot{
      \nabla \lcgnc
    }
    {
      \nabla \lcgnc
    }
    +
    \frac
    {
      \kappa \absnorm{\nabla \temp}^2
    }
    {
      \temp
    }
  \right\}
  ,
\end{equation}
where we have used the notation
\begin{equation}
  \label{eq:214}
  \tensorddot{
    \nabla \lcgnc
  }
  {
    \nabla \lcgnc
  }
  =_{\bydefinition}
  \pd{\tensor{\lcgncc}{_i_j}}{x_m}
  \pd{\tensor{\lcgncc}{_i_j}}{x_m}
  .
\end{equation}
(Note that the additional term $  \tensorddot{
    \nabla \lcgnc
  }
  {
    \nabla \lcgnc
  }
$
is nonnegative.) The material parameter $\tilde{\mu}$ that will be later identified as a stress diffusion coefficient can be a nonnegative function of the temperature. The other material parameters $\nu$, $\nu_1$, $2\nu + 3\lambda$ and $\kappa$ are, for the sake of simplicity, assumed to be nonnegative constants.

The \emph{ansatz} is motivated by the fact that we want to model a fluid that behaves almost as a Maxwell/Oldroyd-B fluid. This is guaranteed by the presence of the first three terms and the last term in~\eqref{eq:214}. See in particular~\eqref{eq:61} and~\eqref{eq:62} for the reformulation of the entropy production \emph{ansatz}~\eqref{eq:44} in a way that motivates~\eqref{eq:213}. The penultimate term in~\eqref{eq:213}, that is
\begin{equation}
  \label{eq:215}
    \frac{\mu \tilde{\mu}(\temp)}{2\nu_1}
    \tensorddot{
      \nabla \lcgnc
    }
    {
      \nabla \lcgnc,
    }
\end{equation}
is, as we shall see later, the entropy production term due to stress diffusion. The entropy production is clearly nonnegative.

The desired entropy evolution equation is
\begin{equation}
  \label{eq:216}
    \rho \dd{\entropy}{t} + \divergence \entfluxc = \widetilde{\entprodc},
\end{equation}
where the entropy production $\widetilde{\entprodc}$ is given by~\eqref{eq:213}. The desired entropy evolution equation/entropy production must be compared with the entropy evolution equation/entropy production specified in~\eqref{eq:212}. Let us now manipulate~\eqref{eq:216} with $\widetilde{\entprodc}$ given by~\eqref{eq:213} into the form comparable to~\eqref{eq:212}.

In~rewriting of~\eqref{eq:216} we use several identities. The first of them is
\begin{subequations}
  \label{eq:217}
  \begin{multline}
    \label{eq:218}
    \frac{1}{2}
    \left[
      \pd{}{x_m}
      \left(
        \tilde{\mu}
        \pd{\tensor{\lcgncc}{_i_j}}{x_m}
      \right)
      \tensor{\lcgncc}{_j_l}
      +
      \pd{}{x_m}
      \left(
        \tilde{\mu}
        \pd{\tensor{\lcgncc}{_l_j}}{x_m}
      \right)
      \tensor{\lcgncc}{_j_i}
    \right]
    \left(
      \kdelta{_l_i}
      -
      \tensor{\left(\inverse{\lcgncc}\right)}{_l_i}
    \right)
    \\
    =
    \frac{1}{2}
    \left[
      \pd{}{x_m}
      \left(
        \tilde{\mu}
        \pd{\tensor{\lcgncc}{_i_j}}{x_m}
      \right)
      \left(
        \tensor{\lcgncc}{_j_i}
        -
        \kdelta{_j_i}
      \right)
      +
      \pd{}{x_m}
      \left(
        \tilde{\mu}
        \pd{\tensor{\lcgncc}{_l_j}}{x_m}
      \right)
      \left(
        \tensor{\lcgncc}{_j_l}
        -
        \kdelta{_j_l}
      \right)
    \right]
    \\
    =
    \frac{1}{2}
    \pd{}{x_m}
    \left[
      \tilde{\mu}
      \left(
        \pd{\tensor{\lcgncc}{_i_j}}{x_m}
        \left(
          \tensor{\lcgncc}{_j_i}
          -
          \kdelta{_j_i}
        \right)
        +
        \pd{\tensor{\lcgncc}{_l_j}}{x_m}
        \left(
          \tensor{\lcgncc}{_j_l}
          -
          \kdelta{_j_l}
        \right)
      \right)
    \right]
    -
    \frac{\tilde{\mu}}{2}
    \left[
      \pd{\tensor{\lcgncc}{_i_j}}{x_m}
      \pd{\tensor{\lcgncc}{_j_i}}{x_m}
      +
      \pd{\tensor{\lcgncc}{_l_j}}{x_m}
      \pd{\tensor{\lcgncc}{_j_l}}{x_m}
    \right]
    ,
  \end{multline}
which can be symbolically written as
\begin{multline}
  \label{eq:219}
  \frac{1}{2}
  \Tr
  \left\{
    \bigg[
      \divergence \left( \tilde{\mu} \nabla \lcgnc \right)
      \lcgnc
      +
      \lcgnc
      \divergence \left( \tilde{\mu} \nabla \lcgnc \right)
    \bigg]
    \left(
      \identity
      -
      \inverse{\lcgnc}
    \right)
  \right\}
  \\
  =
  \frac{1}{2}
  \divergence
  \left\{
    \tilde{\mu}
    \Tr
    \bigg[
    \left(\nabla \lcgnc \right)
    \left(
      \lcgnc
      -
      \identity
    \right)
    +
    \left(
      \lcgnc
      -
      \identity
    \right)
    \left(\nabla \lcgnc \right)
    \bigg]
  \right\}
  -
  \tilde{\mu}
  \tensorddot{\nabla \lcgnc}{\nabla \lcgnc}
  .
\end{multline}
\end{subequations}
(We have again exploited the symmetry of $\lcgnc$.) Further, the term $\Tr \lcgnc + \Tr \inverse{\lcgnc} - 6$ can be rewritten as
\begin{equation}
  \label{eq:220}
  \Tr \lcgnc + \Tr \inverse{\lcgnc} - 6
  =
  \Tr
  \left[
    \left(
      \lcgnc - \identity
    \right)
    \left(
      \identity
      -
      \inverse{\lcgnc}
    \right)
  \right]
  .
\end{equation}
Using~\eqref{eq:219} and \eqref{eq:220} in~\eqref{eq:216} with $\tilde{\entprodc}$ specified via~\eqref{eq:213} yields after some manipulation
\begin{multline}
  \label{eq:222}
  \rho \dd{\entropy}{t}
  +
  \divergence
  \left\{
    \entfluxc
    -
    \frac{
      \frac{\mu\tilde{\mu}}{4\nu_1}
      \Tr
      \bigg[
      \left(\nabla \lcgnc \right)
      \left(
        \lcgnc
        -
        \identity
      \right)
      +
      \tilde{\mu}
      \left(
        \lcgnc
        -
        \identity
      \right)
      \left(\nabla \lcgnc \right)
      \bigg]
    }
    {
      \temp
    }
  \right\}
  \\
  =
  \frac{1}{\temp}
  \bigg\{
    \frac{2 \nu + 3 \lambda}{3} \left(\divergence \vec{v}\right)^2
    +
    2 \nu \tensordot{\traceless{\gradsym}}{\traceless{\gradsym}}
    +
    \frac{\mu}{2\nu_1}
    \Tr
    \left[
      \bigg[
        \mu
        \left(
          \lcgnc - \identity
        \right)
        -
        \frac{1}{2}
        \bigg(
        \divergence \left( \tilde{\mu} \nabla \lcgnc \right)
        \lcgnc
        +
        \lcgnc
        \divergence \left( \tilde{\mu} \nabla \lcgnc \right)
        \bigg)
      \bigg]
      \left(
        \identity
        -
        \inverse{\lcgnc}
      \right)
    \right]
  \bigg\}
  \\
  +
  \frac
  {
    \vectordot{
      \left\{
        \kappa
        \nabla \temp
        +
        \frac{\mu\tilde{\mu}}{4\nu_1}
        \Tr
        \bigg[
        \left(\nabla \lcgnc \right)
        \left(
          \lcgnc
          -
          \identity
        \right)
        +
        \tilde{\mu}
        \left(
          \lcgnc
          -
          \identity
        \right)
        \left(\nabla \lcgnc \right)
        \bigg]
      \right\}
    }
    {
      \nabla \temp
    }
  }
  {
    \temp^2
  }
  .
\end{multline}

The expression for the entropy evolution~\eqref{eq:222} that follows from the \emph{ansatz} for the entropy production can now be compared with the entropy evolution~\eqref{eq:212} implied by the chosen \emph{ansatz} for the Helmholtz free energy and the underlying kinematics. Clearly, the two equations will coincide if we set
\begin{subequations}
  \label{eq:223}
  \begin{align}
    \label{eq:224}
    \mns
    +
    \thpressuredMdiff
    -
    \frac{\mu}{3}
    \Tr \lcgnc
    +
    \mu
    &=
    \frac{2 \nu + 3 \lambda}{3} \divergence \vec{v}
    ,
    \\
    \label{eq:225}
    \traceless{\cstress}
    -
    \mu \traceless{\left( \lcgnc \right)}
    &=
    2 \nu \traceless{\gradsym}
    ,
    \\
    \label{eq:226}
    -\nu_1 \fid{\overline{\lcgnc}}
    &=
    \mu
    \left(
      \lcgnc - \identity
    \right)
    -
    \frac{1}{2}
    \bigg[
    \divergence \left( \tilde{\mu} \nabla \lcgnc \right)
    \lcgnc
    +
    \lcgnc
    \divergence \left( \tilde{\mu} \nabla \lcgnc \right)
    \bigg]
    ,
    \\
    \label{eq:227}
    \efluxc
    &=
    -
    \kappa
    \nabla \temp
    -
    \frac{\mu\tilde{\mu}}{4\nu_1}
    \Tr
    \bigg[
    \left(\nabla \lcgnc \right)
    \left(
      \lcgnc
      -
      \identity
    \right)
    +
    \left(
      \lcgnc
      -
      \identity
    \right)
    \left(\nabla \lcgnc \right)
    \bigg]
    ,
    \\
    \label{eq:228}
    \entfluxc
    &=
    -
    \frac{
      \kappa \nabla \temp
    }
    {
      \temp
    }
    .
  \end{align}
  These are the sought constitutive relations for the Cauchy stress tensor~$\cstress = \mns \identity + \traceless{\cstress}$, the energy flux~$\efluxc$ and the entropy flux~$\entfluxc$.
\end{subequations}

\subsection{Evolution equation for the temperature}
\label{sec:evol-equat-temp-1}
Having identified the entropy flux $\entfluxc$ we are ready to formulate the governing equation for the temperature. Following the same steps as in Section~\ref{sec:evol-equat-temp} we see that the Helmholtz free energy \emph{ansatz}~\eqref{eq:201} and the entropy production \emph{ansatz}~\eqref{eq:213} imply that the evolution equation for the temperature $\temp$ reads
\begin{multline}
  \label{eq:229}
  \rho
  \cheatvolNSE
  \dd{\temp}{t}
  +
  \temp
  \pd{\thpressuredMdiff}{\temp}
  \divergence{\vec{v}}
  =
  \divergence \left(\kappa \nabla \temp \right)
  +
  \frac{2 \nu + 3 \lambda}{3} \left(\divergence \vec{v}\right)^2
  +
  2 \nu \tensordot{\traceless{\gradsym}}{\traceless{\gradsym}}
  \\
  +
  \frac{\mu^2}{2\nu_1}
  \left(
    \Tr \lcgnc
    +
    \Tr \inverse{\lcgnc}
    -
    6
  \right)
  +
  \frac{\mu \tilde{\mu}(\temp)}{2\nu_1}
  \tensorddot{
    \nabla \lcgnc
  }
  {
    \nabla \lcgnc
  }
  ,
\end{multline}
where the specific heat at constant volume $\cheatvolNSE$ is obtained via differentiation of the Helmholtz free energy, see~\eqref{eq:65}.

This equation can be further rewritten as follows. Using the constitutive relations for the Cauchy stress tensor $\cstress = \mns \identity + \traceless{\cstress}$, where the mean normal stress $\mns$ and the traceless part $\traceless{\cstress}$ are given by~\eqref{eq:224} and~\eqref{eq:225}, we see that $\cstress$ can be decomposed as
\begin{equation}
  \label{eq:230}
  \cstress
  =
  -
  p_{\mathrm{eq}}
  \identity
  -
  \tensorq{P}_{\mathrm{eq}}
  +
  \cstress_{\mathrm{vis}}
  ,
\end{equation}
where
\begin{subequations}
  \label{eq:231}
  \begin{align}
    \label{eq:232}
    \tensorq{P}_{\mathrm{eq}}
    &=_{\bydefinition}
    -
    \mu \traceless{\left( \lcgnc \right)}
    ,
    \\
    \label{eq:233}
    p_{\mathrm{eq}}
    &=_{\bydefinition}
    \thpressuredMdiff
    -
    \frac{\mu}{3}
    \Tr \lcgnc
    +
    \mu
    ,
    \\
    \label{eq:234}
    \cstress_{\mathrm{vis}}
    &=_{\bydefinition}
    \lambda \left( \divergence \vec{v} \right)
    \identity
    +
    2
    \nu
    \gradsym
    .
  \end{align}
\end{subequations}
Using this notation, the evolution equation for the temperature can be rewritten as
\begin{equation}
  \label{eq:235}
  \rho
  \cheatvolNSE
  \dd{\temp}{t}
  =
  \tensordot{\cstress_{\mathrm{vis}}}{\gradsym}
  -
  \temp
  \pd{p_{\mathrm{eq}}}{\temp} \divergence \vec{v}
  +
  \divergence\left(\kappa \nabla \temp \right)
  +
  \frac{\mu^2}{2\nu_1}
  \left(
    \Tr \lcgnc
    +
    \Tr \inverse{\lcgnc}
    -
    6
  \right)
  +
  \frac{\mu \tilde{\mu}(\temp)}{2\nu_1}
  \tensorddot{
    \nabla \lcgnc
  }
  {
    \nabla \lcgnc
  }
  ,
\end{equation}
which resembles the standard formula for a compressible Navier--Stokes--Fourier fluid~\eqref{eq:114}, and allows one to identify the additional terms due to viscoelasticity and stress diffusion. The first three terms on the right hand side of~\eqref{eq:235} are the standard terms known for the compressible Navier--Stokes--Fourier fluid, see for example~\cite{gurtin.me.fried.e.ea:mechanics}. The next to last term corresponds to an additional ``viscous heating'' that takes place in viscoelastic fluids, see for example~\cite{hron.j.milos.v.ea:on} and also~\cite{dressler.m.edwards.bj.ea:macroscopic}, while the last term is a new ``viscous heating'' mechanism due to stress diffusion.

\subsection{Full system of governing equations for primitive variables -- compressible fluid}
\label{sec:full-syst-govern}

The Helmholtz free energy \emph{ansatz}~\eqref{eq:201} and entropy production \emph{ansatz}~\eqref{eq:213} then lead to the following governing equations for $\vec{v}$, $\rho$ and $\lcgnc$:
\begin{subequations}
  \label{eq:236}
  \begin{align}
    \label{eq:237}
    \dd{\rho}{t} + \rho \divergence \vec{v} &= 0, \\
    \label{eq:238}
    \rho \dd{\vec{v}}{t} &= \divergence \cstress + \rho \vec{b}, \\
    \label{eq:239}
    \nu_1 \fid{\overline{\lcgnc}}
    +
    \mu \left(\lcgnc - \identity\right)
    &=
    \frac{1}{2}
    \bigg[
    \divergence \left( \tilde{\mu} \nabla \lcgnc \right)
    \lcgnc
    +
    \lcgnc
    \divergence \left( \tilde{\mu} \nabla \lcgnc \right)
    \bigg]
  \end{align}
  and the temperature $\temp$:
  \begin{multline}
    \label{eq:240}
    \rho
    \cheatvolNSE
    \dd{\temp}{t}
    +
    \temp
    \pd{\thpressuredMdiff}{\temp}
    \divergence{\vec{v}}
    =
    \divergence \left(\kappa \nabla \temp \right)
    +
    \frac{2 \nu + 3 \lambda}{3} \left(\divergence \vec{v}\right)^2
    +
    2 \nu \tensordot{\traceless{\gradsym}}{\traceless{\gradsym}}
    \\
    +
    \frac{\mu^2}{2\nu_1}
    \left(
      \Tr \lcgnc
      +
      \Tr \inverse{\lcgnc}
      -
      6
    \right)
    +
    \frac{\mu \tilde{\mu}(\temp)}{2\nu_1}
    \tensorddot{
      \nabla \lcgnc
    }
    {
      \nabla \lcgnc
    },
  \end{multline}
  where the constitutive relation for the Cauchy stress tensor $\cstress$ reads
  \begin{equation}
    \label{eq:241}
    \cstress
    =
    -
    \thpressuredMdiff
    \identity
    +
    \lambda \left( \divergence \vec{v} \right)
    \identity
    +
    2 \nu \gradsym
    +
    \mu
    \left(
      \lcgnc
      -
      \identity
    \right).
  \end{equation}
\end{subequations}
The specific heat at constant volume $\cheatvol$ and the thermodynamic pressure $\thpressuredMdiff$ are calculated from the Helmholtz free energy \emph{ansatz} via formula~\eqref{eq:65} and~\eqref{eq:204} respectively. The final full system of governing equations is shown in Summary~\ref{summary:compressible-diff}.

\input{text/summary-c}

\subsection{Incompressible fluid}
\label{sec:incompressible-fluid}

As in the previous case, it is again possible to develop an incompressible variant of the model discussed in Section~\ref{sec:full-syst-govern}. The counterpart of the Helmholtz free energy \emph{ansatz}~\eqref{eq:201} reads
\begin{subequations}
  \label{eq:74}
  \begin{equation}
    \label{eq:261}
    \fenergy
    =_{\bydefinition}
    \widetilde{\fenergy} \left(\temp\right)
    +
    \frac{\mu}{2\rho}
    \left(
      \Tr \lcgnc
      -
      3
      -
      \ln \det \lcgnc
    \right)
    ,
  \end{equation}
  where $\rho$ is a constant. The entropy evolution equation remains almost the same as in~\eqref{eq:206} except for the term
  $
  \left( J_{\divergence \vec{v}} \right)
  \divergence{\vec{v}}
  $
  that vanishes by virtue of the incompressibility of the fluid. The counterpart of the entropy production \emph{ansatz}~\eqref{eq:213} is in the incompressible case
  \begin{equation}
    \label{eq:262}
    \widetilde{\entprodc}
    =
    \frac{1}{\temp}
    \left(
      2 \nu \tensordot{\traceless{\gradsym}}{\traceless{\gradsym}}
      +
      \frac{\mu^2}{2\nu_1}
      \left(
        \Tr \lcgnc
        +
        \Tr \inverse{\lcgnc}
        -
        6
      \right)
      +
      \frac{\mu \tilde{\mu}(\temp)}{2\nu_1}
      \tensorddot{
        \nabla \lcgnc
      }
      {
        \nabla \lcgnc
      }
      +
      \frac
      {
        \kappa \absnorm{\nabla \temp}^2
      }
      {
        \temp
      }
    \right)
    .
  \end{equation}
  (Note that for an incompressible fluid one has $\traceless{\gradsym} = \gradsym$.)
\end{subequations}

\subsection{Full system of governing equations for primitive variables -- incompressible fluid}
\label{sec:full-syst-govern-3}

The Helmholtz free energy \emph{ansatz}~\eqref{eq:261} and entropy production \emph{ansatz}~\eqref{eq:262} then lead to the following governing equations for $\vec{v}$, $\mns$, $\lcgnc$ and $\temp$:
\begin{subequations}
  \label{eq:263}
  \begin{align}
    \label{eq:264}
    \divergence \vec{v} &= 0, \\
    \label{eq:265}
    \rho \dd{\vec{v}}{t} &= \divergence \cstress + \rho \vec{b}, \\
    \label{eq:266}
    \nu_1 \fid{\overline{\lcgnc}}
    +
    \mu \left(\lcgnc - \identity\right)
    &=
    \frac{1}{2}
    \bigg[
    \divergence \left( \tilde{\mu} \nabla \lcgnc \right)
    \lcgnc
    +
    \lcgnc
    \divergence \left( \tilde{\mu} \nabla \lcgnc \right)
    \bigg]
    ,
    \\
    \label{eq:267}
    \rho
    \cheatvolNSE
    \dd{\temp}{t}
    &=
    \divergence \left(\kappa \nabla \temp \right)
    +
    2 \nu \tensordot{\traceless{\gradsym}}{\traceless{\gradsym}}
    +
    \frac{\mu^2}{2\nu_1}
    \left(
      \Tr \lcgnc
      +
      \Tr \inverse{\lcgnc}
      -
      6
    \right)
    +
    \frac{\mu \tilde{\mu}(\temp)}{2\nu_1}
    \tensorddot{
      \nabla \lcgnc
    }
    {
      \nabla \lcgnc
    }
    .
  \end{align}
  (Recall that in the incompressible case one has $\traceless{\gradsym} = \gradsym$.) The constitutive relation for the Cauchy stress tensor $\cstress$ reads $\cstress = \mns \identity + 2 \nu \gradsym + \mu \traceless{\left( \lcgnc \right)}$, which can be rewritten as
  \begin{equation}
    \label{eq:268}
    \cstress = \phi \identity + 2 \nu \gradsym + \mu \lcgnc,
  \end{equation}
  where $\phi$ denotes the spherical stress, $\phi =_{\bydefinition} \mns + \frac{\mu}{3} \Tr \lcgnc$.
\end{subequations}
The mean normal stress $\mns$/spherical stress $\phi$ is in the incompressible case a \emph{primitive quantity that must be solved for}; it is not given by a constitutive relation. The specific heat at constant volume $\cheatvol$ is again calculated from the Helmholtz free energy \emph{ansatz} via the formula~\eqref{eq:65}. The final full system of governing equations is shown in Summary~\ref{summary:incompressible-diff}. Apparently, if $\tilde{\mu}=0$, then~\eqref{eq:264}, \eqref{eq:265}, \eqref{eq:266} and \eqref{eq:268} coincide with the standard governing equations for an incompressible Maxwell/Oldroyd-B viscoelastic fluid without stress diffusion, see also~\cite{hron.j.milos.v.ea:on}.

\input{text/summary-d}

\subsection{Remarks}
\label{sec:remarks-1}
Let us now focus on the model introduced in Section~\ref{sec:full-syst-govern-3} (incompressible fluid, Maxwell/Oldroyd-B, where stress diffusion is a consequence of a non-standard entropy production mechanism). A few remarks are in order.

First, we have obtained explicit formulae for the entropy flux $\entfluxc$ and energy flux $\efluxc$. This means, that we can easily identify the boundary conditions that make the system of interest isolated with respect to the energy and entropy exchange via the boundary.

Second, if we consider~$\lcgnc$ in the form $\lcgnc = \identity + \tensorq{b}$, and if we  assume that $\tensorq{b}$ is a small quantity, then the right-hand side of~\eqref{eq:266} can be approximated as
\begin{equation}
  \label{eq:221}
  \frac{1}{2}
    \bigg[
    \divergence \left( \tilde{\mu} \nabla \lcgnc \right)
    \lcgnc
    +
    \lcgnc
    \divergence \left( \tilde{\mu} \nabla \lcgnc \right)
    \bigg]
    \approx
    \divergence \left( \tilde{\mu} \nabla \tensorq{b} \right).
\end{equation}
(Taking $\lcgnc = \identity + \tensorq{b}$, where $\tensorq{b}$ is a small quantity means that we are investigating close-to-the-equilibrium flows, see Section~\ref{sec:init-value-probl} for details.) In this regime, we therefore obtain the stress diffusion term as the Laplace operator acting on the extra stress tensor $\tensorq{b}$. This is the frequently used \emph{ad hoc} form of the stress diffusion term.

\section{Models based on a nonstandard energy storage mechanism versus models based on a nonstandard entropy production mechanism}
\label{sec:models-based-nonst}

Concerning the difference between the models based on a nonstandard energy storage mechanism and models based on a nonstandard entropy production mechanism, we can in particular observe that the evolution equations for the temperature $\temp$ are qualitatively different.

Let us for example compare the temperature evolution equations in the case of incompressible Maxwell/Oldroyd-B type models, see~\eqref{eq:172} and~\eqref{eq:267}. (See Table~\ref{tab:models} for an overview of the governing equations for specific models.) In the case of the model based on a nonstandard energy storage mechanism the left-hand side of the evolution equation for the temperature reads
\begin{subequations}
  \label{eq:104}
  \begin{equation}
    \label{eq:121}
    \left(
      \rho
      \cheatvolNSE
      -
      \frac{\temp}{2}
      \ddd{\tilde{\mu}}{\temp}
      \absnorm{\nabla \Tr \lcgnc}^2
    \right)
    \dd{\temp}{t}
    ,
  \end{equation}
see~\eqref{eq:172}, while in the case of the model based on the nonstandard entropy production mechanism the left hand side of the evolution equation for the temperature reads
\begin{equation}
  \label{eq:122}
  \rho
  \cheatvolNSE
  \dd{\temp}{t}.
\end{equation}
\end{subequations}
Consequently, if the generalised specific heat capacity is understood as the coefficient multiplying the time derivative of the temperature in the temperature evolution equation, then we see that in the former case the specific heat capacity depends on the gradient of $\lcgnc$, while this is not true in the latter case. This can in principle help one to distinguish between the two alternative models.

In fact, the situation is similar to that in the case of standard polymeric liquids, where the conformational structure of the polymer may or may not contribute to the heat capacity, see for example \cite{astarita.g:thermodynamics}, \cite{sarti.gc.esposito.n:testing}, \cite{hutter.m.luap.c.ea:energy} and especially~\cite{ionescu.tc.edwards.bj.ea:energetic}. 

\section{Initial/boundary value problems for viscoelastic rate type fluids with stress diffusion}
\label{sec:init-value-probl}

Having identified the complete sets of governing equations for compressible/incompressible Maxwell/Oldroyd-B type models with stress diffusion, see Table~\ref{tab:models}, one can proceed with the solution of initial/boundary value problems.

\input{text/models-table-summary} 

\subsection{Externally driven flows}
\label{sec:extern-driv-flows}
The derived governing equations are nonlinear, hence difficult to solve especially in the nonequilibrium setting such as flows driven by an external pressure gradient and so forth. However, solutions to several \emph{mechanical} viscoelastic rate type models with a stress diffusion term have been already investigated by numerical or semi-analytical methods, see for example \cite{olmsted.pd.radulescu.o.ea:johnson-segalman}, \cite{thomases.b:analysis}, \cite{chupin.l.martin.s:stationary}, \cite{biello.ja.thomases.b:equilibrium}, \cite{chupin.l.ichim.a.ea:stationary} and~\cite{cheng.p.burroughs.mc.ea:distinguishing}. The same holds for \emph{thermomechanical} viscoelastic rate type models without stress diffusion, see \cite{hron.j.milos.v.ea:on} and also \cite{hutter.m.luap.c.ea:energy}, and partially also for thermomechanical models with stress diffusion, see~\cite{ireka.ie.chinyoka.t:analysis}. 

Consequently, we see that the solution of the proposed models is within the reach of current numerical methods, and that the models derived can be prospectively used in the analysis of complex flows. Such analysis and comparison with experiments would also enable one to decide which of the alternative models (nonstandard energy storage mechanism/nonstandard entropy production mechanism) is suitable for the particular fluid of interest. This is however beyond the scope of the current contribution.

\subsection{Container flows}
\label{sec:container-flows}
On the other hand, once the thermodynamical background of viscoelastic rate-type models with stress diffusion is identified, certain qualitative features of flows of these fluids are easy to establish. 

For example, the natural property of a fluid flow in an isolated vessel is that it in the long run reaches a spatially uniform equilibrium rest state. The question is whether this property is indeed implied by the corresponding governing equations. This important question can be hardly answered for the models ``derived'' by the \emph{ad hoc} addition of a stress diffusion term into the standard governing equations for viscoelastic rate-type fluids. However, once the models are developed from scratch using thermodynamical arguments, they naturally have this desired property.

We document the last statement by the analysis of the incompressible variants of the models, see Table~\ref{tab:models}. In particular, we consider a rigid isolated vessel occupied by an incompressible fluid described by the governing equations~\eqref{eq:168} or \eqref{eq:263}, that is an isolated vessel occupied by an incompressible Maxwell/Oldroyd-B fluid with stress diffusion induced by a nonstandard energy storage/entropy production mechanism respectively. The phrase \emph{isolated vessel} means that no energy exchange with the surroundings is allowed. Further, for the sake of simplicity we assume that $\widetilde{\fenergy} \left(\temp\right)$ in the corresponding Helmholtz free energy \emph{ansatz}~\eqref{eq:142} and \eqref{eq:261} respectively takes the form
\begin{equation}
  \label{eq:81}
  \widetilde{\fenergy} \left(\temp\right)
  =_{\bydefinition}
  - 
  \cheatvolNSE \temp
  \left(
    \ln 
    \left(
      \frac{\temp}{\tempref}
    \right)
    -
    1
  \right)
  ,
\end{equation}
where $\tempref$ is a constant arbitrary reference temperature. This leads to constant specific heat capacity at constant volume~$\cheatvolNSE$, see the standard formula~\eqref{eq:65}, which greatly simplifies the ongoing algebraic manipulations. 

We introduce the (time dependent) functional
\begin{subequations}
  \begin{equation}
    \label{eq:77}
    {\mathcal{V}}_{\mathrm{eq}}
    =_{\bydefinition}
    -
    \left(
      \netentropy - \frac{1}{\tempref} \left( \nettenergy - \widehat{\nettenergy} \right)
    \right)
    ,
  \end{equation}
  where $\netentropy$ and $\nettenergy$ denote the net entropy and the net total energy of the fluid inside the vessel occupying the domain $\Omega$, that is
  \begin{align}
    \label{eq:78}
    \netentropy &=_{\bydefinition} \int_{\Omega} \rho \entropy \, \cvolumee, \\
    \label{eq:79}
    \nettenergy &=_{\bydefinition} \int_{\Omega} \left( \rho \frac{1}{2} \absnorm{\vec{v}}^2 + \rho \ienergy \right) \, \cvolumee,
  \end{align}
\end{subequations}
and $\widehat{\nettenergy}$ denotes a fixed constant net total energy value. (This energy value will be specified later.) Note that once we know the specific Helmholtz free energy $\fenergy$, which is the case, then we can easily write down the explicit formulae for the specific entropy $\entropy$ and the specific internal energy $\ienergy$ by appealing to the standard thermodynamical identities, see the formulae~\eqref{eq:16}, \eqref{eq:123} and \eqref{eq:124}.

Following \cite{coleman.bd:on} and \cite{gurtin.me:thermodynamics}, we claim that the functional ${\mathcal{V}}_{\mathrm{eq}}$ is a Lyapunov functional\footnote{See \cite{la-salle.j.lefschetz.s:stability}, \cite{yoshizawa.t:stability} or \cite{henry.d:geometric} for the stability analysis theory of infinite-dimensional dynamical systems based on the Lyapunov method. The proposal for using \eqref{eq:77} as a Lyapunov functional characterising the stability of the equilibrium rest state is also articulated in other works on continuum thermodynamics, see for example~\cite{grmela.m.ottinger.hc:dynamics} or~\cite{silhavy.m:mechanics}. See also~\cite{bul-cek.m.malek.j.ea:thermodynamics} for a proposal concerning the possible extension of this procedure to stability analysis of non-equilibrium steady states in thermodynamically open systems.} for the spatially uniform equilibrium rest state defined by the triple $[\vec{v}, \temp, \lcgnc] = [ \vec{0}, \tempeq, \identity]$, where $\tempeq$ is a spatially uniform temperature field. Let us now show that ${\mathcal{V}}_{\mathrm{eq}}$ indeed has all the properties of the Lyapunov functional -- it is nonnegative, it vanishes only at the spatially uniform equilibrium rest state and it decreases in time. Once these properties are verified, we can conclude that the spatially uniform equilibrium rest state is unconditionally asymptotically stable, which means that the fluid in a closed vessel indeed has a natural tendency to reach the equilibrium rest state.

First, it is straightforward to check that the spatially uniform equilibrium rest state $[\vec{v}, \temp, \lcgnc] = [ \vec{0}, \tempeq, \identity]$ is indeed a solution to the governing equations, hence it is meaningful to investigate its stability.

Second, if no energy exchange with surroundings is allowed, then we need to set
\begin{equation}
  \label{eq:88}
  \left. \vec{v} \right|_{\partial \Omega} = 0,
\end{equation}
which prohibits the \emph{mechanical} energy exchange with the surroundings. Further, the requirement of no energy exchange implies that the \emph{energy flux} $\efluxc$ must also vanish on the boundary $\partial \Omega$ as well. In both models we have an explicit formula for the energy flux $\efluxc$, see~\eqref{eq:198} and~\eqref{eq:227} respectively, 
\begin{subequations}
  \label{eq:83}
  \begin{align}
    \label{eq:82}
    \efluxc
    &=
    -
    \kappa \nabla \temp
    -
    \tilde{\mu}
    \left( \nabla \Tr \lcgnc \right)  \dd{}{t} \left( \Tr \lcgnc \right)
    ,
    \\
    \label{eq:87}
    \efluxc
    &=
    -
    \kappa
    \nabla \temp
    -
    \frac{\mu\tilde{\mu}}{4\nu_1}
    \Tr
    \bigg[
    \left(\nabla \lcgnc \right)
    \left(
      \lcgnc
      -
      \identity
    \right)
    +
    \left(
      \lcgnc
      -
      \identity
    \right)
    \left(\nabla \lcgnc \right)
    \bigg]
    .
  \end{align}
\end{subequations}
This implies that the energy flux $\efluxc$ in both cases vanishes provided that for example we set
\begin{subequations}
  \label{eq:84}
  \begin{align}
    \label{eq:85}
    \left.
      \vectordot{\nabla \temp}{\vec{n}}
    \right|_{\partial \Omega}
    &=
    0
    ,
    \\
    \label{eq:86}
    \left.
      \vectordot{\nabla \Tr \lcgnc}{\vec{n}}
    \right|_{\partial \Omega}
    &=
    0
    .
  \end{align}
\end{subequations}
(See also the discussion in Section~\ref{sec:remarks}. Recall that~\eqref{eq:87} is in terms of components given by the expressions introduced in~\eqref{eq:217}.) Consequently, if we enforce the boundary conditions~\eqref{eq:88} and \eqref{eq:84} then the vessel is indeed a thermodynamically isolated system. Note that~\eqref{eq:85} also implies that the \emph{entropy flux} $\entfluxc$ vanishes on the boundary.

If we apply the boundary conditions~\eqref{eq:88} and~\eqref{eq:84}, then the net total energy $\nettenergy$ is conserved,
\begin{subequations}
  \label{eq:90}
  \begin{equation}
    \label{eq:80}
    \dd{\nettenergy}{t} = 0.
  \end{equation}
  (This observation is a straightforward consequence of the evolution equation for the total energy, see~\eqref{eq:195}, and the Stokes theorem.) Consequently, we know, that the net total energy $\nettenergy$ remains constant, and that it is equal to the net total energy $\widehat{\nettenergy}$ at the initial time. Further, the net entropy $\netentropy$ is a nondecreasing function of time,
  \begin{equation}
    \label{eq:89}
    \dd{\netentropy}{t} = \int_{\Omega} \entprodc \, \cvolumee \geq 0.
  \end{equation}
\end{subequations}
This observation is a straightforward consequence of the evolution equation for the specific entropy, see \eqref{eq:24}, the Stokes theorem, and the fact that the entropy flux $\entfluxc$ vanishes on the boundary of $\Omega$. The sign of the time derivative is determined by the sign of the entropy production $\entprodc = \widetilde{\entprodc}$ \emph{ansatz}, where $\widetilde{\entprodc}$ is given by
\begin{subequations}
  \label{eq:93}
  \begin{align}
    \label{eq:94}
    \widetilde{\entprodc}
    &=
      \frac{1}{\temp}
      \left(
      2 \nu \tensordot{\gradsym}{\gradsym}
      +
      \frac{\nu_1}{2}
      \Tr
      \left(
      \fid{\overline{\lcgnc}}\inverse{\lcgnc}\fid{\overline{\lcgnc}}
      \right)
      +
      \frac
      {
      \kappa \absnorm{\nabla \temp}^2
      }
      {
      \temp
      }
      \right)
      ,
    \\
    \label{eq:95}
    \widetilde{\entprodc}
    &=
      \frac{1}{\temp}
      \left(
      2 \nu \tensordot{\gradsym}{\gradsym}
      +
      \frac{\mu^2}{2\nu_1}
      \left(
      \Tr \lcgnc
      +
      \Tr \inverse{\lcgnc}
      -
      6
      \right)
      +
      \frac{\mu \tilde{\mu}(\temp)}{2\nu_1}
      \tensorddot{
      \nabla \lcgnc
      }
      {
      \nabla \lcgnc
      }
      +
      \frac
      {
      \kappa \absnorm{\nabla \temp}^2
      }
      {
      \temp
      }
      \right)
      ,
  \end{align}
\end{subequations}
where the first formula holds in the case of a fluid with the nonstandard energy storage mechanism, while the second formula holds for a fluid with the nonstandard entropy production mechanism, see \eqref{eq:143} and \eqref{eq:262} respectively\footnote{See also identity~\eqref{eq:57}, and the entropy production \emph{ansatz} \eqref{eq:22} and~\eqref{eq:23} respectively.}.

The entropy production is in both cases a nonnegative quantity. Moreover, we see that the entropy production in both cases vanishes if and only if the velocity field $\vec{v}$, the temperature field $\temp$ and the $\lcgnc$ field are spatially homogeneous fields, and if $\lcgnc = \identity$, $\vec{v}= \vec{0}$ and $\temp = \tempref$. This means that the entropy production vanishes if and only if the spatially uniform equilibrium rest state is reached. Using~\eqref{eq:90} and the definition of the candidate for a Lyapunov functional ${\mathcal{V}}_{\mathrm{eq}}$, we see that ${\mathcal{V}}_{\mathrm{eq}}$ decreases in time, 
\begin{equation}
  \label{eq:91}
  \dd{{\mathcal{V}}_{\mathrm{eq}}}{t} =  - \int_{\Omega} \entprodc \, \cvolumee \leq 0,
\end{equation}
and that the time derivative vanishes at the equilibrium rest state. This concludes the discussion on the time derivative of the proposed Lyapunov functional.

Third, let us fix the reference temperature $\tempref$ as
\begin{equation}
  \label{eq:92}
  \tempref 
  =
  \frac{\widehat{\nettenergy}}{\rho \cheatvolNSE \absnorm{\Omega}},
\end{equation}
where $\widehat{\nettenergy}$ denotes the value of the net total energy at the initial time, and $\absnorm{\Omega}$ denotes the volume of the vessel. By virtue of the conservation of the net total energy, we see that~\eqref{eq:92} implies
\begin{equation*}
  \tempeq = \tempref.
\end{equation*}
In other words, the temperature attained at the equilibrium rest state must correspond to the initial net total energy value in the vessel.

Finally, let us investigate the proposed Lyapunov functional and its nonnegativity. Using the particular formula~\eqref{eq:81} for the purely thermal part of the Helmholtz free energy $\widetilde{\fenergy}$, we see that the Helmholtz free energy $\fenergy$ is given by the formulae
\begin{subequations}
  \label{eq:96}
  \begin{align}
    \label{eq:97}
    \fenergy
    &=
    - 
    \cheatvolNSE \temp
    \left(
    \ln 
    \left(
    \frac{\temp}{\tempref}
    -
    1
    \right)
    \right)
    +
    \frac{\mu}{2\rho}
    \left(
      \Tr \lcgnc
      -
      3
      -
      \ln \det \lcgnc
    \right)
    +
    \frac{\tilde{\mu}(\temp)}{2\rho}
    \absnorm{ \nabla \Tr \lcgnc}^2
    ,
    \\
    \label{eq:98}
    \fenergy
    &=
    - 
    \cheatvolNSE \temp
    \left(
    \ln 
    \left(
    \frac{\temp}{\tempref}
    -
    1
    \right)
    \right)
    +
    \frac{\mu}{2\rho}
    \left(
      \Tr \lcgnc
      -
      3
      -
      \ln \det \lcgnc
    \right)
    .
  \end{align}
\end{subequations}
The first formula holds for a fluid with a nonstandard energy storage mechanism, while the second formula holds for a fluid with a nonstandard entropy production mechanism, see~\eqref{eq:142} and \eqref{eq:261}~respectively. Consequently, in the subsequent analysis of the proposed Lyapunov functional we can work with~\eqref{eq:97} since \eqref{eq:98} is a special case of~\eqref{eq:97} for $\tilde{\mu}=0$.

Using the definition of the proposed Lyapunov functional, see~\eqref{eq:77}, we see that
\begin{multline}
  \label{eq:99}
  {\mathcal{V}}_{\mathrm{eq}}
  =
  -
  \int_{\Omega}
  \left[
    \rho \entropy
    -
    \frac{1}{\tempref}
    \left(
      \frac{1}{2} \rho \absnorm{\vec{v}}^2
      +
      \rho \ienergy
    \right)
    +
    \rho \cheatvolNSE
  \right]
  \,
  \cvolumee
  =
  \frac{1}{\tempref}
  \int_{\Omega}
  \left[
    \rho
    \left(
      \ienergy
      -
      \temp \entropy
    \right)
    +
    \rho
    \left(
      \temp - \tempref
    \right)
    \entropy
    +
    \frac{1}{2} \rho \absnorm{\vec{v}}^2
    -
    \rho \cheatvolNSE \tempref
  \right]
  \,
  \cvolumee
  \\
  =
  \frac{1}{\tempref}
  \int_{\Omega}
  \left[
    \rho
    \fenergy
    -
    \rho
    \pd{\fenergy}{\temp}
    \left(
      \temp - \tempref
    \right)
    -
    \rho \cheatvolNSE \tempref
    +
    \frac{1}{2} \rho \absnorm{\vec{v}}^2
  \right]
  \,
  \cvolumee
  ,
\end{multline}
where we have used the standard relations between the Helmholtz free energy, the internal energy and the entropy, see for example Section~\ref{sec:evol-equat-entr-1}. A straightforward calculation for the Helmholtz free energy $\fenergy$ given by~\eqref{eq:97} reveals that
\begin{multline}
  \label{eq:100}
  \rho
  \fenergy
  -
  \rho
  \pd{\fenergy}{\temp}
  \left(
    \temp - \tempref
  \right)
  -
  \rho \cheatvolNSE \tempref
  \\
  =
  \rho \cheatvolNSE
  \left[
    \frac{\temp}{\tempref}
    -
    \ln
    \left(
      \frac{\temp}{\tempref}
    \right)
    -
    1
  \right]
  +
  \frac{\mu}{2}
  \left(
    \Tr \lcgnc
    -
    3
    -
    \ln \det \lcgnc
  \right)
  +
  \frac{1}{2}
  \absnorm{ \nabla \Tr \lcgnc}^2
  \left[
    \tilde{\mu}(\temp)
    +
    \dd{\tilde{\mu}}{\temp}(\temp)
    \left(
      \tempref - \temp
    \right)
  \right]
  .
\end{multline}
Using the Lagrange form of the remainder of Taylor expansion of $\tilde{\mu}(\tempref)$, we see that
\begin{equation}
  \label{eq:101}
  \tilde{\mu}(\tempref)
  =
  \tilde{\mu}(\temp)
  +
  \left.
    \dd{\tilde{\mu}}{\temp}
  \right|_{\temp = \temp + s (\tempref - \temp) }
  \left(
    \tempref - \temp
  \right)
  +
  \frac{1}{2}
  \left.
    \ddd{\tilde{\mu}}{\temp}
  \right|_{\temp = \temp + s (\tempref - \temp) }
  \left(
    \tempref - \temp
  \right)^2
  ,
\end{equation}
where $s \in [0,1]$, which means that we can rewrite~\eqref{eq:99} as
\begin{multline}
  \label{eq:102}
  {\mathcal{V}}_{\mathrm{eq}}
  =
  \frac{1}{\tempref}
  \int_{\Omega}
  \frac{1}{2}
  \rho
  \absnorm{\vec{v}}^2
  \,
  \cvolumee
  +
  \frac{1}{\tempref}
  \int_{\Omega}
  \rho \cheatvolNSE
  \left[
    \frac{\temp}{\tempref}
    -
    \ln
    \left(
      \frac{\temp}{\tempref}
    \right)
    -
    1
  \right]
  \,
  \cvolumee
  +
  \frac{1}{\tempref}
  \int_{\Omega}
  \frac{\mu}{2}
  \left(
    \Tr \lcgnc
    -
    3
    -
    \ln \det \lcgnc
  \right)
  \,
  \cvolumee
  \\
  +
  \frac{1}{\tempref}
  \int_{\Omega}
  \frac{1}{2}
  \absnorm{ \nabla \Tr \lcgnc}^2
  \tilde{\mu}(\tempref)
  \,
  \cvolumee
  -
  \frac{1}{\tempref}
  \int_{\Omega}
  \frac{1}{4}
  \absnorm{ \nabla \Tr \lcgnc}^2
  \left.
    \ddd{\tilde{\mu}}{\temp}
  \right|_{\temp = \temp + s (\tempref - \temp) }
  \left(
    \tempref - \temp
  \right)^2
  \,
  \cvolumee
  .
\end{multline}

The function
\begin{equation}
  \label{eq:103}
  g(\temp)
  =_{\bydefinition}
  \frac{\temp}{\tempref}
  -
  \ln
  \left(
    \frac{\temp}{\tempref}
  \right)
  -
  1
\end{equation}
is for $\temp>0$ nonegative, and it vanishes if and only if $\temp=\tempref$. Further, the function
$
\Tr \lcgnc
-
3
-
\ln \det \lcgnc
$
is nonnegative for any symmetric positive definite matrix $\lcgnc$, and it vanishes if and only if $\lcgnc = \identity$. Consequently, if we assume that $\tilde{\mu}$ is a concave function, that is $\ddd{\tilde{\mu}}{\temp} \leq 0$, then we immediately see that the functional ${\mathcal{V}}_{\mathrm{eq}}$ is nonnegative. (The concavity of $\tilde{\mu}$ seems to be a plausible assumption. It is definitely valid for the fluid investigated by~\cite{mohammadigoushki.h.muller.sj:flow}, see Figure~\ref{fig:temeprature-stress-diffusion}. In this case the experimental data indicate that $\tilde{\mu}$ is a linear function of the temperature. Note, however, that~\cite{mohammadigoushki.h.muller.sj:flow} have used the diffusive Johnson--Segalman model, which is a more complex viscoelastic rate-type model than the Maxwell/Oldroyd-B model.) Moreover, the functional ${\mathcal{V}}_{\mathrm{eq}}$ vanishes if and only if $\temp = \tempref = \tempeq$, $\vec{v} = 0$ and $\lcgnc = \identity$, that is at the equilibrium rest state.

Therefore, we can conclude that ${\mathcal{V}}_{\mathrm{eq}}$ is indeed a Lyapunov functional characterising the (nonlinear) asymptotic stability of the spatially uniform equilibrium rest state both for a fluid with a nonstandard energy storage mechanism and a fluid with a nonstandard entropy production mechanism. Note that the stability result holds with respect to \emph{any} initial disturbance, hence we have in fact shown \emph{unconditional asymptotic stability of the equilibrium rest state}. Such a result would be difficult, if not impossible, to obtain without the proper understanding of the thermodynamical background of the corresponding governing equations. 

Finally, we note that the analysis outlined above provides a justification for the notation $\temp$ for the temperature. Once the system approaches the spatially uniform equilibrium rest state, that is the state where the temperature is defined without any controversy, we see that the quantity $\temp$ coincides with the equilibrium temperature $\tempeq$ of the fluid inside the vessel.

\section{Conclusion}
\label{sec:conclusion-2}
We have derived thermodynamically consistent models for compressible/incompressible Maxwell/Oldroyd-B type fluids with a stress diffusion term. Following~\cite{rajagopal.kr.srinivasa.ar:on*7} we have shown that the governing equations for all primitive variables can be derived via the specification of two scalar quantities, namely the Helmholtz free energy and the entropy production. In particular, we have identified the corresponding temperature evolution equation that must supplement the governing equations for mechanical variables, and that must be used if one is interested in thermomechanical coupling.

The stress diffusion term has been interpreted as a symptom of either a \emph{nonstandard energy storage mechanism} or a \emph{nonstandard entropy production mechanism}. In both cases the \emph{ansatz} for the Helmholtz free energy and the \emph{ansatz} for the entropy production respectively included a gradient (nonlocal) term, which subsequently resulted in a stress diffusion term in the evolution equation for the extra stress tensor.

If the stress diffusion is interpreted as a consequence of a \emph{nonstandard energy storage mechanism}, then the resulting governing equations include, besides the stress diffusion term, other additional terms. In particular the Cauchy stress tensor contains Korteweg type terms. On the other hand, if the stress diffusion is interpreted as a consequence of a \emph{nonstandard entropy production mechanism}, then the stress diffusion term is the only additional term in the evolution equations for the mechanical quantities (compared to the standard Maxwell/Oldroyd-B model). The combination of the two approaches, and more elaborate choices of the \emph{ansatz} for the Helmholtz free energy or the entropy production, can be further exploited in the development of more complex models that go beyond Maxwell/Oldroyd-B type models. Diffusive Johnson--Segalman model, see~\cite{lu.cyd.olmsted.pd.ea:effects} and \cite{olmsted.pd.radulescu.o.ea:johnson-segalman} or diffusive Giesekus model, see~\cite{helgeson.me.vasquez.pa.ea:rheology} and \cite{helgeson.me.reichert.md.ea:relating}, are in this respect natural candidates.

The key thermodynamical relations, including the entropic equation of state, has been identified as a byproduct of the derivation of the governing equations. This means that the complete arsenal of generic thermodynamics-based methods for the investigation of the dynamics of non-equilibrium systems, see for example~\cite{glansdorff.p.prigogine:thermodynamic} and subsequent works, is unlocked for possible future applications. In particular, such thermodynamics-based methods may provide an interesting insight into the stability of viscoelastic fluids of interest, and they may also provide a helpful guide in the mathematical analysis of the corresponding governing equations.


%% file: text/summary-a.tex
\begin{summary}[Compressible viscoelastic rate-type fluid with stress diffusion and temperature-dependent stress diffusion coefficient $\tilde{\mu}$, splitting parameters $\alpha=1$, $\beta=0$]
  \label{summary:compressible-Tr-alt}
  Helmholtz free energy $\fenergy$, \emph{ansatz}~\eqref{eq:12}:
  \begin{subequations}
    \label{eq:144}
    \begin{align}
      \label{eq:145}
      \fenergy
      &=
      \widetilde{\fenergy} \left(\temp, \rho\right)
      +
      \frac{1}{\rho}\widetilde{\widetilde{\fenergy}},
      \\
      \widetilde{\widetilde{\fenergy}}
      &=
      \frac{\mu}{2}
      \left(
        \Tr \lcgnc
        -
        3
        -
        \ln \det \lcgnc
      \right)
      +
      \frac{\tilde{\mu}(\temp)}{2}
      \absnorm{ \nabla \Tr \lcgnc}^2.
    \end{align}
  \end{subequations}

  Entropy production $\entprodc$, \emph{ansatz}~\eqref{eq:44} rewritten in terms of primitive quantities, see~\eqref{eq:61}:
  \begin{equation}
    \label{eq:146}
    \widetilde{\entprodc}
    =
    \frac{1}{\temp}
    \left(
      \frac{2 \nu + 3 \lambda}{3} \left(\divergence \vec{v}\right)^2
      +
      2 \nu \tensordot{\traceless{\gradsym}}{\traceless{\gradsym}}
      +
      \frac{\nu_1}{2}
      \Tr
      \left(
        \fid{\overline{\lcgnc}}\inverse{\lcgnc}\fid{\overline{\lcgnc}}
      \right)
      +
      \frac
      {
        \kappa \absnorm{\nabla \temp}^2
      }
      {
        \temp
      }
    \right).
  \end{equation}

  Material parameters (constants): $\mu$, $\nu$, $\lambda$, $\nu_1$, $\kappa$.

  Material parameters (temperature-dependent): $\tilde{\mu}$.

  Evolution equations for $\rho$, $\vec{v}$, $\lcgnc$:
  \begin{subequations}
    \label{eq:147}
    \begin{align}
      \label{eq:125}
      \dd{\rho}{t} + \rho \divergence \vec{v} &= 0, \\
      \label{eq:126}
      \rho \dd{\vec{v}}{t} &= \divergence \cstress + \rho \vec{b}, \\
      \label{eq:127}
      \nu_1 \fid{\overline{\lcgnc}}
      +
      \mu \left(\lcgnc - \identity\right)
      &=
      2
      \tilde{\mu}
      \left( \Delta \Tr \lcgnc \right)
      \lcgnc
      +
      2
      \dd{\tilde{\mu}}{\temp}
      \left[
        \vectordot{
          \left(\nabla \Tr \lcgnc\right)
        }
        {
          \nabla \temp
        }
      \right]
      \lcgnc.
    \end{align}

    Evolution equation for $\temp$:
    \begin{multline}
      \label{eq:128}
      \left(
        \rho
        \cheatvolNSE
        -
        \frac{\temp}{2}
        \ddd{\tilde{\mu}}{\temp}
        \absnorm{\nabla \Tr \lcgnc}^2
      \right)
      \dd{\temp}{t}
      =
      \tensordot{\cstress_{\mathrm{vis}}}{\gradsym}
      -
      \temp
      \pd{p_{\mathrm{eq}}}{\temp} \divergence \vec{v}
      +
      \divergence\left(\kappa \nabla \temp \right)
      -
      \temp
      \tensordot{
        \pd{
          \tensorq{P}_{\mathrm{eq}}
        }
        {
          \temp
        }
      }
      {
        \traceless{\gradsym}
      }
      +
      \frac{\nu_1}{2}
      \Tr
      \left(
        \fid{\overline{\lcgnc}}\inverse{\lcgnc}\fid{\overline{\lcgnc}}
      \right)
      \\
      -
      2
      \temp
      \dd{\tilde{\mu}}{\temp}
      \left( \Delta \Tr \lcgnc \right)
      \left(
        \tensordot{
          \rcgnc
        }
        {
          \gradsymrn
        }
      \right)
      +
      \temp
      \dd{\tilde{\mu}}{\temp}
      \divergence
      \left[
        \left(\nabla \Tr \lcgnc \right)
        \dd{}{t}
        \Tr \lcgnc
      \right].
    \end{multline}
  \end{subequations}

  Auxiliary terms:
  \begin{align}
    \label{eq:129}
    \dd{}{t} \Tr \lcgnc
    &=
    2 \tensordot{\lcgnc}{\gradsym}
    -
    \frac{\mu}{\nu_1}
    \left(
      \Tr \lcgnc - 3
    \right)
    +
    \frac{2 \tilde{\mu}}{\nu_1}
    \left(
      \Delta \Tr \lcgnc
    \right)
    \Tr \lcgnc
    +
    \frac{2}{\nu_1}
    \dd{\tilde{\mu}}{\temp}
    \left[
      \vectordot{\left(\nabla \Tr \lcgnc\right)}{\nabla \temp}
    \right]
    \Tr \lcgnc,
    \\
    \label{eq:130}
    \tensordot{\rcgnc}{\gradsymrn}
    &=
    \frac{\mu}{2\nu_1}
    \left(
      \Tr \lcgnc - 3
    \right)
    -
    \frac{\tilde{\mu}}{\nu_1}
    \left(
      \Delta \Tr \lcgnc
    \right)
    \Tr \lcgnc
    -
    \frac{1}{\nu_1}
    \dd{\tilde{\mu}}{\temp}
    \left[
      \vectordot{\left(\nabla \Tr \lcgnc\right)}{\nabla \temp}
    \right]
    \Tr \lcgnc.
  \end{align}

Constitutive relation for the Cauchy stress tensor $\cstress$:
\begin{equation}
  \label{eq:131}
  \cstress
  =
  -
  p_{\mathrm{eq}}
  \identity
  -
  \tensorq{P}_{\mathrm{eq}}
  +
  \cstress_{\mathrm{vis}}
  -
  2
  \dd{\tilde{\mu}}{\temp}
  \left[
    \vectordot{\left(\nabla \Tr \lcgnc\right)}{\nabla \temp}
  \right]
  \lcgnc.
\end{equation}

Definitions of $p_{\mathrm{eq}}$, $\tensorq{P}_{\mathrm{eq}}$ and $\cstress_{\mathrm{vis}}$:
\begin{subequations}
  \label{eq:132}
  \begin{align}
    \label{eq:133}
    \tensorq{P}_{\mathrm{eq}}
    &=
    -
    \mu \traceless{\left( \lcgnc \right)}
    +
    \tilde{\mu}
    \traceless{
      \left[
        \tensortensor{\left( \nabla  \Tr \lcgnc \right)}{\left( \nabla  \Tr \lcgnc \right)}
      \right]
    }
    +
    2
    \tilde{\mu}
    \left(\Delta \Tr \lcgnc\right)
    \traceless{
      \left(
        \lcgnc
      \right)
    }
    ,
    \\
    \label{eq:134}
    p_{\mathrm{eq}}
    &=
    \thpressuredMTr
    -
    \frac{\mu}{3}
    \Tr \lcgnc
    +
    \mu
    +
    \frac{\tilde{\mu}}{3}
    \Tr
    \left[
      \tensortensor{\left( \nabla  \Tr \lcgnc \right)}{\left( \nabla  \Tr \lcgnc \right)}
    \right]
    +
    \frac{2}{3}\tilde{\mu}
    \Tr \lcgnc
    \left(
      \Delta
      \Tr \lcgnc
    \right),
    \\
    \label{eq:135}
    \cstress_{\mathrm{vis}}
    &=
    \lambda \left( \divergence \vec{v} \right)
    \identity
    +
    2
    \nu
    \gradsym.
  \end{align}

Quantities derived from the Helmholtz free energy:
\begin{align}
  \label{eq:136}
  \thpressureNSE &= \rho^2 \pd{\widetilde{\fenergy}}{\rho}, \\
  \label{eq:137}
  \thpressuredMTr &= \thpressureNSE - \widetilde{\widetilde{\fenergy}},\\
  \label{eq:138}
  \cheatvolNSE
  &=
  -
  \temp
  \ppd{
    \widetilde{\fenergy}
  }{\temp}.
\end{align}
\end{subequations}

Constitutive relation for the energy flux $\efluxc$, has been already used in~\eqref{eq:128}:
\begin{equation}
  \label{eq:139}
  \efluxc
  =
  -
  \kappa \nabla \temp
  -
  \tilde{\mu}
  \left( \nabla \Tr \lcgnc \right)  \dd{}{t} \left( \Tr \lcgnc \right).
\end{equation}

Constitutive relation for the entropy flux $\entfluxc$:
\begin{equation}
  \label{eq:140}
  \entfluxc
  =
  -
  \frac{
    \kappa \nabla \temp
  }
  {
    \temp
  }.
\end{equation}
\end{summary}


%% file: text/summary-b.tex
\begin{summary}[Incompressible viscoelastic rate-type fluid with stress diffusion and temperature-dependent stress diffusion coefficient $\tilde{\mu}$, splitting parameters $\alpha=1$, $\beta=0$]
  \label{summary:incompressible-Tr-alt}
  Helmholtz free energy $\fenergy$, incompressible variant of \emph{ansatz}~\eqref{eq:12}, see~\eqref{eq:142}:
  \begin{subequations}
    \label{eq:176}
    \begin{align}
      \label{eq:177}
      \fenergy
      &=
      \widetilde{\fenergy} \left(\temp\right)
      +
      \frac{1}{\rho}\widetilde{\widetilde{\fenergy}},
      \\
      \widetilde{\widetilde{\fenergy}}
      &=
      \frac{\mu}{2}
      \left(
        \Tr \lcgnc
        -
        3
        -
        \ln \det \lcgnc
      \right)
      +
      \frac{\tilde{\mu}(\temp)}{2}
      \absnorm{ \nabla \Tr \lcgnc}^2.
    \end{align}
  \end{subequations}

  Entropy production $\entprodc$, incompressible variant of \emph{ansatz}~\eqref{eq:44}, see~\eqref{eq:143}, rewritten in terms of primitive quantities:
  \begin{equation}
    \label{eq:178}
    \widetilde{\entprodc}
    =
    \frac{1}{\temp}
    \left(
      2 \nu \tensordot{\traceless{\gradsym}}{\traceless{\gradsym}}
      +
      \frac{\nu_1}{2}
      \Tr
      \left(
        \fid{\overline{\lcgnc}}\inverse{\lcgnc}\fid{\overline{\lcgnc}}
      \right)
      +
      \frac
      {
        \kappa \absnorm{\nabla \temp}^2
      }
      {
        \temp
      }
    \right).
  \end{equation}

  Material parameters (constants): $\mu$, $\nu$, $\lambda$, $\nu_1$, $\kappa$.

  Material parameters (temperature-dependent): $\tilde{\mu}$.

  Evolution equations for $\mns$, $\vec{v}$, $\lcgnc$:
  \begin{subequations}
    \label{eq:179}
    \begin{align}
      \label{eq:180}
      \divergence \vec{v} &= 0, \\
      \label{eq:181}
      \rho \dd{\vec{v}}{t} &= \divergence \cstress + \rho \vec{b}, \\
      \label{eq:182}
      \nu_1 \fid{\overline{\lcgnc}}
      +
      \mu \left(\lcgnc - \identity\right)
      &=
      2
      \tilde{\mu}
      \left( \Delta \Tr \lcgnc \right)
      \lcgnc
      +
      2
      \dd{\tilde{\mu}}{\temp}
      \left[
        \vectordot{
          \left(\nabla \Tr \lcgnc\right)
        }
        {
          \nabla \temp
        }
      \right]
      \lcgnc.
    \end{align}

    Evolution equation for $\temp$:
    \begin{multline}
      \label{eq:183}
      \left(
        \rho
        \cheatvolNSE
        -
        \frac{\temp}{2}
        \ddd{\tilde{\mu}}{\temp}
        \absnorm{\nabla \Tr \lcgnc}^2
      \right)
      \dd{\temp}{t}
      =
      \tensordot{\tensorq{S}}{\gradsym}
      +
      \divergence\left(\kappa \nabla \temp \right)
      -
      \temp
      \tensordot{
        \pd{
          \tensorq{P}
        }
        {
          \temp
        }
      }
      {
        \gradsym
      }
      +
      \frac{\nu_1}{2}
      \Tr
      \left(
        \fid{\overline{\lcgnc}}\inverse{\lcgnc}\fid{\overline{\lcgnc}}
      \right)
      \\
      -
      2
      \temp
      \dd{\tilde{\mu}}{\temp}
      \left( \Delta \Tr \lcgnc \right)
      \left(
        \tensordot{
          \rcgnc
        }
        {
          \gradsymrn
        }
      \right)
      +
      \temp
      \dd{\tilde{\mu}}{\temp}
      \divergence
      \left[
        \left(\nabla \Tr \lcgnc \right)
        \dd{}{t}
        \Tr \lcgnc
      \right].
    \end{multline}
  \end{subequations}

  Auxiliary terms:
  \begin{align}
    \label{eq:184}
    \dd{}{t} \Tr \lcgnc
    &=
    2 \tensordot{\lcgnc}{\gradsym}
    -
    \frac{\mu}{\nu_1}
    \left(
      \Tr \lcgnc - 3
    \right)
    +
    \frac{2 \tilde{\mu}}{\nu_1}
    \left(
      \Delta \Tr \lcgnc
    \right)
    \Tr \lcgnc
    +
    \frac{2}{\nu_1}
    \dd{\tilde{\mu}}{\temp}
    \left[
      \vectordot{\left(\nabla \Tr \lcgnc\right)}{\nabla \temp}
    \right]
    \Tr \lcgnc,
    \\
    \label{eq:185}
    \tensordot{\rcgnc}{\gradsymrn}
    &=
    \frac{\mu}{2\nu_1}
    \left(
      \Tr \lcgnc - 3
    \right)
    -
    \frac{\tilde{\mu}}{\nu_1}
    \left(
      \Delta \Tr \lcgnc
    \right)
    \Tr \lcgnc
    -
    \frac{1}{\nu_1}
    \dd{\tilde{\mu}}{\temp}
    \left[
      \vectordot{\left(\nabla \Tr \lcgnc\right)}{\nabla \temp}
    \right]
    \Tr \lcgnc.
  \end{align}

  Constitutive relation for the Cauchy stress tensor $\cstress$:
  \begin{equation}
    \label{eq:186}
    \cstress
    =
    \mns \identity + \tensorq{S}
    -
    2
    \dd{\tilde{\mu}}{\temp}
    \left[
      \vectordot{\left(\nabla \Tr \lcgnc\right)}{\nabla \temp}
    \right]
    \lcgnc.
  \end{equation}

  Definitions of $\traceless{\cstress}$ and $\tensorq{P}$:
  \begin{subequations}
    \label{eq:187}
    \begin{align}
      \label{eq:188}
      \tensorq{S}
      &=
      2 \nu \traceless{\gradsym}
      -
      \tensorq{P}
      ,
      \\
      \label{eq:189}
      \tensorq{P}
      &=
      -
      \mu \traceless{\left( \lcgnc \right)}
      +
      \tilde{\mu}
      \traceless{
        \left[
          \tensortensor{\left( \nabla  \Tr \lcgnc \right)}{\left( \nabla  \Tr \lcgnc \right)}
        \right]
      }
      +
      2
      \tilde{\mu}
      \left(\Delta \Tr \lcgnc\right)
      \traceless{
        \left(
          \lcgnc
        \right)
      }.
    \end{align}

    Quantities derived from the Helmholtz free energy:
    \begin{equation}
      \label{eq:190}
      \cheatvolNSE
      =
      -
      \temp
      \ppd{
        \widetilde{\fenergy}
      }{\temp}.
    \end{equation}
  \end{subequations}

  Constitutive relation for the energy flux $\efluxc$, has been already used in~\eqref{eq:183}:
  \begin{equation}
    \label{eq:191}
    \efluxc
    =
    -
    \kappa \nabla \temp
    -
    \tilde{\mu}
    \left( \nabla \Tr \lcgnc \right)  \dd{}{t} \left( \Tr \lcgnc \right).
  \end{equation}

  Constitutive relation for the entropy flux $\entfluxc$:
  \begin{equation}
    \label{eq:192}
    \entfluxc
    =
    -
    \frac{\kappa \nabla \temp}{\temp}.
\end{equation}
\end{summary}


%% file: text/summary-c.tex
\begin{summary}[Compressible viscoelastic rate-type fluid with stress diffusion and temperature-dependent stress diffusion coefficient $\tilde{\mu}$; stress diffusion interpreted as an entropy producing mechanism]
  \label{summary:compressible-diff}
  Helmholtz free energy $\fenergy$, \emph{ansatz}~\eqref{eq:201}:
  \begin{subequations}
    \label{eq:244}
    \begin{align}
      \label{eq:245}
      \fenergy
      &=
      \widetilde{\fenergy} \left(\temp, \rho\right)
      +
      \frac{1}{\rho}\widetilde{\widetilde{\fenergy}},
      \\
      \widetilde{\widetilde{\fenergy}}
      &=
      \frac{\mu}{2}
      \left(
        \Tr \lcgnc
        -
        3
        -
        \ln \det \lcgnc
      \right).
    \end{align}
  \end{subequations}

  Entropy production $\entprodc$, \emph{ansatz}~\eqref{eq:213}, see~\eqref{eq:217} for notation:
  \begin{equation}
    \label{eq:246}
    \widetilde{\entprodc}
    =
    \frac{1}{\temp}
    \left(
      \frac{2 \nu + 3 \lambda}{3} \left(\divergence \vec{v}\right)^2
      +
      2 \nu \tensordot{\traceless{\gradsym}}{\traceless{\gradsym}}
      +
      \frac{\mu^2}{2\nu_1}
      \left(
        \Tr \lcgnc
        +
        \Tr \inverse{\lcgnc}
        -
        6
      \right)
      +
      \frac{\mu \tilde{\mu}(\temp)}{2\nu_1}
      \tensorddot{
        \nabla \lcgnc
      }
      {
        \nabla \lcgnc
      }
      +
      \frac
      {
        \kappa \absnorm{\nabla \temp}^2
      }
      {
        \temp
      }
    \right).
  \end{equation}

  Material parameters (constants): $\mu$, $\nu$, $\lambda$, $\nu_1$, $\kappa$.

  Material parameters (temperature-dependent): $\tilde{\mu}$.

  Evolution equations for $\rho$, $\vec{v}$, $\lcgnc$:
  \begin{subequations}
    \label{eq:247}
    \begin{align}
      \label{eq:248}
      \dd{\rho}{t} + \rho \divergence \vec{v} &= 0, \\
      \label{eq:249}
      \rho \dd{\vec{v}}{t} &= \divergence \cstress + \rho \vec{b}, \\
      \label{eq:250}
      \nu_1 \fid{\overline{\lcgnc}}
      +
      \mu \left(\lcgnc - \identity\right)
      &=
      \frac{1}{2}
      \bigg[
        \divergence \left( \tilde{\mu} \nabla \lcgnc \right)
        \lcgnc
        +
        \lcgnc
        \divergence \left( \tilde{\mu} \nabla \lcgnc \right)
      \bigg].
    \end{align}

    Evolution equation for $\temp$:
    \begin{multline}
      \label{eq:251}
      \rho
      \cheatvolNSE
      \dd{\temp}{t}
      +
      \temp
      \pd{\thpressuredMdiff}{\temp}
      \divergence{\vec{v}}
      =
      \divergence \left(\kappa \nabla \temp \right)
      +
      \frac{2 \nu + 3 \lambda}{3} \left(\divergence \vec{v}\right)^2
      +
      2 \nu \tensordot{\traceless{\gradsym}}{\traceless{\gradsym}}
      \\
      +
      \frac{\mu^2}{2\nu_1}
      \left(
        \Tr \lcgnc
        +
        \Tr \inverse{\lcgnc}
        -
        6
      \right)
      +
      \frac{\mu \tilde{\mu}(\temp)}{2\nu_1}
      \tensorddot{
        \nabla \lcgnc
      }
      {
        \nabla \lcgnc
      }.
    \end{multline}
  \end{subequations}

  Constitutive relation for the Cauchy stress tensor $\cstress$:
  \begin{equation}
    \label{eq:252}
    \cstress = \mns \identity + \traceless{\cstress}.
  \end{equation}

  Definitions of $\mns$ and $\traceless{\cstress}$:
  \begin{subequations}
    \label{eq:253}
    \begin{align}
      \label{eq:254}
      \traceless{\cstress}
      &=
      2 \nu \traceless{\gradsym}
      +
      \mu \traceless{\left( \lcgnc \right)}
      ,
      \\
      \label{eq:255}
      \mns
      &=
      -
      \thpressuredMdiff
      +
      \frac{\mu}{3}
      \Tr \lcgnc
      -
      \mu
      +
      \frac{2 \nu + 3 \lambda}{3} \divergence \vec{v}.
    \end{align}

  Quantities derived from the Helmholtz free energy:
  \begin{align}
    \label{eq:256}
    \thpressureNSE &= \rho^2 \pd{\widetilde{\fenergy}}{\rho}, \\
    \label{eq:257}
    \thpressuredMdiff &= \thpressureNSE - \widetilde{\widetilde{\fenergy}},\\
    \label{eq:258}
    \cheatvolNSE
    &=
    -
    \temp
    \ppd{
      \widetilde{\fenergy}
    }{\temp}.
  \end{align}
\end{subequations}

Constitutive relation for the energy flux $\efluxc$, has been already used in~\eqref{eq:251}:
\begin{equation}
  \label{eq:259}
  \efluxc
  =
  -
  \kappa
  \nabla \temp
  -
  \frac{\mu\tilde{\mu}}{4\nu_1}
  \Tr
  \bigg[
  \left(\nabla \lcgnc \right)
  \left(
    \lcgnc
    -
    \identity
  \right)
  +
  \left(
    \lcgnc
    -
    \identity
  \right)
  \left(\nabla \lcgnc \right)
  \bigg].
\end{equation}

Constitutive relation for the entropy flux $\entfluxc$:
\begin{equation}
  \label{eq:260}
  \entfluxc
  =
  -
  \frac{
    \kappa \nabla \temp
  }
  {
    \temp
  }.
\end{equation}
\end{summary}


%% file: text/summary-d.tex
\begin{summary}[Incompressible viscoelastic rate-type fluid with stress diffusion and temperature-dependent stress diffusion coefficient $\tilde{\mu}$; stress diffusion interpreted as an entropy producing mechanism]
  \label{summary:incompressible-diff}
  Helmholtz free energy $\fenergy$, incompressible variant of \emph{ansatz}~\eqref{eq:201}, see~\eqref{eq:261}:
  \begin{subequations}
    \label{eq:270}
    \begin{align}
      \label{eq:271}
      \fenergy
      &=
      \widetilde{\fenergy} \left(\temp\right)
      +
      \frac{1}{\rho}\widetilde{\widetilde{\fenergy}},
      \\
      \widetilde{\widetilde{\fenergy}}
      &=
      \frac{\mu}{2}
      \left(
        \Tr \lcgnc
        -
        3
        -
        \ln \det \lcgnc
      \right).
    \end{align}
  \end{subequations}

  Entropy production $\entprodc$, incompressible variant of~\emph{ansatz}~\eqref{eq:262}, see~\eqref{eq:261}, notation explained in~\eqref{eq:217}:
  \begin{equation}
    \label{eq:272}
    \widetilde{\entprodc}
    =
    \frac{1}{\temp}
    \left(
      2 \nu \tensordot{\traceless{\gradsym}}{\traceless{\gradsym}}
      +
      \frac{\mu^2}{2\nu_1}
      \left(
        \Tr \lcgnc
        +
        \Tr \inverse{\lcgnc}
        -
        6
      \right)
      +
      \frac{\mu \tilde{\mu}(\temp)}{2\nu_1}
      \tensorddot{
        \nabla \lcgnc
      }
      {
        \nabla \lcgnc
      }
      +
      \frac
      {
        \kappa \absnorm{\nabla \temp}^2
      }
      {
        \temp
      }
    \right).
  \end{equation}

  Material parameters (constants): $\mu$, $\nu$, $\lambda$, $\nu_1$, $\kappa$.

  Material parameters (temperature-dependent): $\tilde{\mu}$.

  Evolution equations for $\mns$, $\vec{v}$, $\lcgnc$ and $\temp$:
  \begin{subequations}
    \label{eq:273}
    \begin{align}
      \label{eq:274}
      \divergence \vec{v} &= 0, \\
      \label{eq:275}
      \rho \dd{\vec{v}}{t} &= \divergence \cstress + \rho \vec{b}, \\
      \label{eq:276}
      \nu_1 \fid{\overline{\lcgnc}}
      +
      \mu \left(\lcgnc - \identity\right)
      &=
      \frac{1}{2}
      \bigg[
      \divergence \left( \tilde{\mu} \nabla \lcgnc \right)
      \lcgnc
      +
      \lcgnc
      \divergence \left( \tilde{\mu} \nabla \lcgnc \right)
      \bigg],
      \\
      \label{eq:277}
      \rho
      \cheatvolNSE
      \dd{\temp}{t}
      &=
      \divergence \left(\kappa \nabla \temp \right)
      +
      2 \nu \tensordot{\traceless{\gradsym}}{\traceless{\gradsym}}
      +
      \frac{\mu^2}{2\nu_1}
      \left(
        \Tr \lcgnc
        +
        \Tr \inverse{\lcgnc}
        -
        6
      \right)
      +
      \frac{\mu \tilde{\mu}(\temp)}{2\nu_1}
      \tensorddot{
        \nabla \lcgnc
      }
      {
        \nabla \lcgnc
      }.
      \end{align}
  \end{subequations}

  Constitutive relation for the Cauchy stress tensor $\cstress$:
  \begin{equation}
    \label{eq:278}
    \cstress = \mns \identity + \traceless{\cstress}.
  \end{equation}

  Definitions of $\traceless{\cstress}$:
  \begin{equation}
    \label{eq:279}
    \traceless{\cstress}
    =
    2 \nu \traceless{\gradsym}
    +
    \mu \traceless{\left( \lcgnc \right)}.
  \end{equation}

  Quantities derived from the Helmholtz free energy:
  \begin{equation}
    \label{eq:280}
    \cheatvolNSE
    =
    -
    \temp
    \ppd{
      \widetilde{\fenergy}
    }{\temp}.
  \end{equation}

Constitutive relation for the energy flux $\efluxc$, has been already used in~\eqref{eq:277}:
\begin{equation}
  \label{eq:281}
  \efluxc
  =
  -
  \kappa
  \nabla \temp
  -
  \frac{\mu\tilde{\mu}}{4\nu_1}
  \Tr
  \bigg[
  \left(\nabla \lcgnc \right)
  \left(
    \lcgnc
    -
    \identity
  \right)
  +
  \left(
    \lcgnc
    -
    \identity
  \right)
  \left(\nabla \lcgnc \right)
  \bigg].
\end{equation}

Constitutive relation for the entropy flux $\entfluxc$:
\begin{equation}
  \label{eq:282}
  \entfluxc
  =
  -
  \frac{
    \kappa \nabla \temp
  }
  {
    \temp
  }.
\end{equation}
\end{summary}


%% file: text/models-table-summary.tex
\begin{table}[h]
  \centering
  \begin{tabular}{lllll}  
    \toprule
    Model  & Helmholtz free energy & Entropy production & Governing equations & Overview\\
    \midrule
    \multicolumn{4}{l}{\emph{Nonstandard energy storage models, Section~\ref{sec:expl-deriv-cons}}} \\[0.5em]
    Compressible Maxwell/Oldroyd-B & \eqref{eq:12} &  \eqref{eq:22} 
                                                        & \eqref{eq:113} and \eqref{eq:117} & Summary~\ref{summary:compressible-Tr-alt}\\
    Incompressible Maxwell/Oldroyd-B & \eqref{eq:142} & \eqref{eq:143} & \eqref{eq:168} & Summary~\ref{summary:incompressible-Tr-alt} \\[0.5em]
    \multicolumn{4}{l}{\emph{Nonstandard entropy production models, Section~\ref{sec:deriv-cons-relat}}} & \\[0.5em]
    Compressible Maxwell/Oldroyd-B & \eqref{eq:13} & \eqref{eq:23} 
                                                        & \eqref{eq:236} & Summary~\ref{summary:compressible-diff} \\
    Incompressible Maxwell/Oldroyd-B & \eqref{eq:261} & \eqref{eq:262} & \eqref{eq:263} & Summary~\ref{summary:incompressible-diff} \\
    \bottomrule
  \end{tabular}
  \bigskip
  \caption{List of thermomechanical models for viscoelastic rate type fluids with stress diffusion.}
  \label{tab:models}
\end{table}
